\documentclass[review]{elsarticle}
\graphicspath{ {./figures/} }
\usepackage{float}
\usepackage{verbatim} 
\usepackage{apalike}
\restylefloat{figure}
\floatstyle{plaintop} 
\restylefloat{table}

\usepackage{lineno}
\modulolinenumbers[5]

\usepackage[colorlinks = true,
            linkcolor = blue,
            urlcolor  = blue,
            citecolor = blue,
            anchorcolor = blue]{hyperref}

\usepackage{enumitem}
\usepackage{graphicx}
\usepackage{caption}
\usepackage{subfig}
\usepackage{multirow}
\usepackage{multicol}
\usepackage{makecell}
\usepackage{longtable}
\usepackage{lipsum}
\usepackage{url}
\usepackage{booktabs}
\usepackage[table,xcdraw]{xcolor}
\usepackage{makecell}
\usepackage{mathtools}
\usepackage[left=2.5cm, right=2.5cm, top=2.5cm]{geometry}

\newcolumntype{L}[1]{>{\raggedright\let\newline\\\arraybackslash\hspace{0pt}}m{#1}}
\newcolumntype{C}[1]{>{\centering\let\newline\\\arraybackslash\hspace{0pt}}m{#1}}
\newcolumntype{R}[1]{>{\raggedleft\let\newline\\\arraybackslash\hspace{0pt}}m{#1}}

\journal{Computers \& Chemical Engineering}

\biboptions{authoryear}

\begin{document}
\begin{frontmatter}











\title{Lessons Learned from Deploying Adaptive Machine Learning Agents with Limited Data for Real-time Cell Culture Process Monitoring}

\author[uts]{Thanh Tung Khuat \corref{cor1}}
\ead{thanhtung.khuat@uts.edu.au}

\author[uts]{Johnny Peng}
\ead{johnny.peng@student.uts.edu.au}

\author[csl]{Robert Bassett}
\ead{robert.bassett@csl.com.au}

\author[csl]{Ellen Otte}
\ead{ellen.otte@csl.com.au}

\author[uts]{Bogdan Gabrys}
\ead{bogdan.gabrys@uts.edu.au}

\cortext[cor1]{Corresponding author.}
\address[uts]{Complex Adaptive Systems Laboratory, The Data Science Institute, University of Technology Sydney, NSW 2007, Australia}
\address[csl]{CSL Innovation, Melbourne, VIC 3000, Australia}

\begin{abstract}
This study explores the deployment of three machine learning (ML) approaches for real-time prediction of glucose, lactate, and ammonium concentrations in cell culture processes, using Raman spectroscopy as input features. The research addresses challenges associated with limited data availability and process variability, providing a comparative analysis of pretrained models, just-in-time learning (JITL), and online learning algorithms. Two industrial case studies are presented to evaluate the impact of varying bioprocess conditions on model performance. The findings highlight the specific conditions under which pretrained models demonstrate superior predictive accuracy and identify scenarios where JITL or online learning approaches are more effective for adaptive process monitoring. This study also highlights the critical importance of updating the deployed models/agents with the latest offline analytical measurements during bioreactor operations to maintain the model performance against the changes in cell growth behaviours and operating conditions throughout the bioreactor run. Additionally, the study confirms the usefulness of a simple mixture-of-experts framework in achieving enhanced accuracy and robustness for real-time predictions of metabolite concentrations based on Raman spectral data. These insights contribute to the development of robust strategies for the efficient deployment of ML models in dynamic and changing biomanufacturing environments.
\end{abstract}


\end{frontmatter}

\section{Introduction} \label{introduction}
The biopharmaceutical industry plays a critical role in addressing the global burden of diseases by developing life-saving therapies such as vaccines, monoclonal antibodies (mAbs), and gene therapies. In recent years, the demand for high-quality and affordable biological products such as mAbs has surged, driven by the need to combat emerging diseases, manage chronic conditions, and improve overall public health \citep{pabu19, khba24}. As reported by \cite{gr22}, the global therapeutic monoclonal antibody market was valued at about US\$185.5 billion in 2021 and is expected to achieve an annual growth rate of 11.30\% from 2022 to 2030 to reach a revenue of around \$500 billion by the end of 2030. To meet these targets and challenges, the industry is increasingly focused on advancing manufacturing processes to enhance efficiency, ensure product consistency, and comply with stringent regulatory standards. Central to these efforts is the ability to monitor and control bioprocesses in real-time, ensuring optimal cell growth and product quality \citep{goum20}. Cell culture processes are inherently complex and dynamic, with critical process parameters like glucose, lactate, and ammonium concentrations directly influencing cell productivity. Raman spectroscopy has emerged as a transformative tool in this domain, offering non-invasive and real-time insights into bioprocesses \citep{wewo21, escu21}. However, the development of predictive models capable of accurately interpreting Raman spectra and adapting to evolving process conditions continues to be a significant challenge \citep{papa21}.

One of the primary obstacles in deploying ML agents for bioprocess monitoring is the limited availability of training data \citep{seva15, khba24a, pekh25a}. Bioreactor experiments are expensive and time-consuming, often creating datasets with insufficient variability to capture the full spectrum of process conditions. Moreover, bioprocesses are subject to both planned changes, such as adjustments to feeding strategies, and unplanned variations, such as anomalies in nutrient addition. These factors necessitate adaptive modeling approaches capable of generalising from limited historical data while responding to new conditions in real-time.

This study assesses the performance of three ML types for predicting critical bioprocess variables using Raman spectra: (1) pretrained models trained on data from 34 historical bioreactor runs, (2) online learning models pretrained on only one historical run and updated incrementally during the new run \citep{khba24b}, and (3) just-in-time learning (JITL) models that dynamically build new local models for each prediction using the most relevant historical data \citep{tusc19}. Two case studies are presented to evaluate these approaches under different conditions. The first case study involves a bioreactor run with the same base and feed culture media as the training data but includes a significant anomaly in glucose addition. The second case study uses two runs with completely new feed culture media and feeding strategies, resulting in distinct cell growth behaviours.

Findings from this study reveal key insights into the suitability of different ML approaches for various scenarios. Pretrained models demonstrate excellent performance for processes closely aligned with historical data, minimizing the need for real-time updates and daily collection of offline measurements. In contrast, JITL and online learning excel in adapting to novel conditions, particularly when new fill/feed media and feeding strategies are employed. Notably, JITL’s ability to rapidly adapt to new Raman samples makes it a valuable tool for dynamic bioprocess environments, where process variability is high. 

The contributions of this work are multifaceted. First, we provide a comparative analysis of three ML approaches under realistic bioprocess conditions, highlighting their strengths and limitations. Second, we offer practical guidelines for deploying ML models in biomanufacturing, emphasizing the importance of adaptability and efficient data utilisation. Third, we propose a simple but effective method to combine the predictive results from different types of ML models aiming to enhance the predictive accuracy of ML solutions in real-time monitoring of metabolite concentrations from cell culture bioreactors based on input Raman spectral data. Lessons learned from our study advance the application of ML in biomanufacturing, paving the way for more resilient and adaptable process monitoring systems.

The remainder of this paper is organised as follows. Section \ref{method} presents the machine learning approaches employed in this study, along with the design of the industrial case studies used to evaluate the performance of various adaptive ML types. The empirical outcomes of the individual ML agents for predicting and real-time monitoring of three cell culture process indicators including glucose, lactate, and ammonium concentrations across two case studies are provided in Section \ref{results_4_agents}. The results of combining all individual adaptive ML agents in the case studies are discussed in Section \ref{results_moe}. Section \ref{discussion} discusses the lessons learned from the deployment of adaptive ML agents for real-time monitoring of biochemical process indicators in industrial cell culture bioreactors. Finally, Section \ref{conclusion} summarises the findings of this study and outlines directions for future research. 
 
\section{Methodology} \label{method}
This section will briefly summarise the main ideas and steps of the various ML types used in our experiments, including online learning, just-in-time learning, and retraining of pretrained models. We also introduce the main characteristics of our practical case studies and metrics used to assess the performance of the different ML models.

\subsection{Introduction to machine learning approaches}
In this study, various ML approaches were developed using Raman spectral data as input features. The outputs of these ML models correspond to the predicted concentrations of key metabolites within the cell culture bioreactor, including \textit{glucose}, \textit{lactate}, and \textit{ammonium}. Separate ML models were constructed for each metabolite to enable precise predictions.

To enhance the robustness of pretrained models under diverse cell culture process conditions, generic Raman-based models are often trained using combined datasets generated across different process environments. This approach aims to ensure high predictive performance across various cell lines with different culture media and compositions \citep{mela15}. However, as highlighted by \cite{tusc19}, several challenges are associated with models pretrained on global datasets: (a) while increasing dataset variability enhances the model's robustness to different conditions, it often reduces overall predictive accuracy, and (b) global models struggle to adapt to abrupt process changes or abnormal operating conditions. Another significant limitation of generic Raman-based models lies in the degradation of pretrained and deployed models over time \citep{tukh18}. This degradation occurs due to the dynamic nature of cell culture processes, which frequently experience changes such as recipe modifications, raw material variability, and process drifts. These factors contribute to the gradual deterioration of pretrained models \citep{zhtu19}. Additionally, the performance of generic pretrained models is heavily based on the quality and representativeness of the Raman training datasets. When process conditions deviate substantially from those reflected in the training data, the pretrained models often exhibit suboptimal performance.

To overcome the challenges associated with the generic pretrained models in real-time monitoring of biomanufacturing processes using Raman spectral data, this study explores adaptive mechanisms for ML models. Specifically, we focus on enabling models to learn adaptively or to be updated using the latest offline analytical data collected during current bioreactor runs. Assessment of these approaches from the industrial case studies is the central objective of this research.

\subsubsection{Online machine learning} \label{online_ml_section}
\begin{figure}[!ht]
    \centering
    \includegraphics[width=0.85\linewidth]{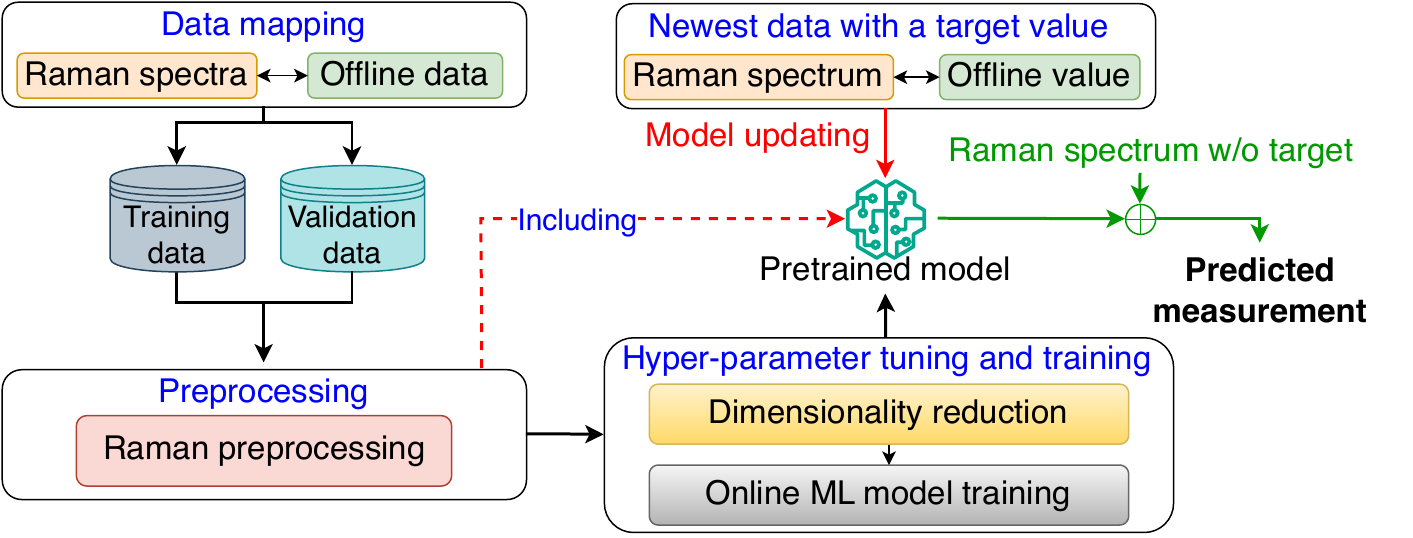}
    \caption{General workflow for online machine learning models from input Raman spectral data and offline analytical metabolites.}
    \label{online_ml}
\end{figure}

A recent study \citep{khba24b} has demonstrated that online learning models are capable of effectively learning from the limited training data while rapidly adapting pretrained models based on updated offline analytical measurements collected during bioreactor operations. Online (incremental) learning algorithms handle individual data instances sequentially, enabling real-time learning and adaptation. The key strength of this approach lies in its ability to adapt dynamically to variations in input data, such as shifts in cellular behaviours within bioreactors caused by changes in operating conditions and nutrient concentrations. However, a main drawback of online learning algorithms is their restricted view of the current state, as they rely on individual data instances, which may inadequately represent broader patterns or be affected by noise. Fortunately, offline analytical metabolite measurements in cell culture bioreactors are typically subjected to rigorous validation by well-trained operators, ensuring the accuracy of the data stored in the database. Consequently, updating the online learning algorithms based on these verified offline measurements is less susceptible to noisy data. The general workflow for training and updating Raman-based online machine learning models is illustrated in Fig. \ref{online_ml}.

Following the method proposed in \cite{khba24b}, this study employed online ML models pretrained on data from a single carefully selected bioreactor run, comprising approximately 30 samples. This pretraining strategy enabled the online algorithms to rapidly adapt to new operating conditions encountered in practical manufacturing scenarios, such as significant increases in nutrient concentrations or variations in base and feed culture media as well as feeding strategies, as demonstrated in two case studies in this paper. Additionally, data from another bioreactor run, using the same cell line and operating conditions, was used as validation data to fine-tune the hyperparameters of the online learning models.

To address the temporal mismatch between online Raman spectra (collected at intervals of 45 to 55 minutes) and offline analytical measurements (recorded twice daily), a mapping procedure was implemented. Each offline analytical measurement was paired with the nearest Raman spectrum collected after the offline sampling time. For analytical measurements taken immediately before feeding events, the Raman spectrum obtained prior to the feeding period was mapped to the respective offline measurement, because no Raman spectra were collected during feeding. In case of where two offline measurements were mapped to the same Raman spectrum, only the pair with the smallest time difference was retained. This mapping process was essential for building the training, validation, and testing datasets. The Raman data in datasets were then passed to the preprocessing phase.

Raw Raman spectra often exhibit noise stemming from environmental influences, instrumentation variability, or sample-specific conditions. To address these challenges, appropriate preprocessing techniques are essential for noise reduction, signal enhancement, and the extraction of meaningful spectral information. In this work, spectral regions between 500 and 3000 $cm^{-1}$ were chosen to exclude silent regions, highly variable spectral slopes, window peaks, and artificial jumps \citep{miwa23,poma22}. The Savitzky-Golay smoothing and derivative filter was then applied, using a filter window size of 25, a polynomial order of 2, and the first derivative. This method effectively attenuates high-frequency noise through its smoothing properties while eliminating low-frequency components, such as offsets and slopes, via differentiation \citep{sago64}. Additionally, the Savitzky-Golay filter helps enhance small spectral contributions \citep{miwa23}. Finally, Standard Normal Variate (SNV) normalisation was applied to each spectrum to minimize inter-batch variability and ensure consistency.

The preprocessed Raman spectra were then employed to train a regressor for each metabolite using online learning algorithms. The training process involved optimising hyperparameters through hold-out validation, applied to both the dimensionality reduction step and the machine learning model. Given the limited availability of training data with a single bioreactor run (about 30 samples) and a similar dataset for validation, dimensionality reduction was crucial to transform the 2500 input features into a lower-dimensional space. In this study, Kernel Principal Component Analysis (KPCA) was deployed for feature extraction, as it is able to capture both linear and nonlinear features effectively \citep{sulv19}. The hyperparameters of KPCA, including the kernel coefficient and the number of components, were jointly optimised with the hyperparameters of the learning algorithms. However, for Partial Least Squares Regression (PLSR), which inherently incorporates dimensionality reduction within its learning process, KPCA was not applied. As highlighted by \cite{koko22}, support vector machines are among the most widely adopted ML models for small datasets. Similarly, PLSR has been extensively utilised for real-time monitoring of critical process parameters and key performance indicators in cell culture bioreactors using Raman spectra \citep{khba24}. As a result, this study employed online support vector regression (OSVR) \citep{math03} and recursive partial least squares regression (RPLSR) \citep{qi98} as online learning algorithms to train the regressors. Hyperparameter tuning for these online learning algorithms, as well as the dimensionality reduction technique, was conducted on the validation set using Bayesian optimisation over 100 iterations with the \textit{Optuna} library \citep{aksa19}. The hyperparameter configurations generating the highest accuracy, defined as the minimum normalised mean absolute error, on the validation set were then applied to train the models on the training set, resulting in the final pretrained models.

In contrast to pretrained batch learning models, pretrained online ML models possess the distinct advantage of dynamically updating their hyperparameters based on newly incoming observations. When offline analytical measurements from the current experimental bioreactor become available, they are mapped to the corresponding Raman spectra. These pairs of input features and target variables are subsequently employed to update the pretrained online learning model during the operational phase of the cell culture bioreactor. As part of this process, the Raman input features will use the same preprocessing and dimensionality reduction procedures as in the training phase. The up-to-dated online learning model is then used to predict glucose, lactate, and ammonium in real-time based on the input Raman spectra. This approach facilitates real-time monitoring of the cell culture process, enabling to enhance the process adaptability and robustness.

\subsubsection{Just-in-time learning}
\begin{figure}[!ht]
\centering
\begin{subfloat}[without updating \label{jitl_wo_update}]{
\includegraphics[width=0.85\textwidth]{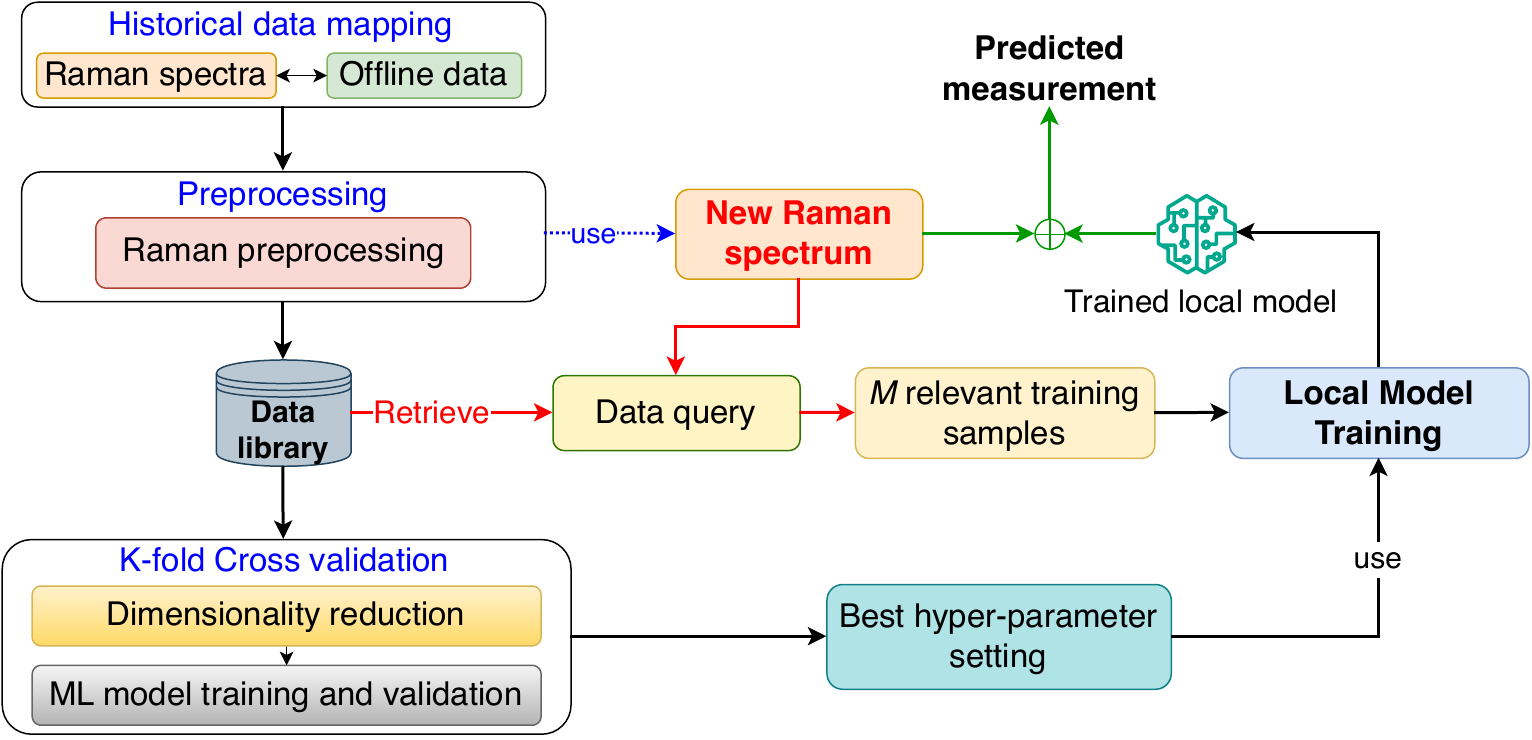}}
\end{subfloat}
\begin{subfloat}[with updating \label{jitl_with_update}]{
\includegraphics[width=0.85\textwidth]{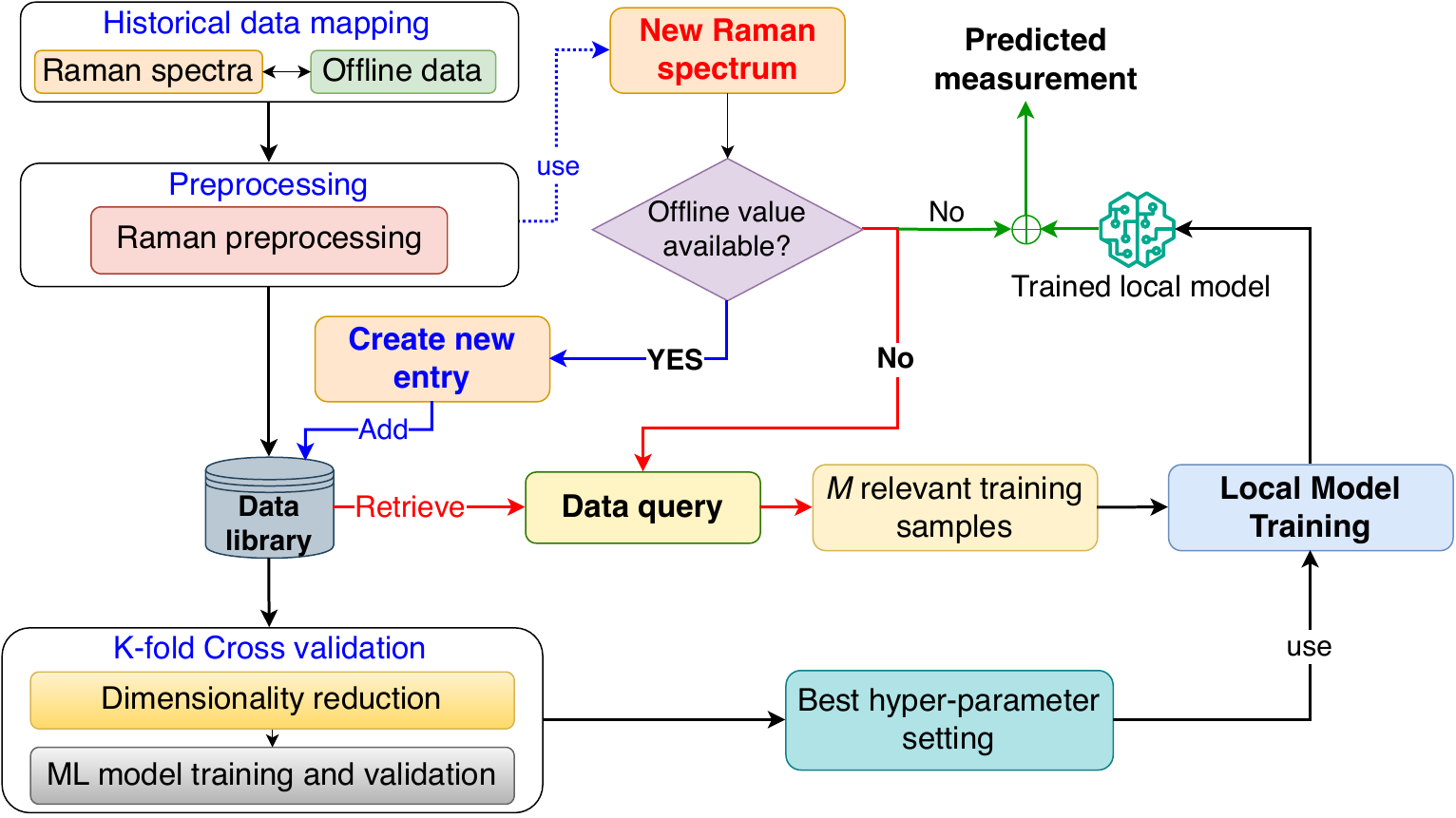}}
\end{subfloat}
\caption{General workflow of Just-in-time learning.} \label{jitl_framework}
\end{figure}

As highlighted by \cite{rakh24}, online ML models offer the capability to address the degradation of model performance by dynamically updating hyperparameters in response to the latest observations from the current bioreactor. However, these models are often inadequate in handling abrupt changes, such as variations in environmental conditions or nutrient availability encountered in industrial processes. Such sudden changes can result in cellular stress, reduce productivity, and significant losses in time and resource. To overcome the limitations of traditional Raman-based models, the just-in-time learning approach has been proposed \citep{tusc19} and its robust, ensemble based version have recently been proposed in the context of robust cell culture process monitoring \citep{pekh25b}. JITL is an adaptive learning technique that selects and learns from a subset of the most relevant historical data. Unlike traditional batch learning methods that rely on pretrained models developed from extensive datasets, JITL constructs a localised, instantaneous model by using a smaller, highly relevant subset of data corresponding to a new Raman scan (query point). This approach is particularly suitable in dynamic environments characterised by continuous data evolution and limited computational resources \citep{kaga10}. As a result, JITL has often been used as a robust solution for developing soft sensors for bioprocesses, offering a competitive alternative to other adaptive and incremental soft sensors reported in the literature for various applications within process industries \citep{kaga09,kagr11}. The primary steps of the JITL workflow are shown in Fig. \ref{jitl_wo_update}.

In this study, a diverse data library was constructed, comprising Raman spectra and corresponding offline analytical measurements from 34 historical bioreactor runs. The same methodologies for acquiring online Raman spectra, offline analytical data, and Raman preprocessing as detailed in Subsection \ref{online_ml_section} were adopted to ensure consistency. From the created data library, a 5-fold cross-validation approach with Bayesian optimisation in the \textit{Optuna} library \citep{aksa19} was employed to identify the optimal hyperparameter configurations for the combined dimensionality reduction and ML models. Our benchmarking analysis revealed that PLSR was the most effective method for constructing predictive models for glucose. For lactate prediction, a combination of Principal Component Analysis (PCA) and Support Vector Regression (SVR) demonstrated superior performance, while KPCA combined with SVR was determined to be the best approach for ammonium prediction. These optimised hyperparameter settings were then employed to train a local model for each query Raman sample.

The selection of the most similar historical samples for each preprocessed input Raman query is typically determined by the similarity between the query point and the features of the historical data. In this study, Euclidean distance \( d(\mathbf{x}_1, \mathbf{x}_2) \) will be used to measure the closeness between two data points, which is calculated as in Eq.~\eqref{eq1}:
\begin{equation}\label{eq1}
    d(\mathbf{x}_1, \mathbf{x}_2) = \sqrt{\sum_{i=1}^{n} (x_{1i} - x_{2i})^2}
\end{equation}
where $\mathbf{x}_1$ and $\mathbf{x}_2$  are two data points represented as n-dimensional vectors:
\[
\mathbf{x}_1 = (x_{11}, x_{12}, \dots, x_{1n})
\]
\[
\mathbf{x}_2 = (x_{21}, x_{22}, \dots, x_{2n})
\]
After computing the Euclidean distance, the obtained value is transformed into a similarity score within the range \( (0, 1] \) using the following equation:
\begin{equation}
    \text{\textbf{similarity}}(\mathbf{x}_1, \mathbf{x}_2) = e^{-d(\mathbf{x}_1, \mathbf{x}_2)}
\end{equation}
$M$ historical data points ($M=30$ in this study) from the library with the highest similarity scores were selected to train a local (single-use) model for each metabolite measurement using the optimal hyperparameter configurations. This local model was subsequently used to predict the Raman query point. 

While the JITL scheme provides significant flexibility in adapting models to abrupt process changes or disturbances for each query point, its reliance on a static data library containing available operating conditions may deteriorate model performance over time due to evolving operating conditions, such as variability in raw materials, cell lines, media compositions, feeding strategies, or nutrient concentrations. Collecting sufficiently diverse datasets to cover nearly all possible operating conditions is challenging in practice, thereby hindering the robustness and generalizability of local models. To address this limitation, the JITL workflow can be enhanced by incorporating a dynamic library that is continually updated with new analytical measurements acquired during the bioreactor run \citep{tuwa20}. This approach enables local models to be trained using the most recent measurements, ensuring they more accurately represent the current process conditions. A schematic flowchart presenting the JITL framework with dynamic updating is provided in Fig. \ref{jitl_with_update}.

\subsubsection{Retraining}
To evaluate the efficiency and applicability of online learning and just-in-time learning paradigms in industrial case studies, a competing pretrained model was developed using batch learning. This model was trained on a data library comprising 34 historical bioreactor runs, including a diverse range of cell lines, culture medium compositions, and feeding strategies. The input features contained only Raman spectral data, while the outputs were glucose, lactate, or ammonium concentrations. Details regarding the mapping between online Raman features and offline analytical measurements, as well as Raman preprocessing procedures, were provided in the previous subsection.

\begin{figure}[!ht]
\centering
\begin{subfloat}[Pretraining \label{pretrain}]{
\includegraphics[width=0.75\textwidth]{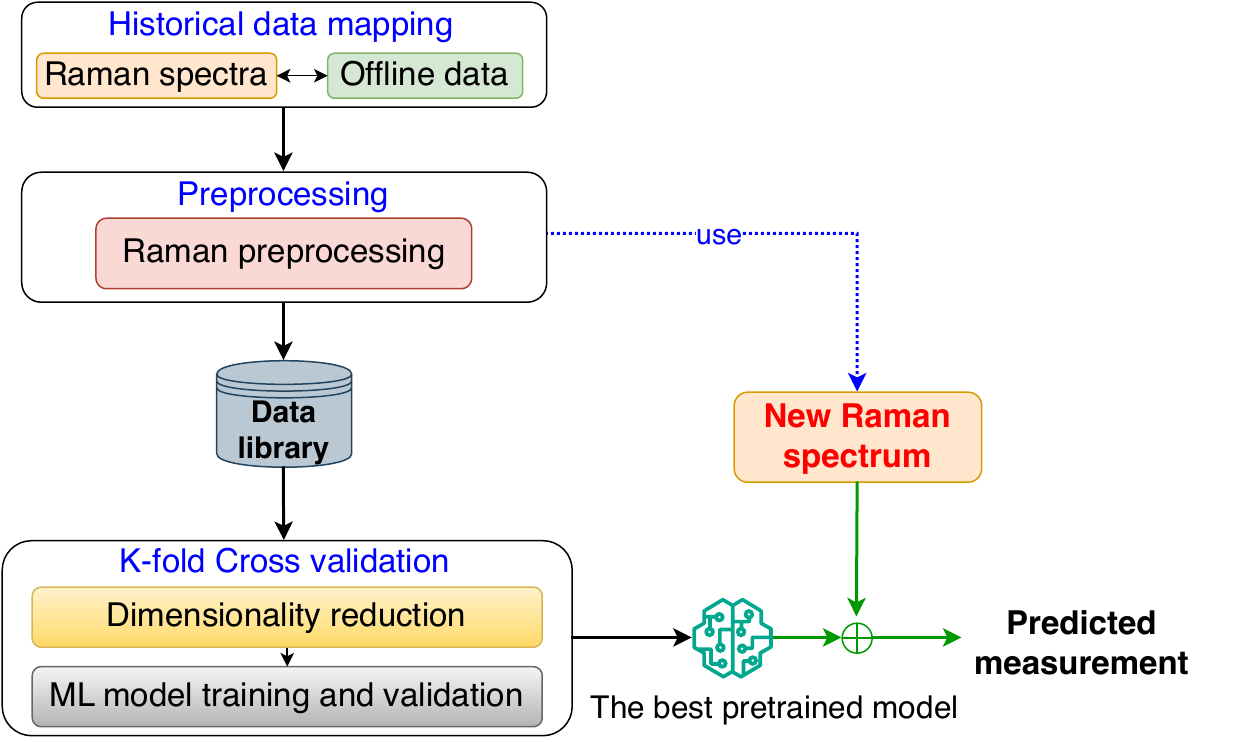}}
\end{subfloat}
\begin{subfloat}[Retraining \label{retrain}]{
\includegraphics[width=0.8\textwidth]{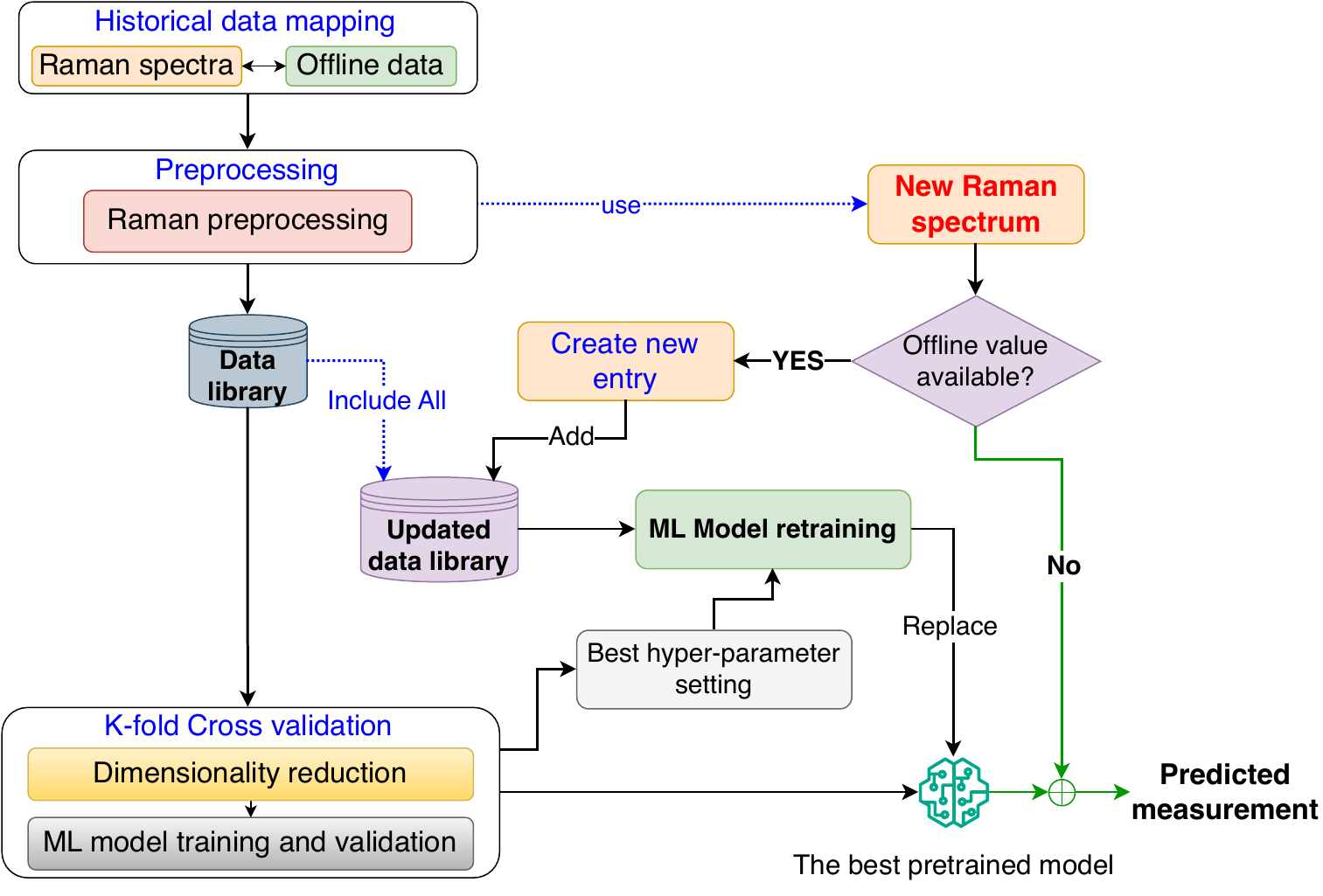}}
\end{subfloat}
\caption{General workflow of pretraining a ML model on all historical data and retraining the model using all historical data and data from the current run.} \label{retraining}
\end{figure}

Similar to the just-in-time learning approach, the pretrained PLSR model was utilised for predicting glucose concentrations. For lactate, a pipeline combining PCA and SVR served as the pretrained model, while KPCA and SVR were used for ammonium prediction. Hyperparameter configurations were optimised through 5-fold cross-validation with the \textit{Optuna} library \citep{aksa19} using 100 iterations and then used to train the learning pipeline on the complete dataset in the library. The primary steps involved in pretraining Raman-based models are depicted in the flow diagram in Fig. \ref{pretrain}.

The performance of pretrained models is often deteriorated when applied to real-time monitoring of manufacturing bioprocesses operating under conditions that deviate from the training data. To address this limitation, we proposed retraining the deployed models using an updated data library containing both the 34 historical bioreactor runs and newly acquired offline analytical measurements from the current bioreactor run. To accelerate the retraining process during operational time, the same optimal hyperparameter configurations identified during the offline fine-tuning process were employed to train the batch learning algorithms on the updated data library. The retrained models were then used to predict metabolite concentrations based on the input Raman spectra. A detailed schematic of the retraining procedure is shown in Fig. \ref{retrain}.

\subsection{Mixture-of-experts solution}
In the industrial case studies presented in this research, four distinct adaptive machine learning agents were deployed including recursive PLSR, online SVR, just-in-time learning models, and ML models retrained during operational time. As highlighted by \cite{ruga05}, an ensemble of diverse learners can outperform individual models. This raises the question of whether the strengths of these different ML approaches can be leveraged to enhance the accuracy of real-time predictions of metabolite concentrations based on input Raman spectral data.

\begin{figure}[!ht]
    \centering
    \includegraphics[width=0.7\linewidth]{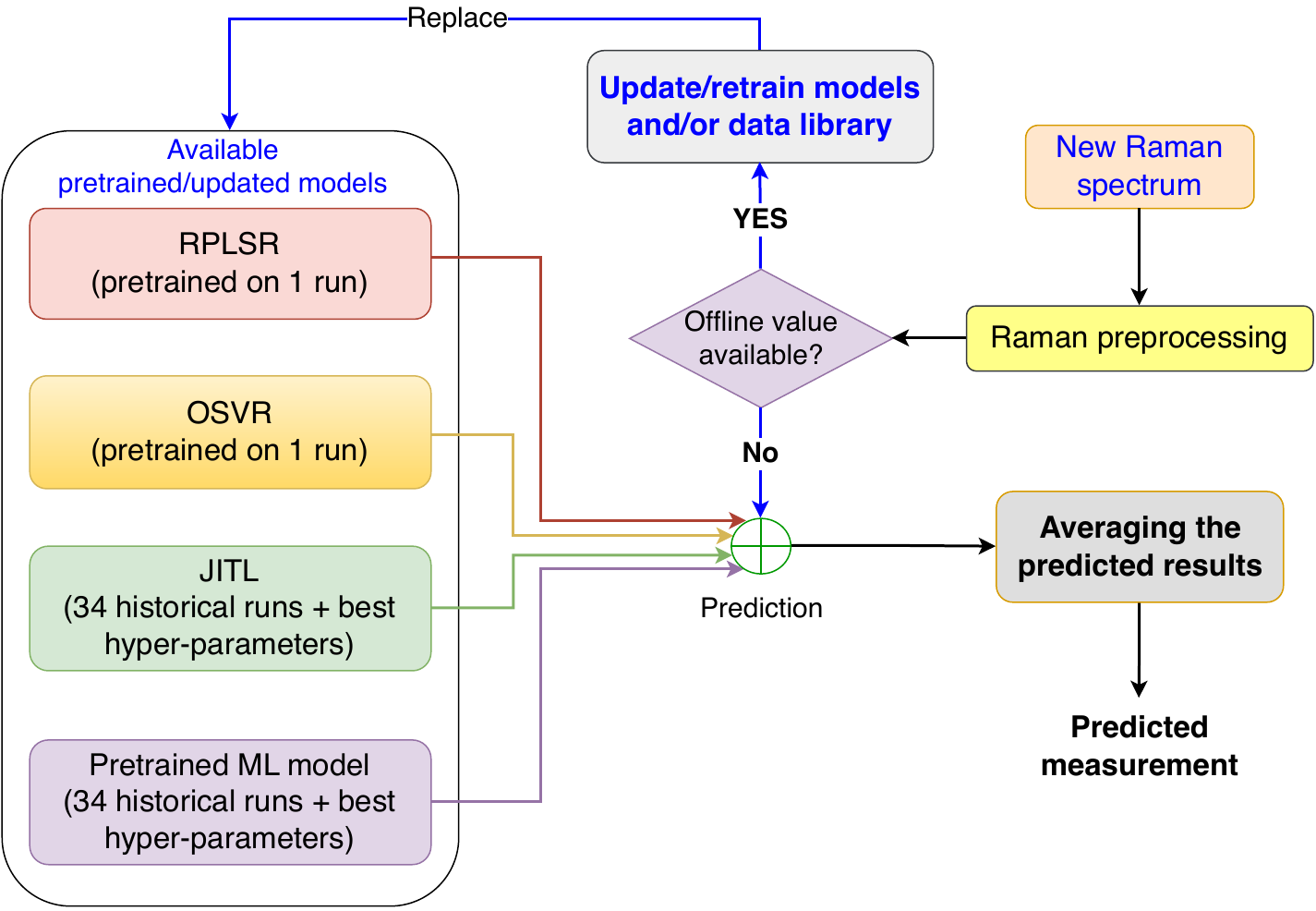}
    \caption{General workflow for the mixture-of-experts approach.}
    \label{moe_workflow}
\end{figure}

To address this question, a simple mixture-of-experts solution was designed. In the proposed method, the final predictive value for each input Raman spectrum is determined as the mean of the predictions from all four ML agents. For input Raman spectra associated with newly collected offline analytical measurements, these pairs of training samples are used to update all four component learning agents. This approach allows us to evaluate the feasibility of using all available deployed models to improve the accuracy of real-time predictions for infrequently collected metabolite concentrations. The details of the mixture-of-experts method are illustrated in Fig. \ref{moe_workflow}.

\subsection{Case study design}\label{case_study}
To evaluate the performance of the adaptive ML models described in the previous subsection, two practical industrial case studies involving cell cultivation within bioreactors were conducted. The first case study focused on a bioreactor run using the same base and feed culture media as the training data. However, an incident during the glucose addition process resulted in glucose concentrations that were approximately four times higher than standard values. This case study aimed to determine whether significant deviations (outliers) in nutrient concentrations from the training data affect the performance of the pretrained models and to assess whether the JITL and online learning models can rapidly adapt to abrupt changes in the target variable after updated with the newly collected values. 

\begin{figure}[!ht]
\centering
\begin{subfloat}[Glucose \label{glucose_casestudy}]{
\includegraphics[width=0.48\textwidth]{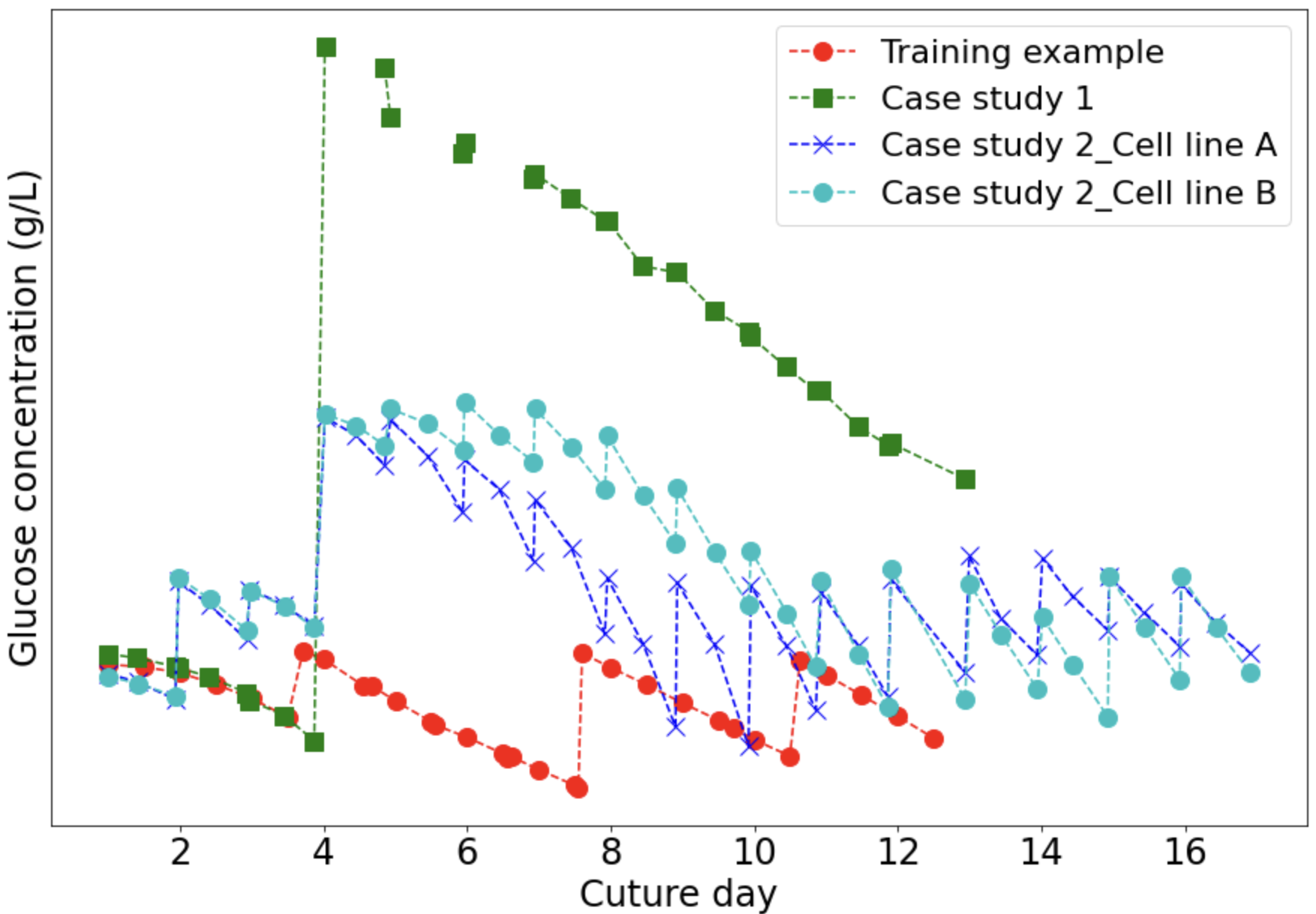}}
\end{subfloat}
\begin{subfloat}[Lactate \label{lactate_casestudy}]{
\includegraphics[width=0.48\textwidth]{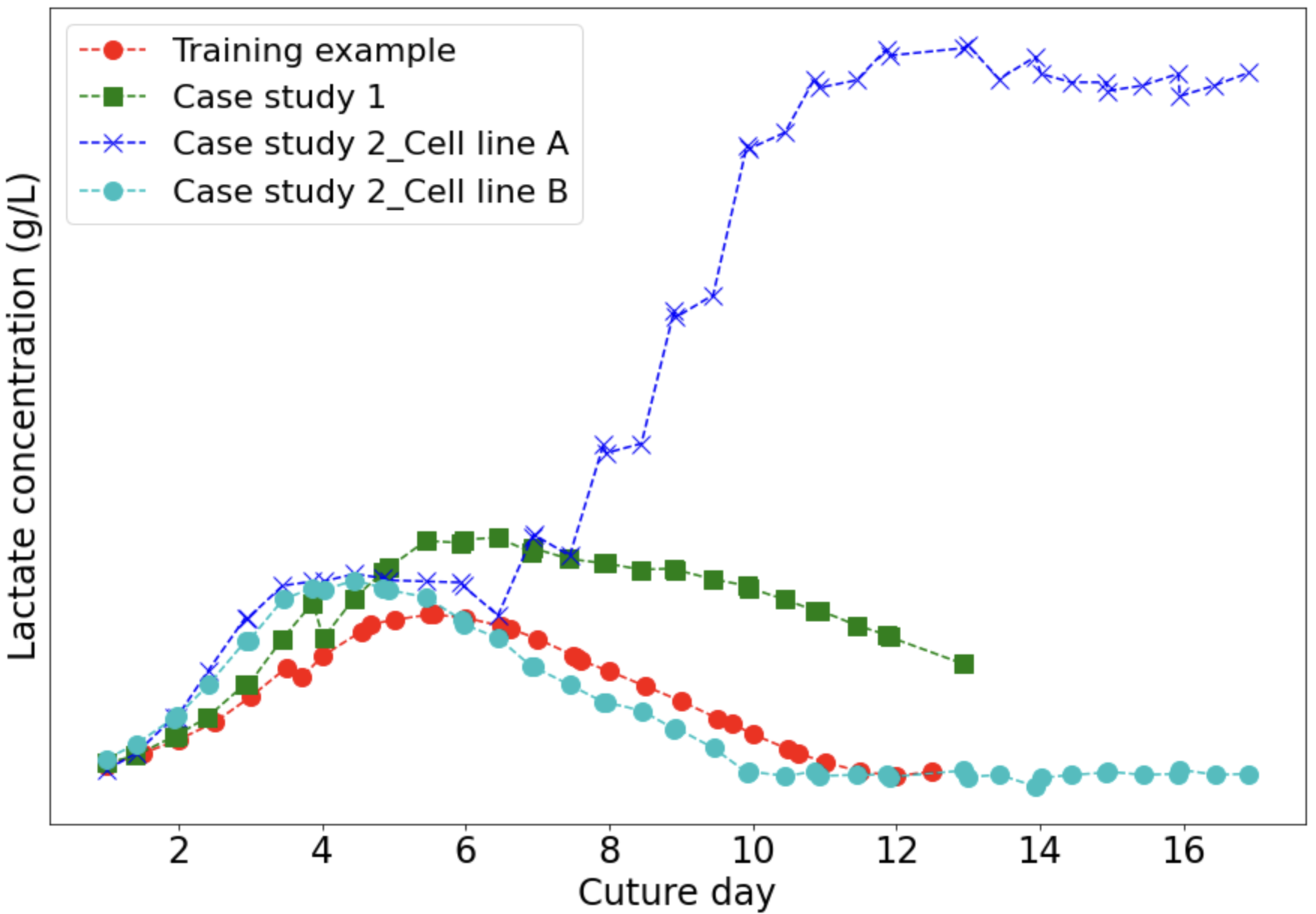}}
\end{subfloat}
\begin{subfloat}[Ammonium \label{ammonium_casestudy}]{
\includegraphics[width=0.48\textwidth]{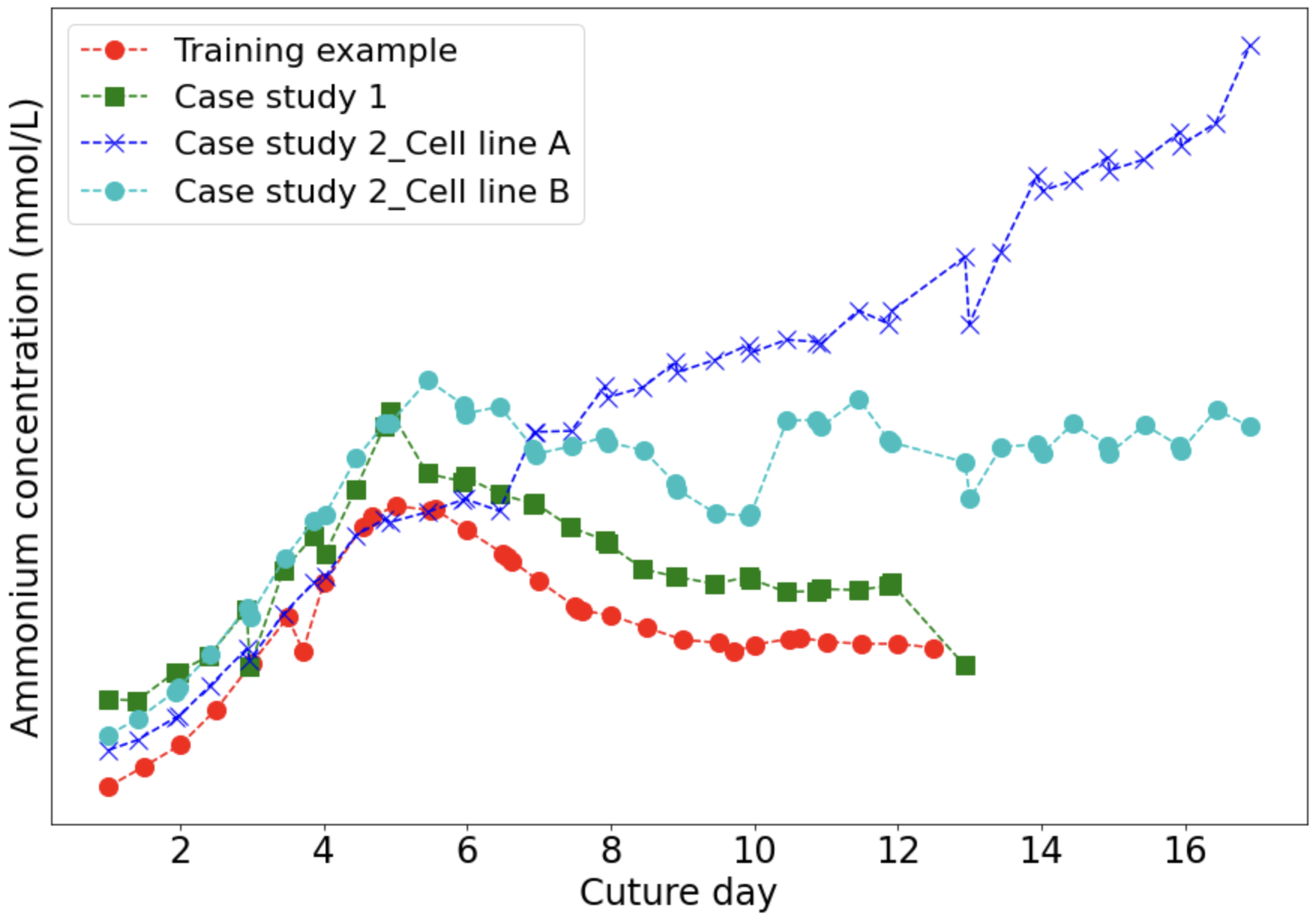}}
\end{subfloat}
\caption{The details of offline analytical measurements in case studies.} \label{case_study_img}
\end{figure}

The second case study comprises two bioreactor runs employing distinct base and feed culture media, as well as differing feeding strategies (including variations in glucose control levels and feeding frequencies) compared to the training data. In one run (cell line A), a significant increase in lactate concentrations was observed, whereas the other run (cell line B) exhibited a dramatic drop in lactate concentrations toward the end of the cell culture process. Additionally, the culture duration in this case study extended by more than four days compared to both the first case study and the training data. The main objective of this case study is to evaluate the performance of pretrained ML models, online learning models, and JITL models under conditions that represent substantial changes in operating parameters in comparison to the training data. This investigation aims to determine whether learning mechanisms capable of real-time model updates are more suitable for dynamic and evolving operational environments. Furthermore, the findings from this case study will provide valuable insights for managers regarding the impact of base and feed culture media on selecting appropriate Raman-based ML models for real-time monitoring of manufacturing bioreactors. A representation of offline analytical measurements from both case studies, together with an example of measurements from the training bioreactor used to train online ML models, is provided in Fig. \ref{case_study_img}. 

\subsection{Evaluation metrics}
Although metrics such as mean absolute percentage error (MAPE), which quantify the percentage deviation of predicted values from actual values, have practical implications and are straightforward for bioreactor operators to understand, MAPE was not used to assess the performance of ML models in this study. This is because metabolite concentrations, such as lactate and ammonia, often include many values much smaller than 1. In such cases, even a small difference between a predicted and an actual value can result in disproportionately large percentage errors, leading to unfair comparisons among ML methods. For instance, if the predicted value is 0.2 and the actual value is 0.1, the absolute percentage error is 100\%, despite the absolute difference being only 0.1. Conversely, for the same absolute difference of 0.1, a predicted value of 1.1 and an actual value of 1 would yield an absolute percentage error of just 10\%.

Another commonly used metric in regression problems is the mean absolute error (MAE). However, MAE is also unsuitable for comparing ML model performance across bioreactors with the same metabolite but different value ranges. For example, as shown in Fig. \ref{lactate_casestudy}, the lactate concentration range in case study 2 for cell line A is approximately four times higher than that for cell line B. Consequently, if an ML model produces the same MAE for both cell lines, it is evident that its performance on cell line A is superior to that on cell line B. To ensure a fair comparison across different metabolite measurements, this study employed normalised MAE (NMAE) as the evaluation metric, as defined in Eq.~\eqref{eq3}:
\begin{equation}\label{eq3}
    NMAE = \cfrac{100}{N} \cdot \cfrac{\sum_{i=1}^N {|y_i - \hat{y}_i|} }{y_{max} - y_{min}}
\end{equation}
where $N$ denotes the number of testing samples, $y_i$ represents the $i$-th actual value, $\hat{y}_i$ is the  $i$-th predicted value, and $y_{max}$ and $y_{min}$ are the maximum and minimum values of all actual data points respectively.

Another widely used evaluation metric for regression models is the coefficient of determination ($\mathbf{R}^2$). This statistical measure quantifies the proportion of variance in the dependent variable that is explained by the independent variable(s) within the regression model. Typically,  values range from 0 to 100\%, where a value of x\% indicates that x\% of the variability in the target variable is accounted for by the regression model. In general, a higher value reflects a greater proportion of variability explained by the model, whereas a low value indicates a poor model performance. Consequently, this study also employed  $\mathbf{R}^2$ as a key metric for evaluating the predictive performance of the models apart from the NMAE metric.

The ML models with updating were assessed using the test-then-train approach \citep{biga23}, where each new incoming pattern is first used to test and record the predicted result, then update the regressor in an online manner (instance by instance). Finally, all predicted values were used along with actual offline analytical values to compute the NMAE and $\mathbf{R}^2$ scores.

\section{Results of adaptive ML agents through two case studies}\label{results_4_agents}
This section presents a comprehensive evaluation of the performance of various ML models, as described in the preceding sections, with and without updates to the models or training data during the operation of the bioreactors in the case studies. Two distinct updating approaches were employed for this purpose: daily updating and real-time updating. In the \textit{daily} updating approach, a predefined timer was set to trigger updates at a specific time each day (e.g., 3:30 PM). At this scheduled time, the ML models or the data library were updated using all samples collected within the preceding 24-hour period. In contrast, the \textit{real-time} updating approach involved the immediate updating of ML models or the data library once new offline analytical measurements became available. This real-time updating ensured that the most recent data was incorporated into the models, enabling rapidly continuous adaptation during bioreactor operation. Although the evaluation metrics were calculated based on the actual values, all values presented in the figures of this paper have been normalised to protect intellectual property. 

\subsection{Case study 1}
\begin{figure}[]
\centering
\begin{subfloat}[Predictions using RPLSR \label{glucose_update_pred_rpls}]{
\includegraphics[width=0.36\textwidth]{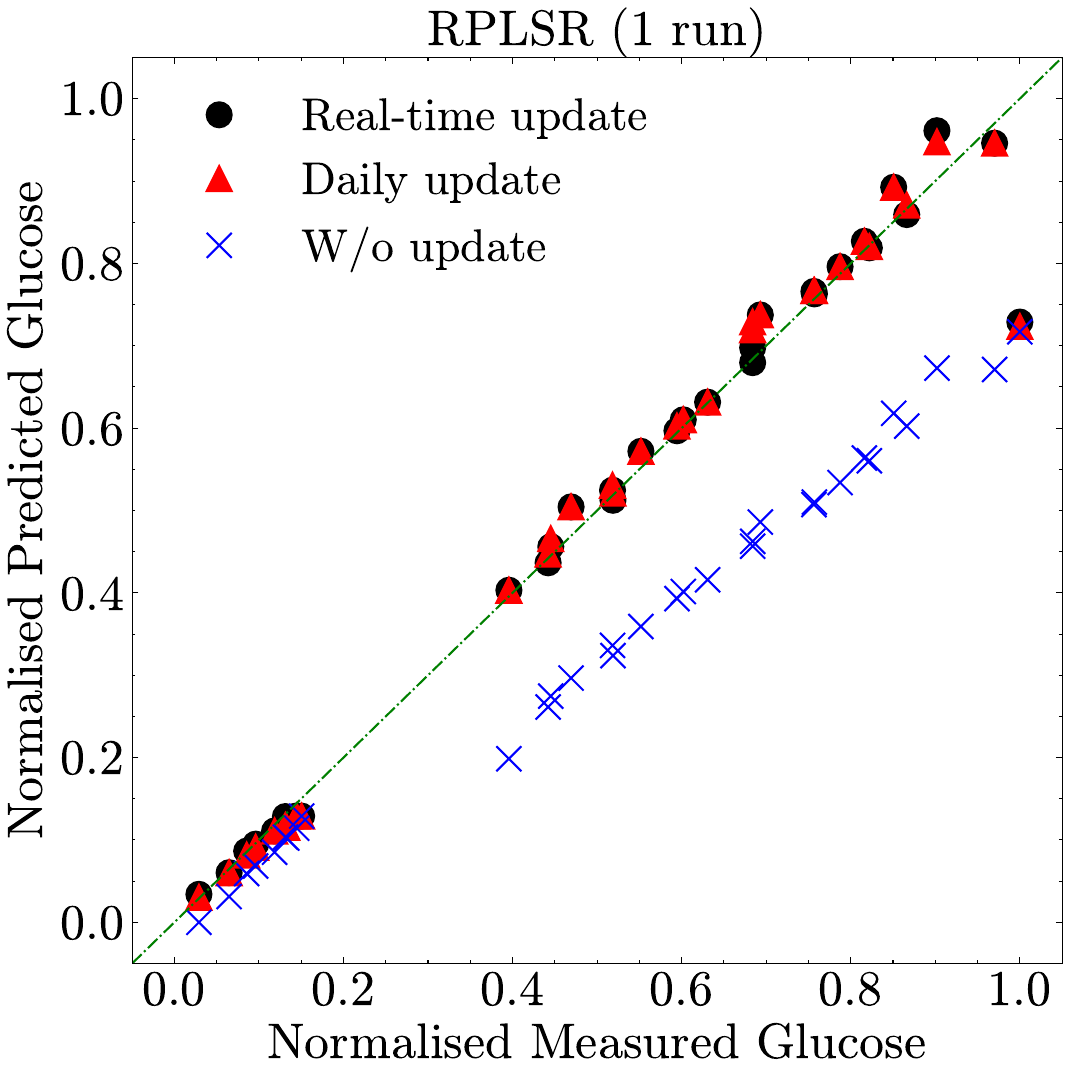}}
\end{subfloat}
\begin{subfloat}[Real-time monitoring using RPLSR \label{glucose_update_realtime_pred_rpls}]{
\includegraphics[width=0.6\textwidth, height=0.36\textwidth]{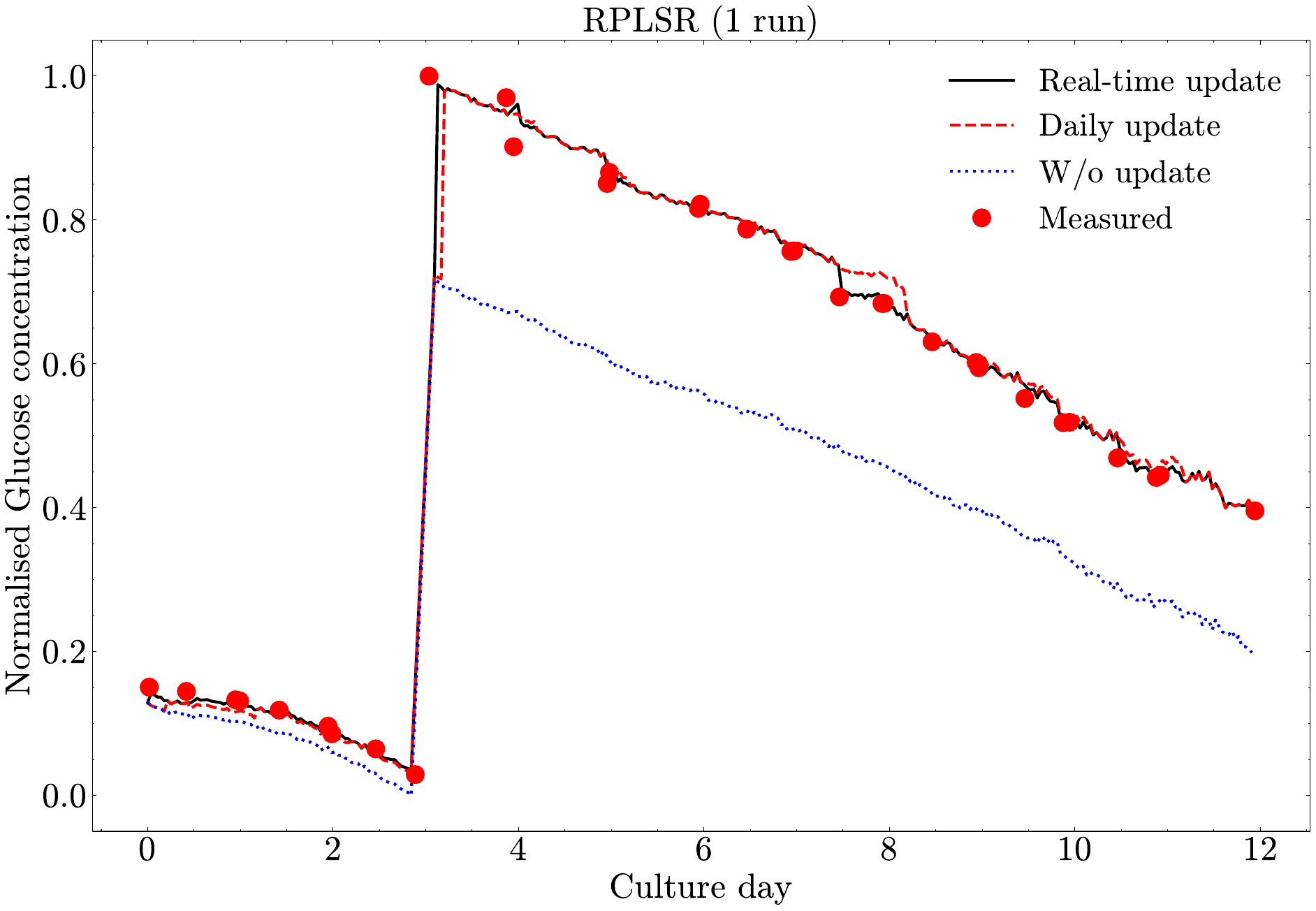}}
\end{subfloat}
\begin{subfloat}[Predictions using JITL (PLSR) \label{glucose_update_pred_jitl}]{
\includegraphics[width=0.36\textwidth]{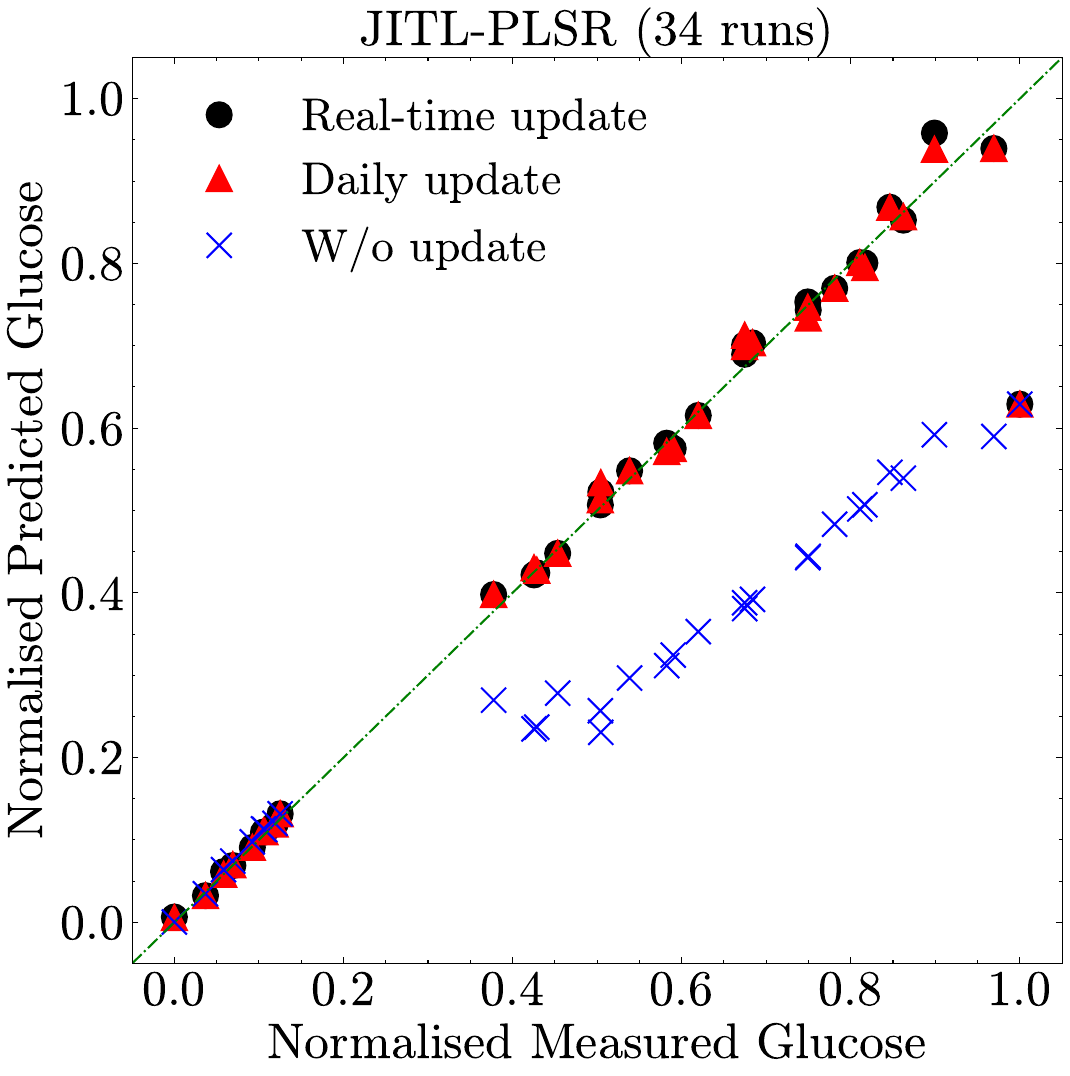}}
\end{subfloat}
\begin{subfloat}[Real-time monitoring using JITL (PLSR) \label{glucose_update_realtime_pred_jitl}]{
\includegraphics[width=0.6\textwidth, height=0.36\textwidth]{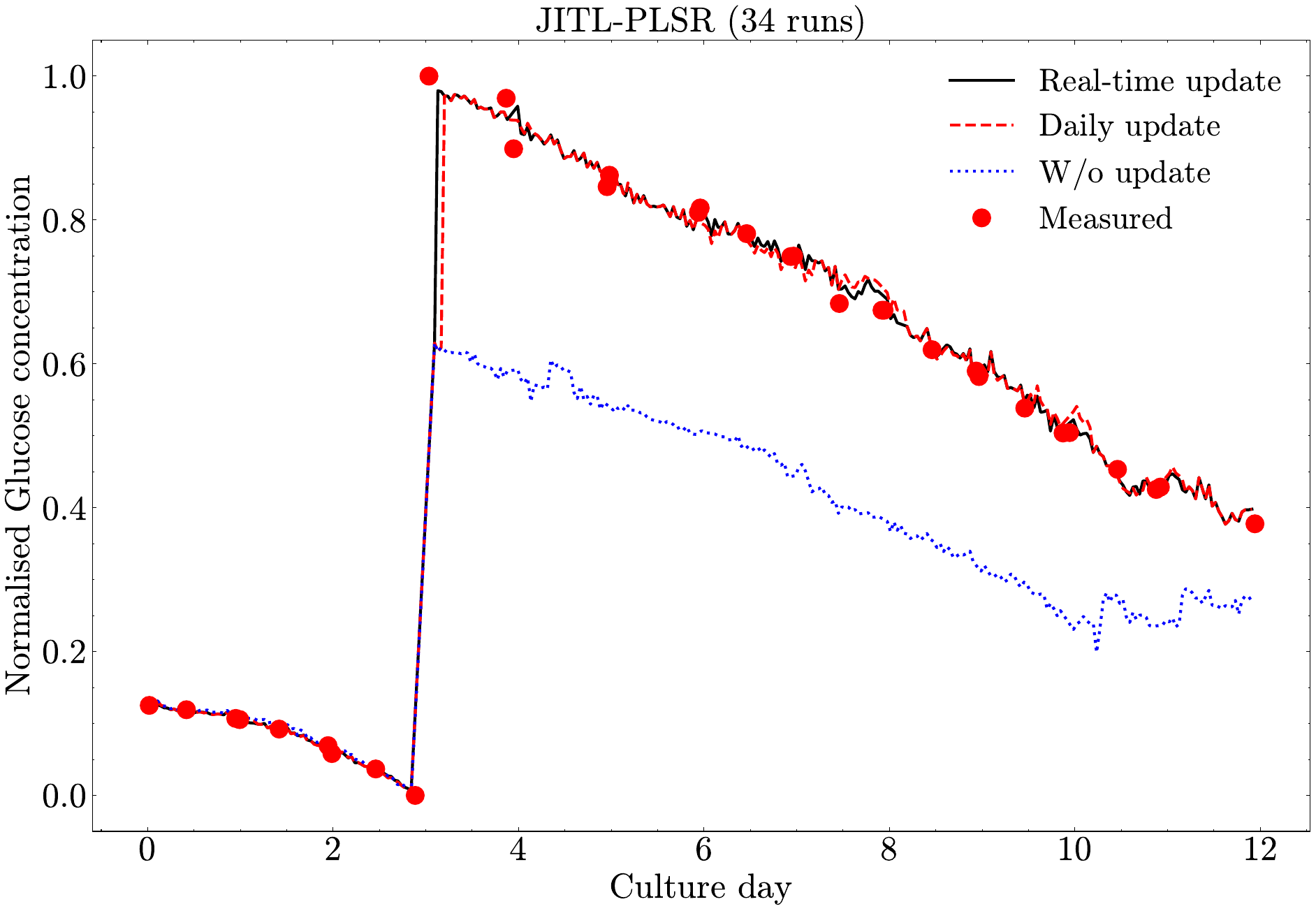}}
\end{subfloat}
\begin{subfloat}[Predictions using the PLSR \label{glucose_update_pred_pretrain}]{
\includegraphics[width=0.36\textwidth]{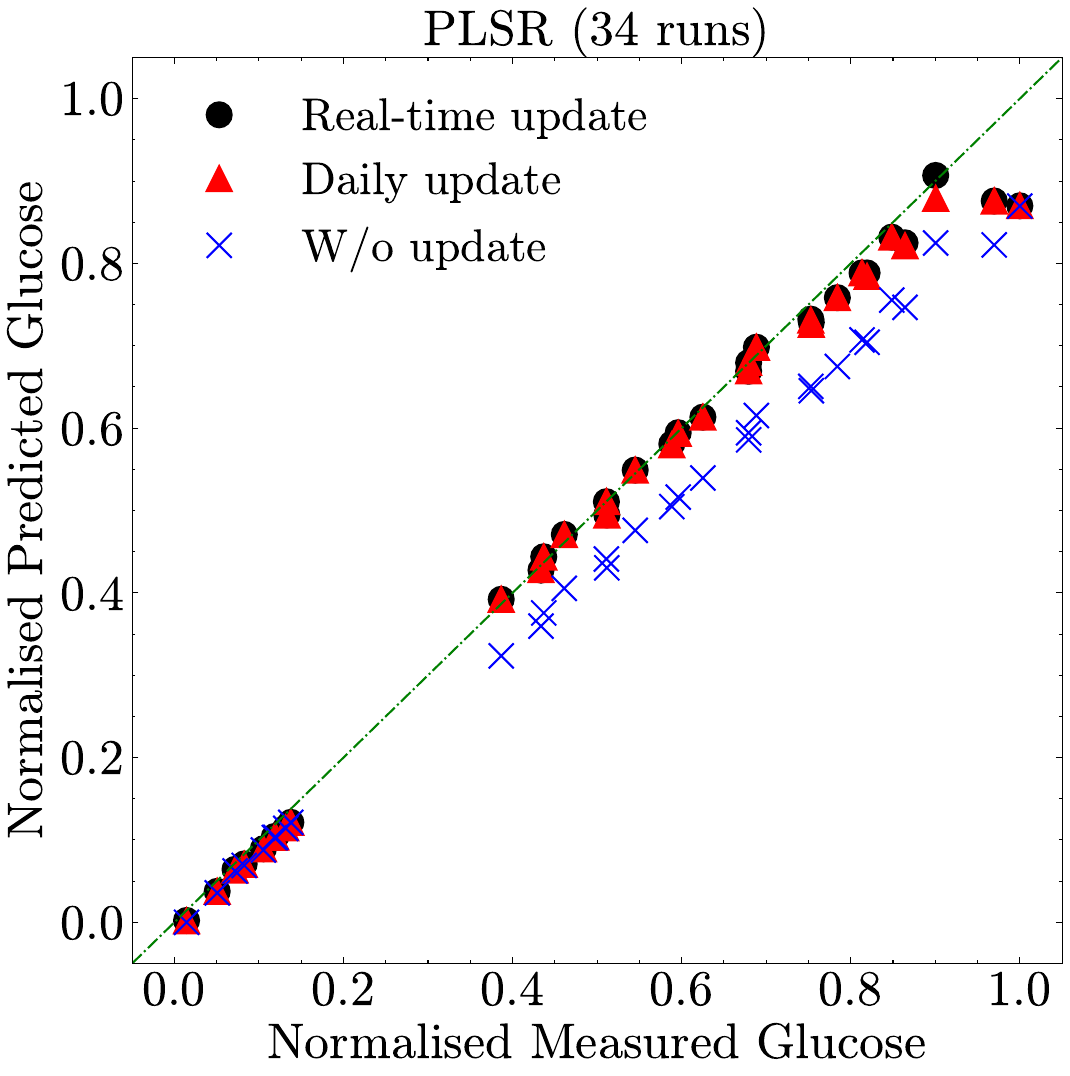}}
\end{subfloat}
\begin{subfloat}[Real-time monitoring using the PLSR \label{glucose_update_realtime_pred_pretrain}]{
\includegraphics[width=0.6\textwidth, height=0.36\textwidth]{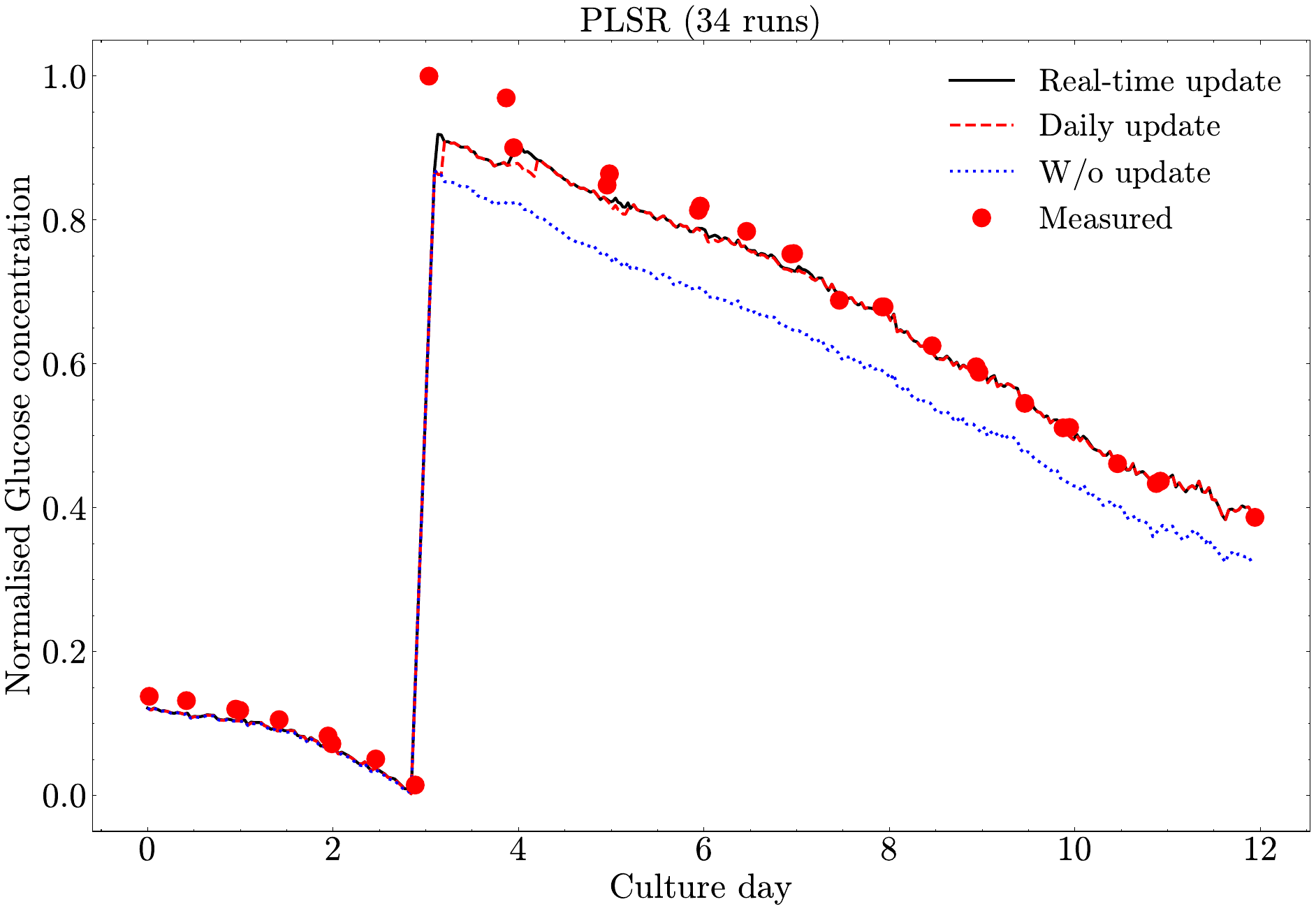}}
\end{subfloat}
\caption{The important roles of updating ML models for predictions of \textit{Glucose} in case study \textbf{1}.} \label{model_update_role_cs1}
\end{figure}

Firstly, the critical role of model and training data updates in the performance of Raman-based models for real-time monitoring of metabolite concentrations was evaluated. Fig. \ref{model_update_role_cs1} presents the predicted glucose concentrations based on offline analytical measurements and real-time monitoring for three learning models including RPLSR, the JITL model employing PLSR for local model building, and the PLSR model trained on 34 historical bioreactor runs. The similar results for lactate and ammonium can be found in Figs. \url{S1} and \url{S2} in the supplemental materials. The figures highlight the differences in model performance with and without updates from newly acquired samples during operational time. It is evident that the performance of online ML models and JITL models improves significantly when the learning model/data library is updated, compared to when no updates are applied. For the pretrained PLSR model trained on all 34 historical runs, the training data included several bioreactor runs with similar base and feed culture medium compositions and feeding strategies as those used in this case study. This similarity was captured in the input Raman patterns, enabling the pretrained PLSR model to produce acceptable predictions and accurately capture the trends in glucose concentration changes, even without retraining. Nevertheless, retraining the PLSR model with newly collected offline analytical measurements further enhanced its predictive performance compared to the non-retrained model.

\begin{figure}[!ht]
\centering
\begin{subfloat}[Glucose prediction \label{glucose_cs1}]{
\includegraphics[width=0.33\textwidth]{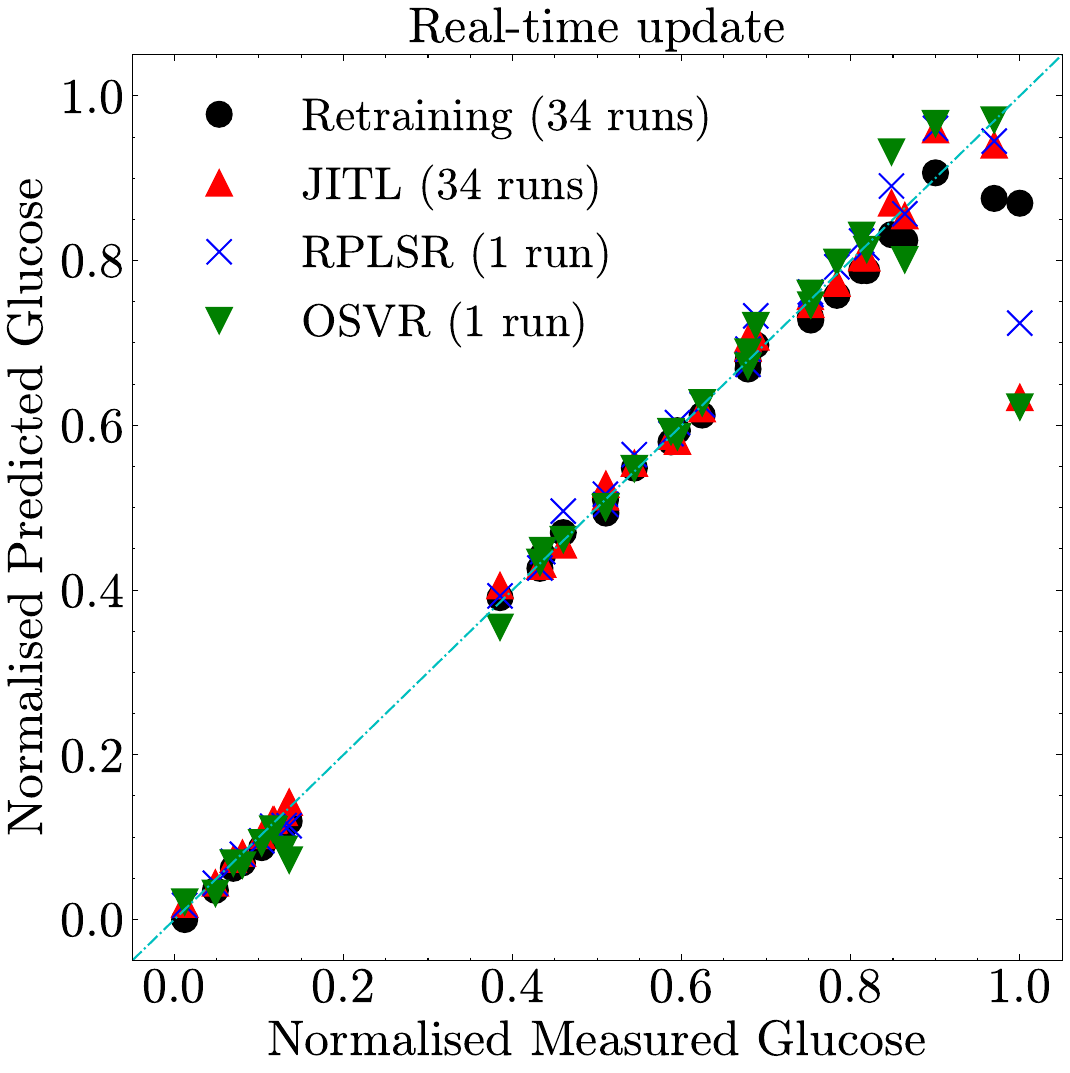}}
\end{subfloat}
\begin{subfloat}[Real-time Glucose monitoring \label{glucose_monitoring_cs1}]{
\includegraphics[width=0.6\textwidth, height=0.33\textwidth]{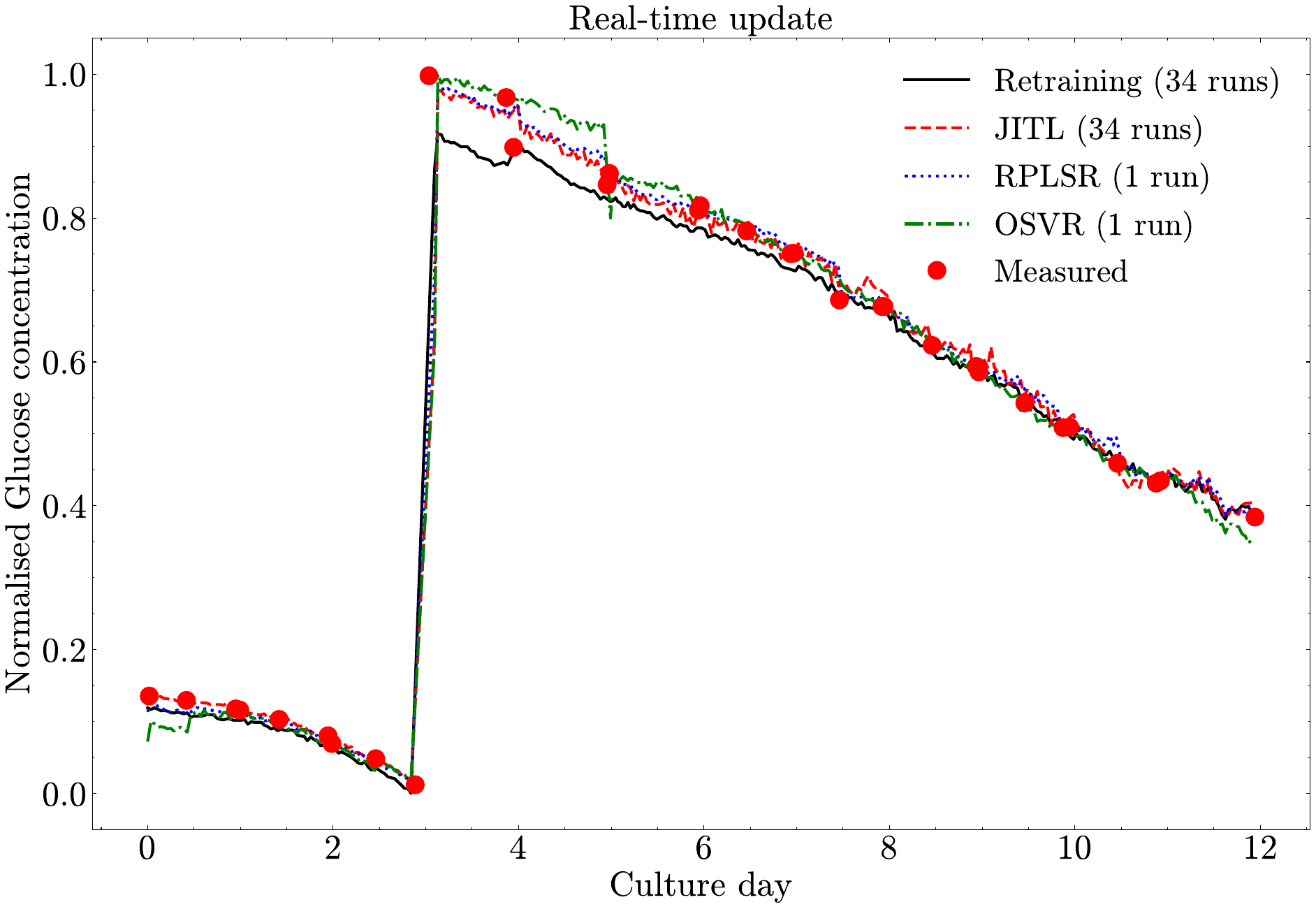}}
\end{subfloat}
\begin{subfloat}[Lactate prediction \label{lactate_cs1}]{
\includegraphics[width=0.33\textwidth]{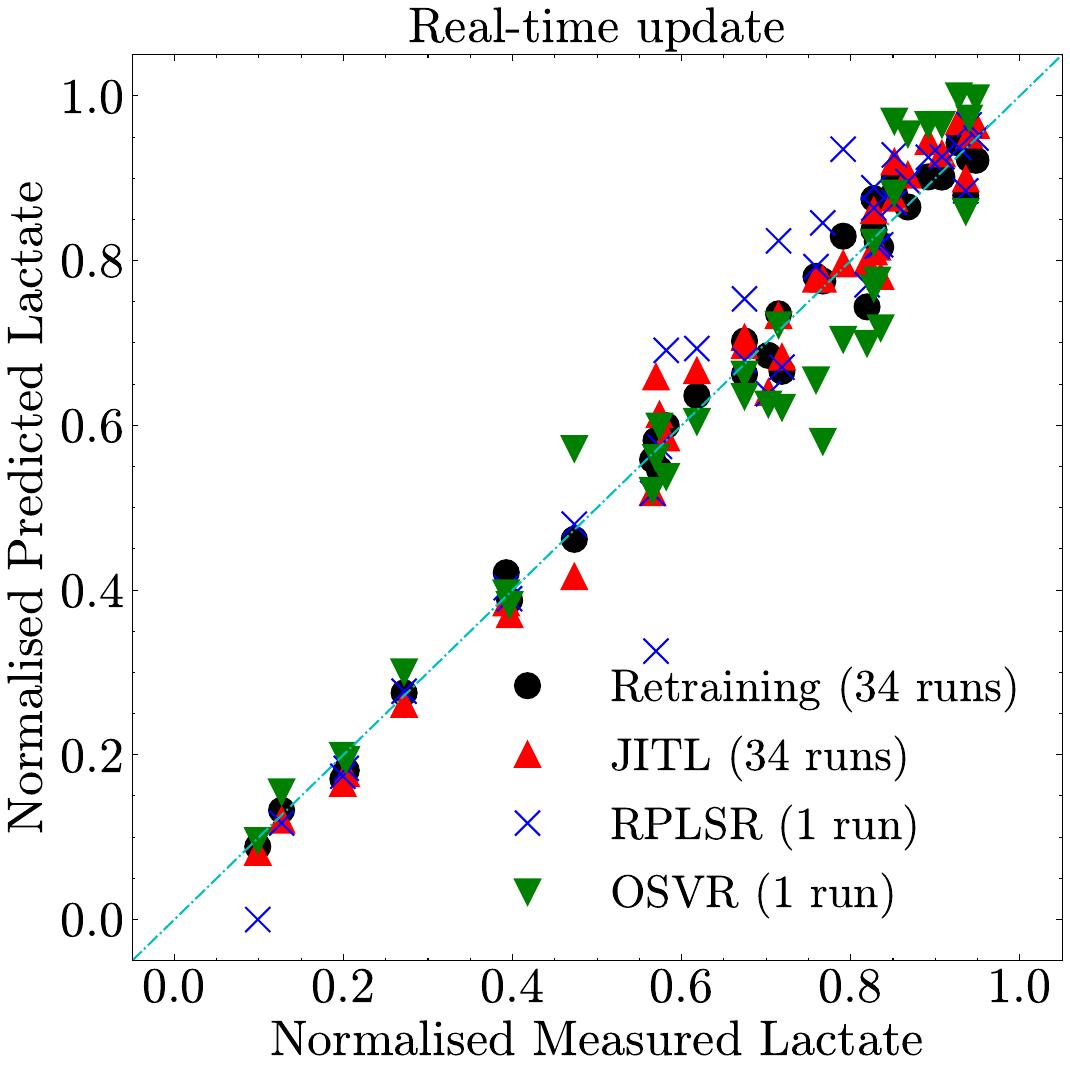}}
\end{subfloat}
\begin{subfloat}[Real-time Lactate monitoring\label{lactate_monitoring_cs1}]{
\includegraphics[width=0.6\textwidth, height=0.33\textwidth]{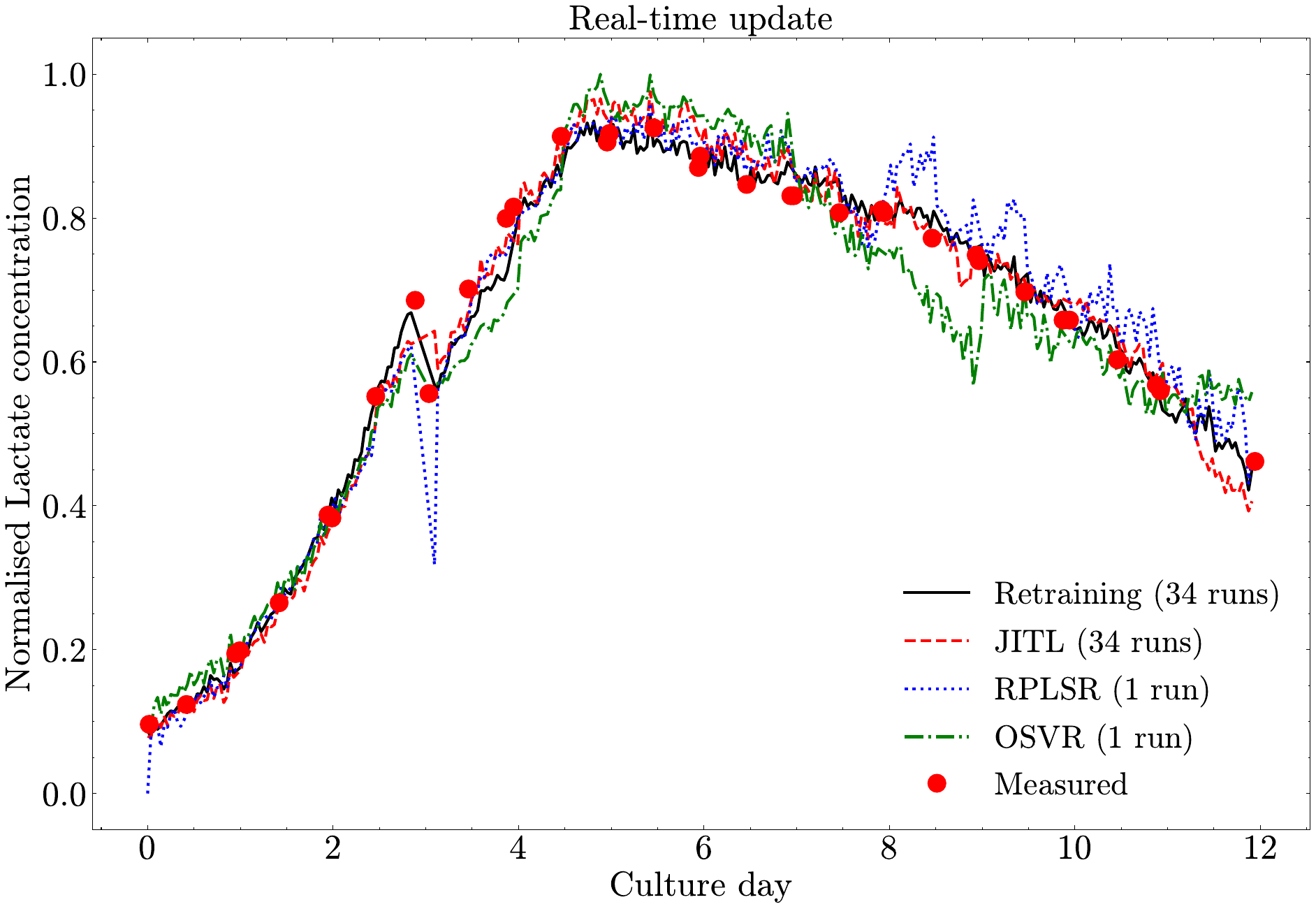}}
\end{subfloat}
\begin{subfloat}[Ammonium prediction \label{ammonium_cs1}]{
\includegraphics[width=0.33\textwidth]{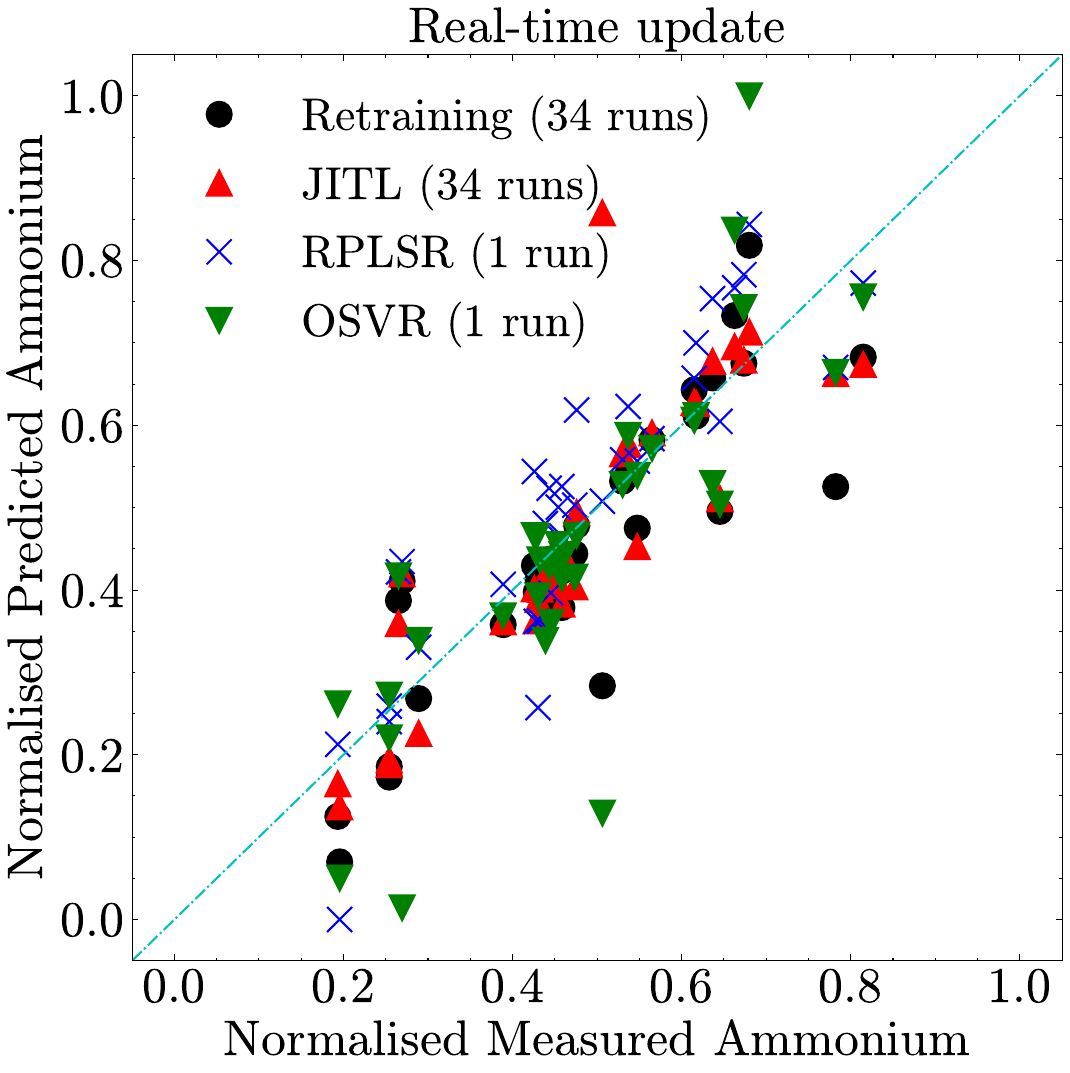}}
\end{subfloat}
\begin{subfloat}[Real-time Ammonium monitoring \label{ammonium_monitoring_cs1}]{
\includegraphics[width=0.6\textwidth, height=0.33\textwidth]{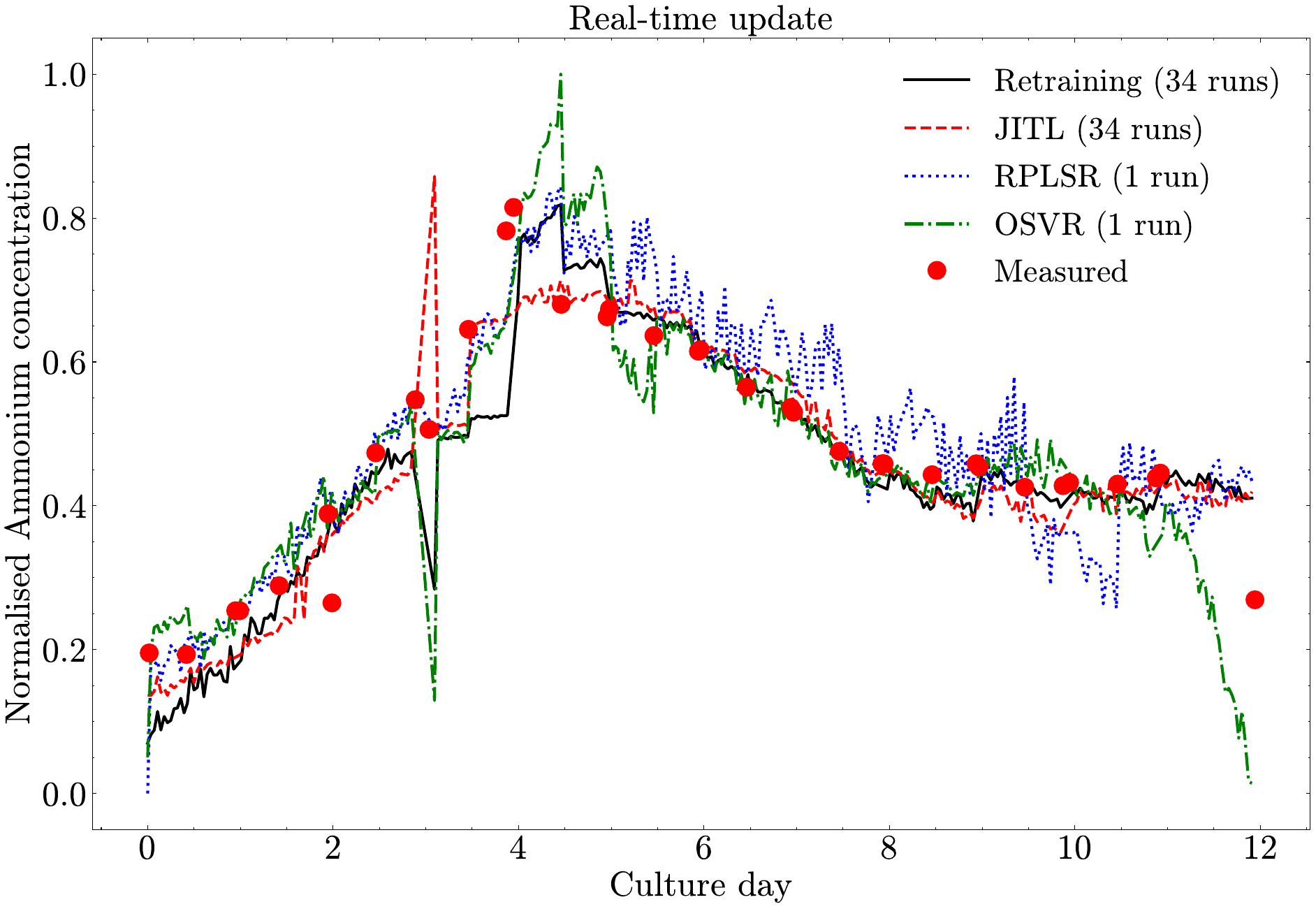}}
\end{subfloat}
\caption{Performance of all ML models with real-time update for different metabolite measurements in case study \textbf{1}.} \label{all_off_results_cs1}
\end{figure}

Fig. \ref{model_update_role_cs1} clearly demonstrates that all ML models with daily updates exhibit some delay in adaptation compared to those employing real-time updates. These findings confirm that real-time updating or retraining of learning models or the data library (in the case of JITL) facilitates superior performance in real-time monitoring of metabolite concentrations during the operation of cell culture bioreactors. This improvement can be attributed to the capability of real-time updates to enable learning models to rapidly adapt to dynamic changes in cellular behaviours and cultivation conditions resulting from external interventions, such as nutrient feeding, pH adjustments, or temperature shifts. Table \ref{case_study_1_table} summarises the performance of all four ML model types, both with and without updating mechanisms, in predicting glucose, lactate, and ammonium concentrations. Furthermore, Fig. \ref{all_off_results_cs1} provides a detailed visualisation of the predicted values and real-time monitoring results of all ML models employing the real-time update mechanism in case study 1.

\begin{table}
\centering
\caption{Performance of different ML models for case study 1.}\label{case_study_1_table}
\resizebox{\textwidth}{!}{
\begin{tabular}{llccccccccc}
\toprule
\textbf{} &
  \textbf{} &
  \multicolumn{2}{c}{\textbf{Glucose}} &
  \multicolumn{2}{c}{\textbf{Lactate}} &
  \multicolumn{2}{c}{\textbf{Ammonium}} \\ 
  \cmidrule(lr){3-8}
\textbf{Model} &
  \textbf{Criteria} &
  \textbf{NMAE (\%)} &
  \textbf{$\mathbf{R}^2$ (\%)} &
  \textbf{NMAE (\%)} &
  \textbf{$\mathbf{R}^2$ (\%)} &
  \textbf{NMAE (\%)} &
  \textbf{$\mathbf{R}^2$ (\%)} \\ 
\midrule
\multirow{3}{*}{\textbf{\makecell[l]{RPLSR \\ (Pretrain: 1 run)}}} &
  \cellcolor[HTML]{D9EAD3}Daily update &
  \cellcolor[HTML]{D9EAD3}2.48 &
  \cellcolor[HTML]{D9EAD3}96.85 &
  \cellcolor[HTML]{D9EAD3}7.68 &
  \cellcolor[HTML]{D9EAD3}87.59 &
  \cellcolor[HTML]{D9EAD3}15.58 &
  \cellcolor[HTML]{D9EAD3}42.54 \\
 &
  \cellcolor[HTML]{FDE9D9}Real-time update &
  \cellcolor[HTML]{FDE9D9}2.17 &
  \cellcolor[HTML]{FDE9D9}97.04 &
  \cellcolor[HTML]{FDE9D9}5.59 &
  \cellcolor[HTML]{FDE9D9}92.27 &
  \cellcolor[HTML]{FDE9D9}11.44 &
  \cellcolor[HTML]{FDE9D9}66.39 \\
 &
  No update &
  17.36 &
  58.76 &
  37.77 &
  -96.55 &
  32.91 &
  -124.48 \\ \midrule
\multirow{3}{*}{\textbf{\makecell[l]{OSVR \\ (Pretrain: 1 run)}}} &
  \cellcolor[HTML]{D9EAD3}Daily update &
  \cellcolor[HTML]{D9EAD3}3.11 &
  \cellcolor[HTML]{D9EAD3}94.76 &
  \cellcolor[HTML]{D9EAD3}8.98 &
  \cellcolor[HTML]{D9EAD3}75.15 &
  \cellcolor[HTML]{D9EAD3}22.37 &
  \cellcolor[HTML]{D9EAD3}-63.15 \\
 &
  \cellcolor[HTML]{FDE9D9}Real-time update &
  \cellcolor[HTML]{FDE9D9}3.01 &
  \cellcolor[HTML]{FDE9D9}94.34 &
  \cellcolor[HTML]{FDE9D9}6.34 &
  \cellcolor[HTML]{FDE9D9}92.09 &
  \cellcolor[HTML]{FDE9D9}12.22 &
  \cellcolor[HTML]{FDE9D9}43.04 \\
 &
  No update &
  20.11 &
  49.41 &
  45.86 &
  -265.13 &
  69.45 &
  -1058.90 \\ \midrule
\multirow{3}{*}{\textbf{\makecell[l]{JITL \\ (Data library: 34 runs)}}} &
  \cellcolor[HTML]{D9EAD3}Daily update data &
  \cellcolor[HTML]{D9EAD3}2.33 &
  \cellcolor[HTML]{D9EAD3}95.20 &
  \cellcolor[HTML]{D9EAD3}4.24 &
  \cellcolor[HTML]{D9EAD3}96.62 &
  \cellcolor[HTML]{D9EAD3}11.99 &
  \cellcolor[HTML]{D9EAD3}52.51 \\
 &
  \cellcolor[HTML]{FDE9D9}Real-time update data &
  \cellcolor[HTML]{FDE9D9}2.22 &
  \cellcolor[HTML]{FDE9D9}95.20 &
  \cellcolor[HTML]{FDE9D9}3.59 &
  \cellcolor[HTML]{FDE9D9}97.80 &
  \cellcolor[HTML]{FDE9D9}\textbf{9.66} &
  \cellcolor[HTML]{FDE9D9}\textbf{68.49} \\
 &
  No update data &
  19.84 &
  40.28 &
  7.99 &
  85.65 &
  15.20 &
  39.56 \\ \midrule
\multirow{3}{*}{\textbf{\makecell[l]{Retraining \\ (Pretrain: 34 runs)}}} &
  \cellcolor[HTML]{D9EAD3}Daily retraining &
  \cellcolor[HTML]{D9EAD3}2.08 &
  \cellcolor[HTML]{D9EAD3}98.82 &
  \cellcolor[HTML]{D9EAD3}2.73 &
  \cellcolor[HTML]{D9EAD3}98.42 &
  \cellcolor[HTML]{D9EAD3}13.51 &
  \cellcolor[HTML]{D9EAD3}44.63 \\
 &
  \cellcolor[HTML]{FDE9D9}Real-time retraining &
  \cellcolor[HTML]{FDE9D9}\textbf{1.98} &
  \cellcolor[HTML]{FDE9D9}\textbf{98.86} &
  \cellcolor[HTML]{FDE9D9}\textbf{2.60} &
  \cellcolor[HTML]{FDE9D9}\textbf{98.72} &
  \cellcolor[HTML]{FDE9D9}9.81 &
  \cellcolor[HTML]{FDE9D9}67.90 \\
 &
  No retraining &
  7.02 &
  93.15 &
  3.91 &
  96.65 &
  28.99 &
  -88.05 \\ 
\bottomrule
\end{tabular}
}
\end{table}

For glucose prediction, the retraining of PLSR models with real-time updates incorporating newly collected glucose concentration values from the current bioreactor run achieved the highest performance, producing the lowest NMAE (1.98\%) and the highest $\mathbf{R}^2$ (98.86\%). Similarly, the real-time update of a recursive PLSR model, pretrained on only a single bioreactor run, also performed well. It recorded the second-lowest NMAE (2.17\%) and the second-highest $\mathbf{R}^2$ (97.04\%) among the four ML models using real-time update approaches.

For lactate prediction, the retraining model (PCA+SVR), trained on 34 historical runs and updated in real time, demonstrated superior performance, achieving the lowest NMAE (2.60\%) and the highest $\mathbf{R}^2$ (98.72\%), thereby surpassing all other methods. The JITL model (PCA+SVR), which updates its data library in real time using newly acquired lactate concentration values from the current run, also performed robustly, with an NMAE of 3.59\% and an $\mathbf{R}^2$ of 97.80\%. These findings indicate that both the retraining and JITL models exhibit exceptional adaptability to dynamic cultivation conditions, particularly for lactate predictions.

For ammonium prediction, the JITL model (KPCA+SVR) with real-time data library updates achieved the highest predictive performance, with the lowest NMAE (9.66\%) and the highest $\mathbf{R}^2$ (68.49\%) among all ML approaches. In contrast to glucose and lactate predictions, the retraining model (KPCA+SVR), while performing well, did not outperform the JITL model with real-time updates. The superior performance of the JITL model is likely attributable to its ability to build local KPCA+SVR models using only the 30 most similar training data points to the input query, effectively reducing noise in Raman spectral patterns. This approach proved advantageous in capturing the complex relationship between Raman signals and ammonium concentrations, which remains challenging for models trained on the full dataset of 34 historical runs containing over 600 samples \citep{khba24b}.

In this case study, the real-time monitoring results demonstrate that retraining and JITL models leveraging 34 historical runs produced smoother predictions compared to online ML models pretrained on a single historical run. These findings confirm that when a new bioreactor run uses base and feed medium compositions similar to those of historical runs, ML models trained on larger datasets generally achieve superior performance compared to models pretrained on a limited number of samples. However, ML models trained on smaller but highly relevant datasets, such as the JITL model, can still generate competitive predictive performance if the training data is dynamically updated with newly collected offline analytical measurements during the current run.

The results also revealed that pretrained online ML and JITL models with limited training data performed poorly in the absence of updates, likely due to their inability to generalise effectively from a small sample size to the variability inherent in the test data. This limitation is particularly evident in lactate and ammonium predictions, where $\mathbf{R}^2$ values were negative, reflecting a lack of alignment with actual measurements. However, for metabolites such as glucose and lactate, the performance of pretrained ML models trained on larger datasets without updates can still achieve relatively high predictive accuracy. Nevertheless, their performance improved significantly when retrained with newly collected data from the current run.

In summary, the results of case study 1 highlight that the pretrained models without updates exhibited poor performance across all evaluation metrics, emphasising the critical importance of model updating, even when historical data is closely aligned with the current run. Furthermore, real-time updates consistently outperformed daily updates, underscoring their capability to adapt more effectively to the dynamic changes inherent in bioreactor processes. Retrained models were shown to be particularly robust for metabolite predictions in new bioreactor runs that share the same base and feed medium compositions and feeding strategies as the training data, especially when large historical datasets were augmented with newly collected samples from the current run.

\subsection{Case study 2}
In contrast to case study 1, the two bioreactor runs in this case study utilised entirely new base and feed medium compositions, along with distinct feeding strategies characterised by varying levels of daily nutrient additions, compared to all 34 historical bioreactor runs. While both two bioreactors in this case study employed identical base and feed medium compositions, they differed in the cell lines used. Both \textit{Cell line A} and \textit{Cell line B} exhibited similar glucose concentration trends; however, their lactate concentration profiles diverged significantly. In the bioreactor using cell line A, lactate concentrations increased dramatically from culture day 6 onward, a trend that was also observed for ammonium levels. Conversely, in the bioreactor employing cell line B, lactate concentrations began to decline after culture day 6 and stabilised at near-zero levels from culture day 11 onwards. These distinct cellular behaviours in nutrient usage and secondary metabolite production present significant challenges for pretrained machine learning models, which lack exposure to these novel conditions in their training datasets.

\begin{figure}
\centering
\begin{subfloat}[Predictions using RPLSR \label{glucose_update_pred_rpls_cs2b}]{
\includegraphics[width=0.36\textwidth]{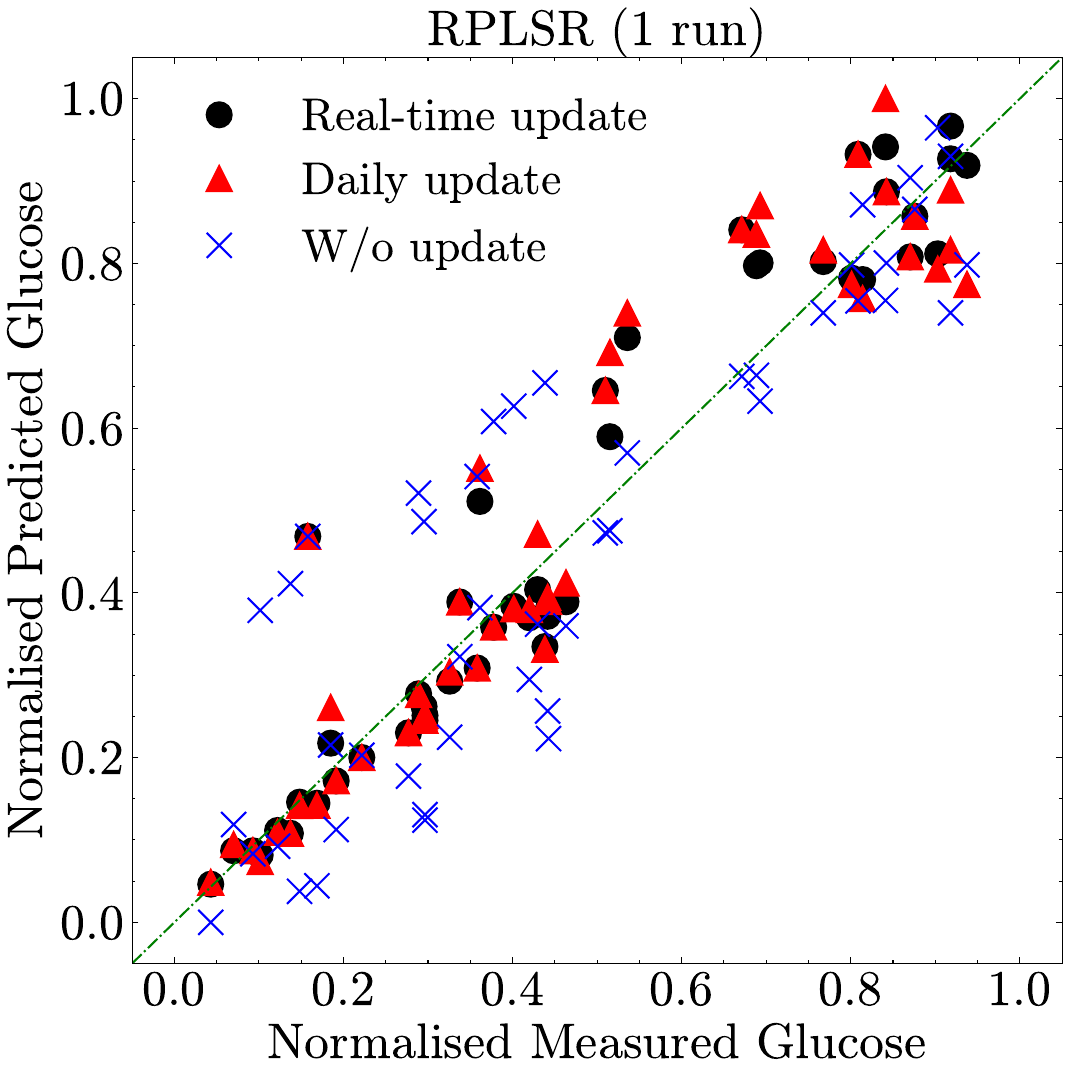}}
\end{subfloat}
\begin{subfloat}[Real-time monitoring using RPLSR \label{glucose_update_realtime_pred_rpls_cs2b}]{
\includegraphics[width=0.6\textwidth, height=0.355\textwidth]{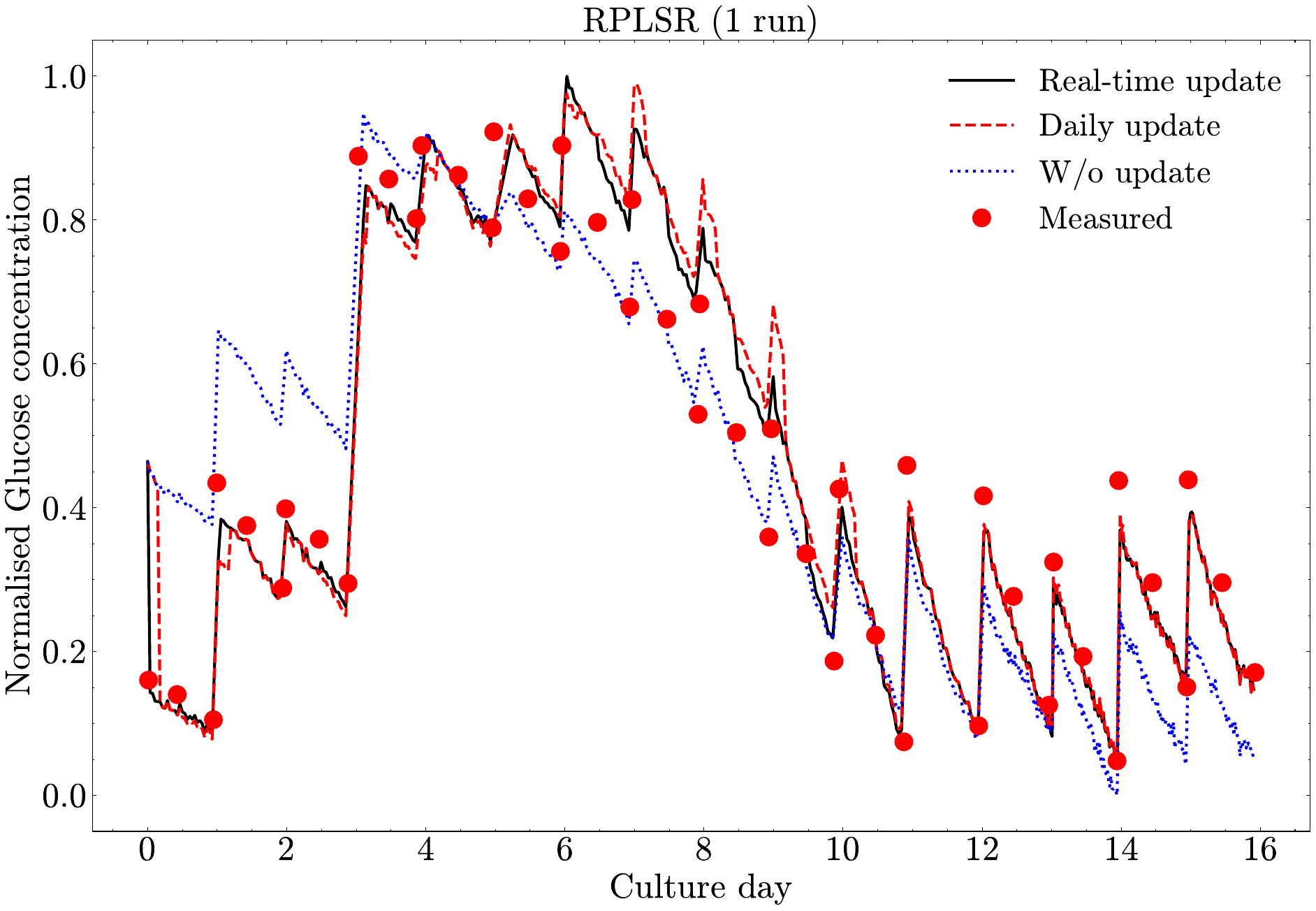}}
\end{subfloat}
\begin{subfloat}[Predictions using JITL (PLSR) \label{glucose_update_pred_jitl_cs2b}]{
\includegraphics[width=0.36\textwidth]{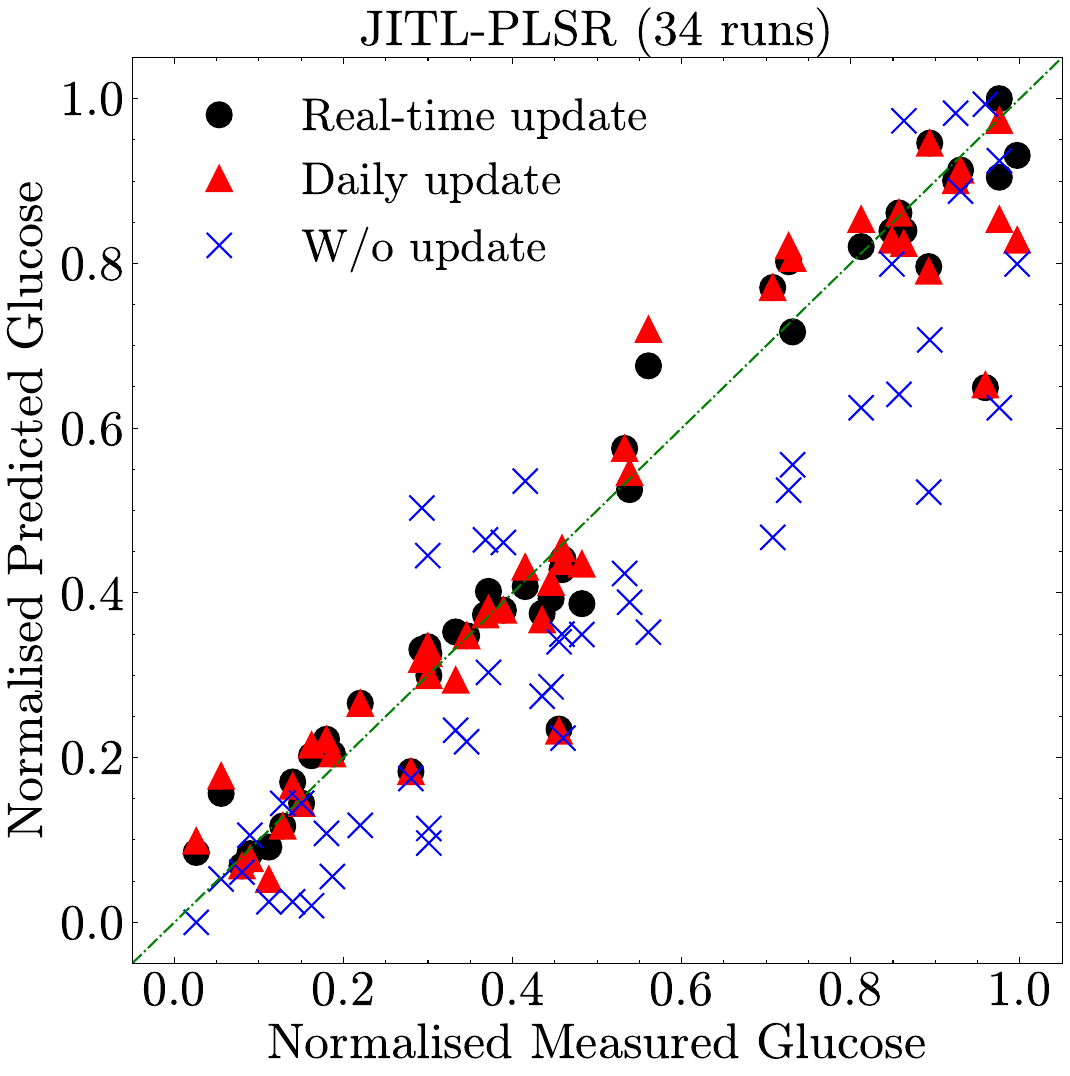}}
\end{subfloat}
\begin{subfloat}[Real-time monitoring using JITL (PLSR) \label{glucose_update_realtime_pred_jitl_cs2b}]{
\includegraphics[width=0.6\textwidth, height=0.355\textwidth]{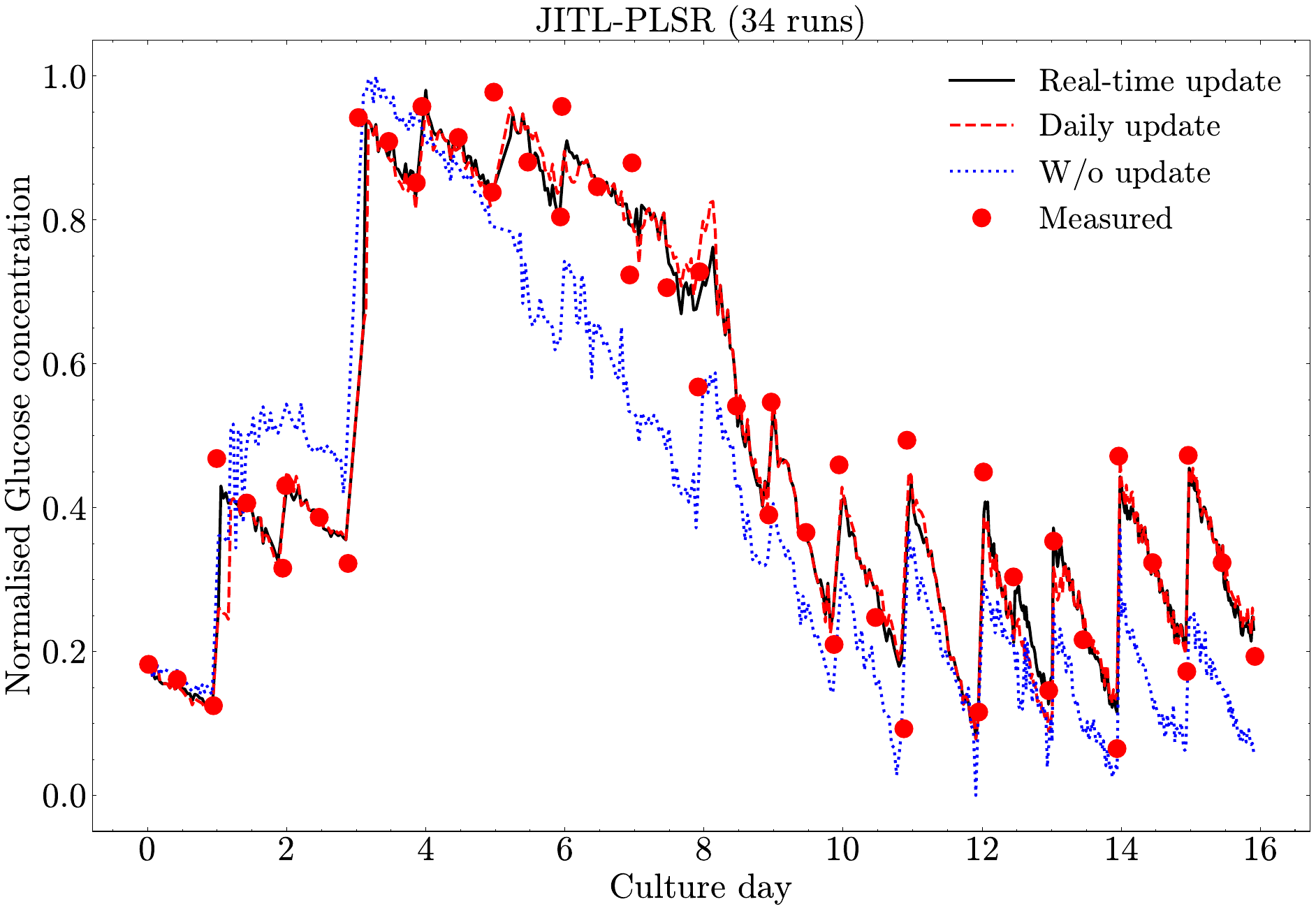}}
\end{subfloat}
\begin{subfloat}[Predictions using the PLSR \label{glucose_update_pred_pretrain_cs2b}]{
\includegraphics[width=0.36\textwidth]{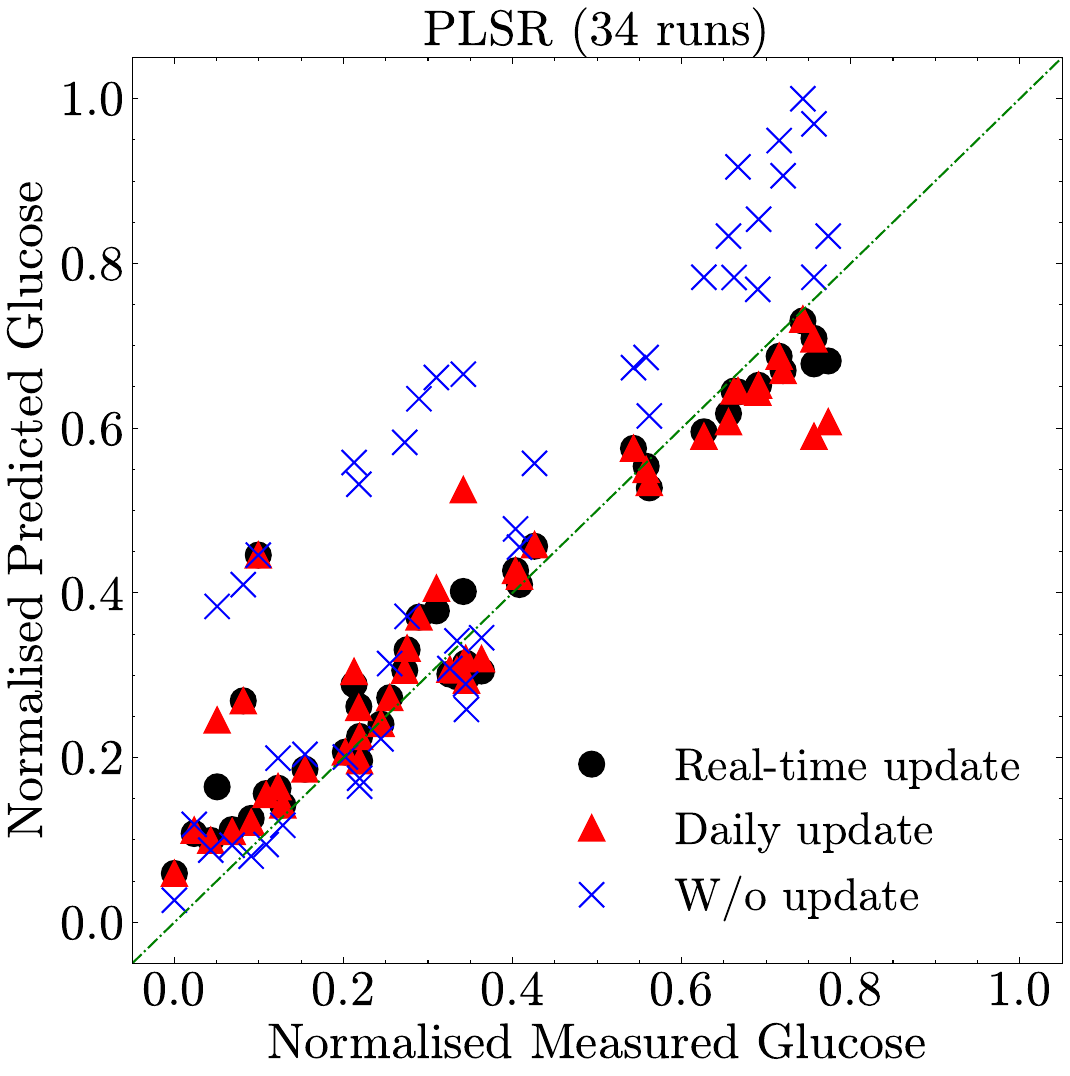}}
\end{subfloat}
\begin{subfloat}[Real-time monitoring using the PLSR \label{glucose_update_realtime_pred_pretrain_cs2b}]{
\includegraphics[width=0.6\textwidth, height=0.355\textwidth]{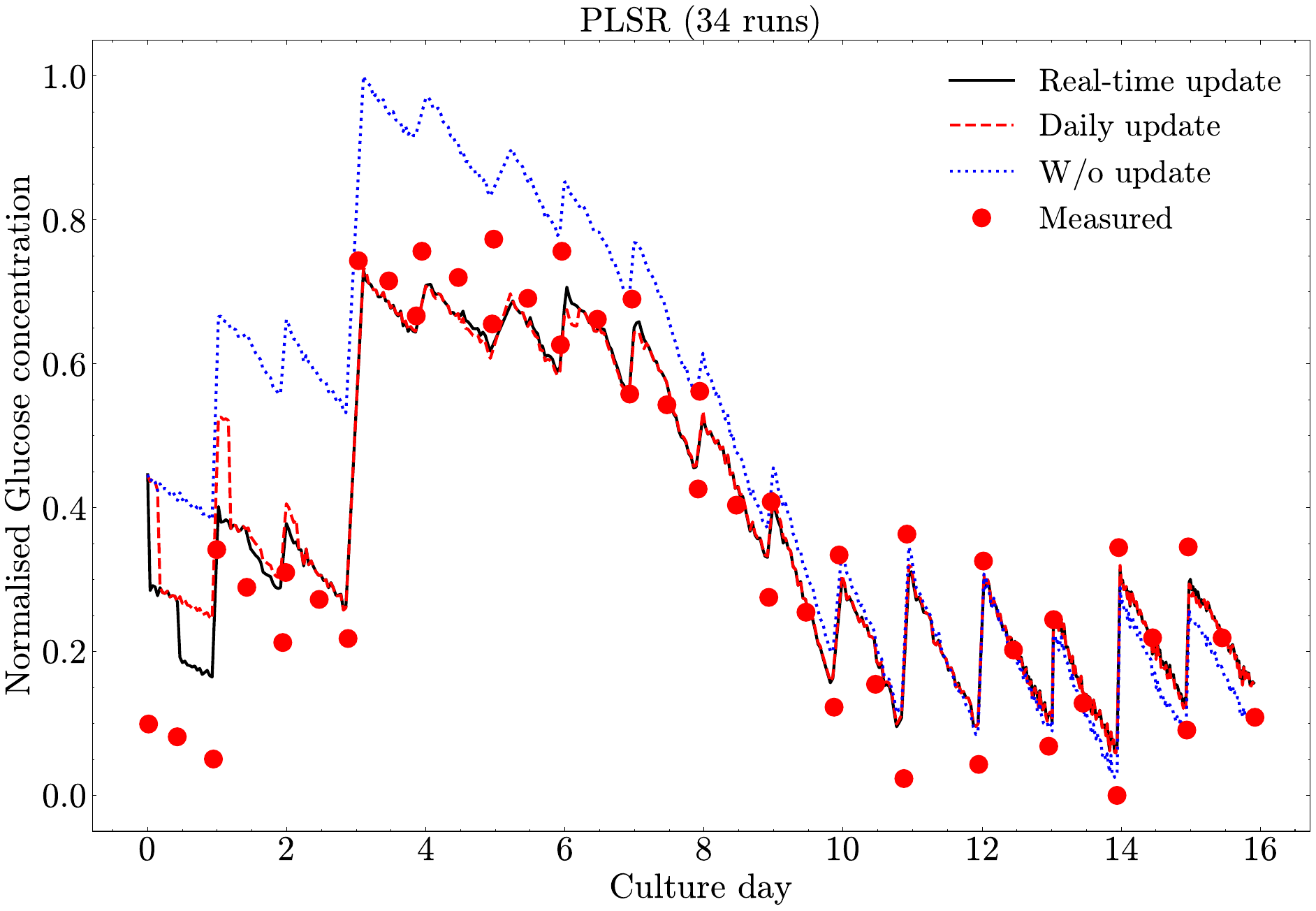}}
\end{subfloat}
\caption{The important roles of updating ML models for predictions of \textit{Glucose} in case study \textbf{2} with Cell Line \textbf{B}.} \label{model_update_role_cs2b}
\end{figure}

\begin{figure}
\centering
\begin{subfloat}[Predictions using OSVR \label{lactate_update_pred_rpls_cs2a}]{
\includegraphics[width=0.36\textwidth]{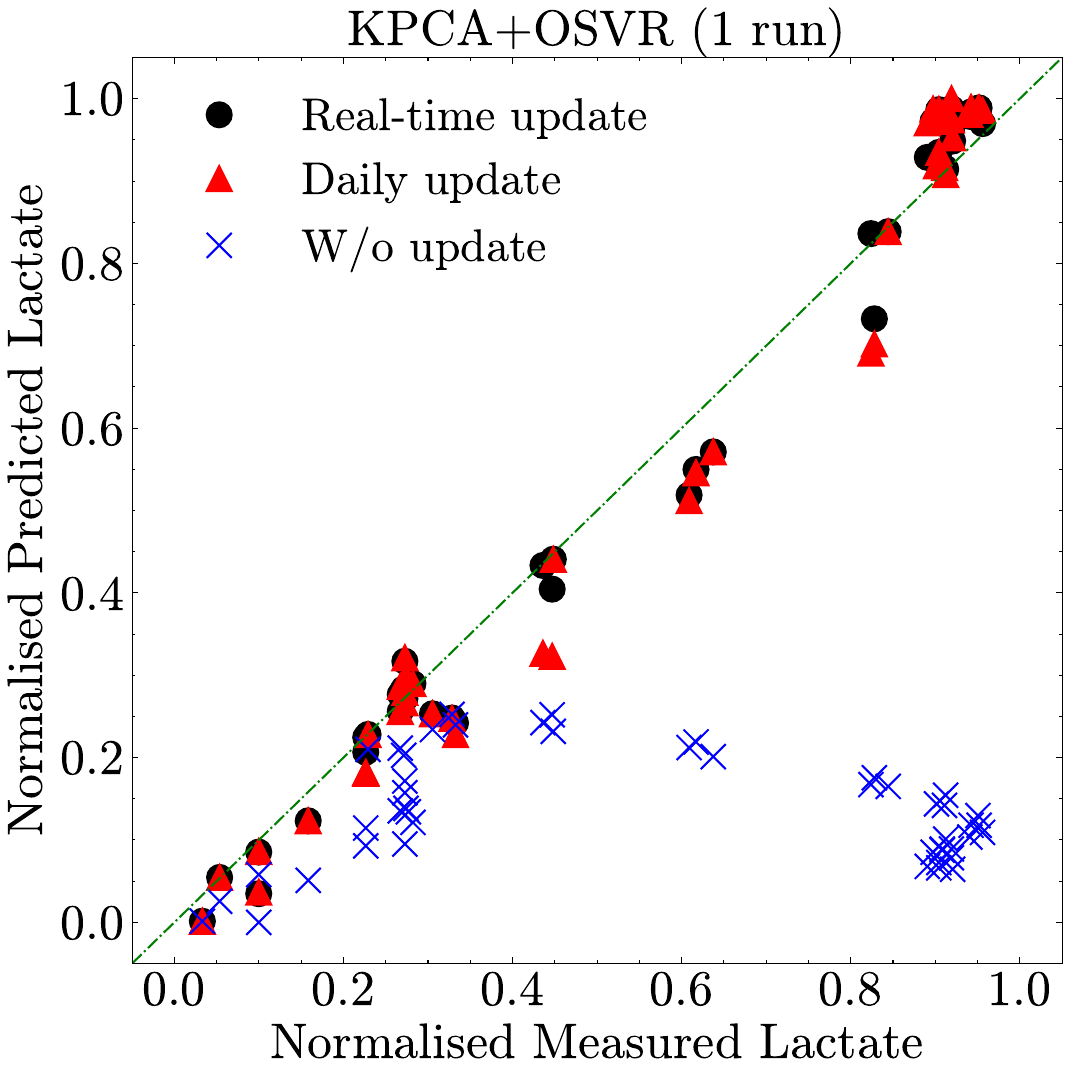}}
\end{subfloat}
\begin{subfloat}[Real-time monitoring using OSVR \label{lactate_update_realtime_pred_rpls_cs2a}]{
\includegraphics[width=0.6\textwidth, height=0.36\textwidth]{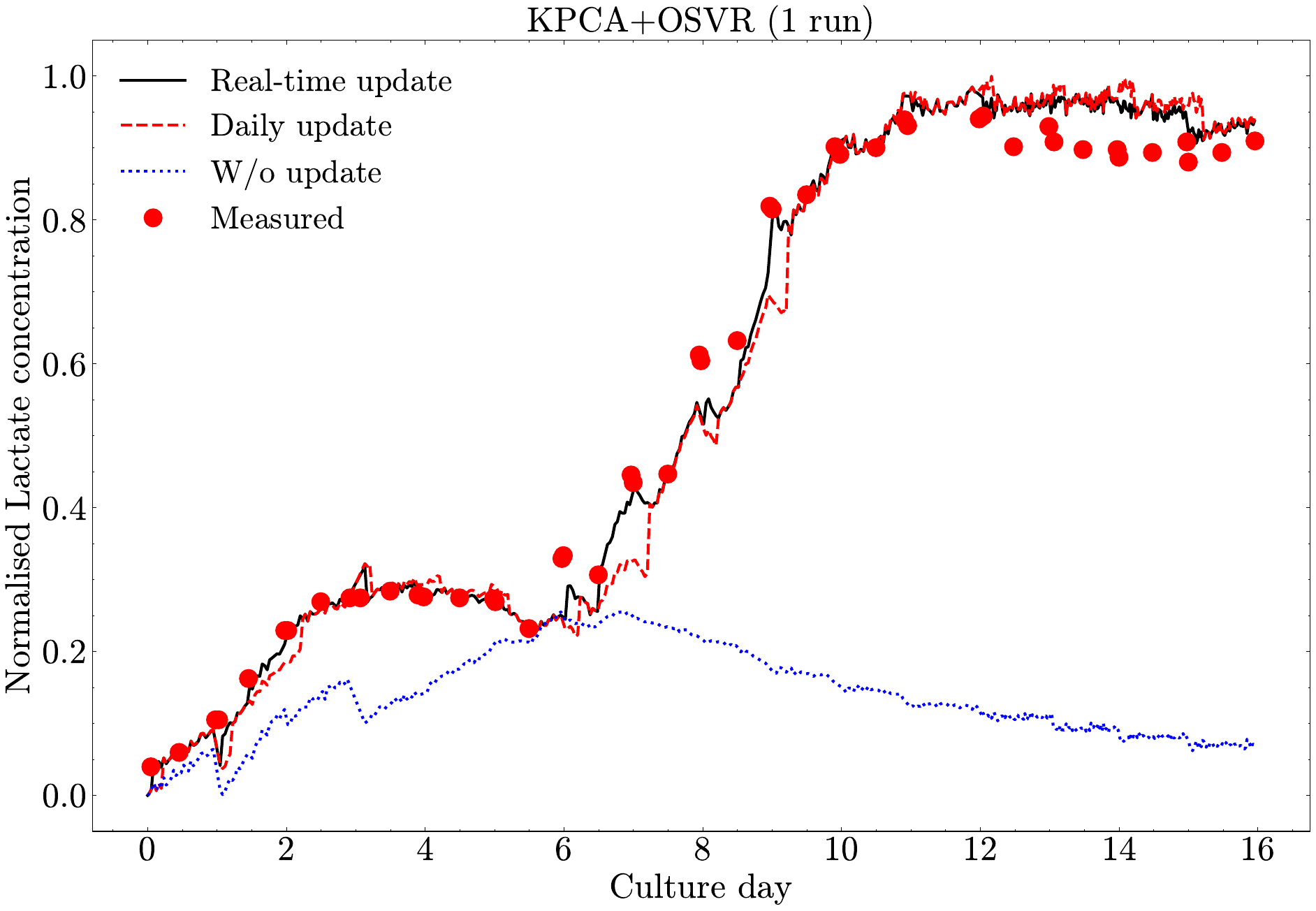}}
\end{subfloat}
\begin{subfloat}[Predictions using JITL (PCA+SVR) \label{lactate_update_pred_jitl_cs2a}]{
\includegraphics[width=0.36\textwidth]{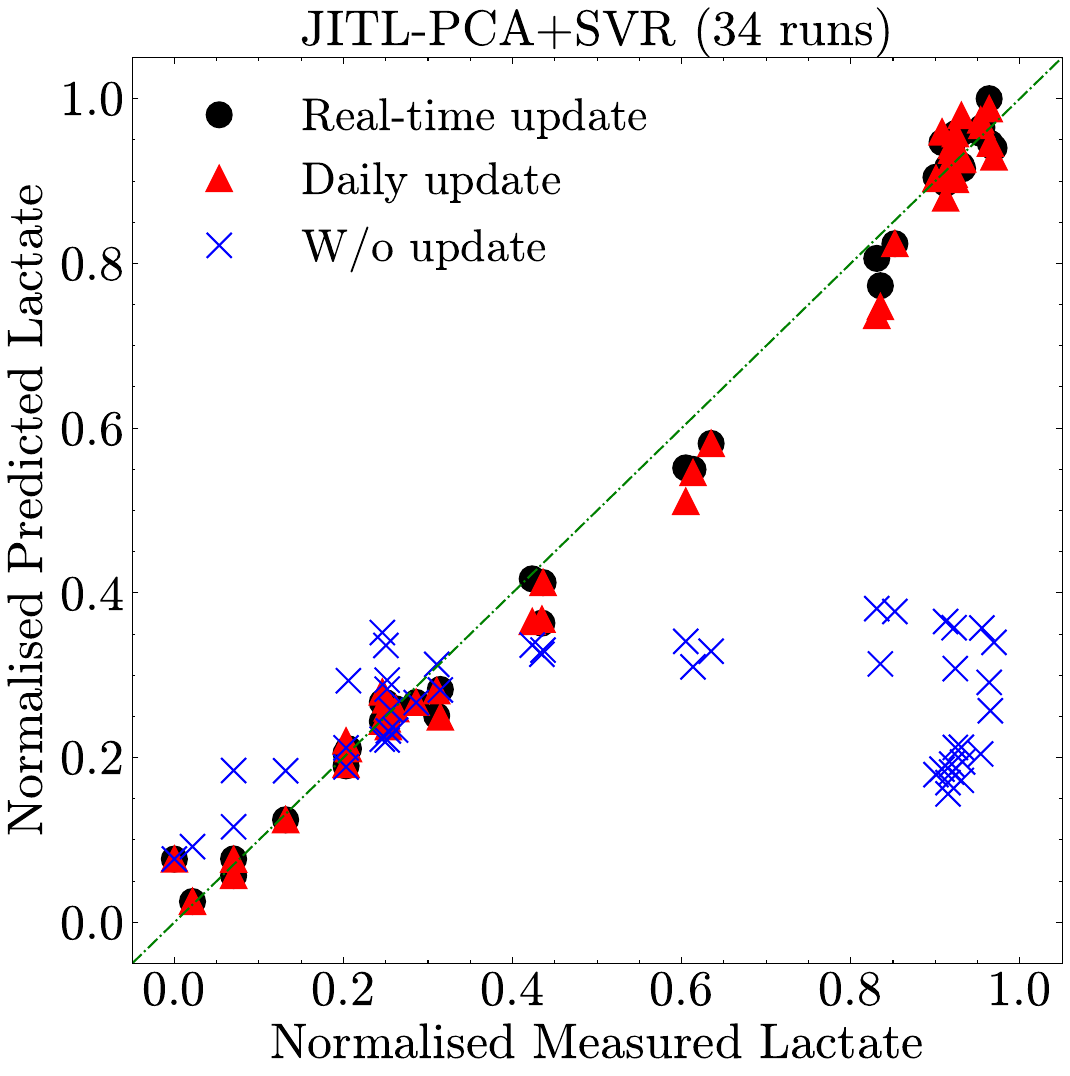}}
\end{subfloat}
\begin{subfloat}[Real-time monitoring using JITL (PCA+SVR) \label{lactate_update_realtime_pred_jitl_cs2a}]{
\includegraphics[width=0.6\textwidth, height=0.36\textwidth]{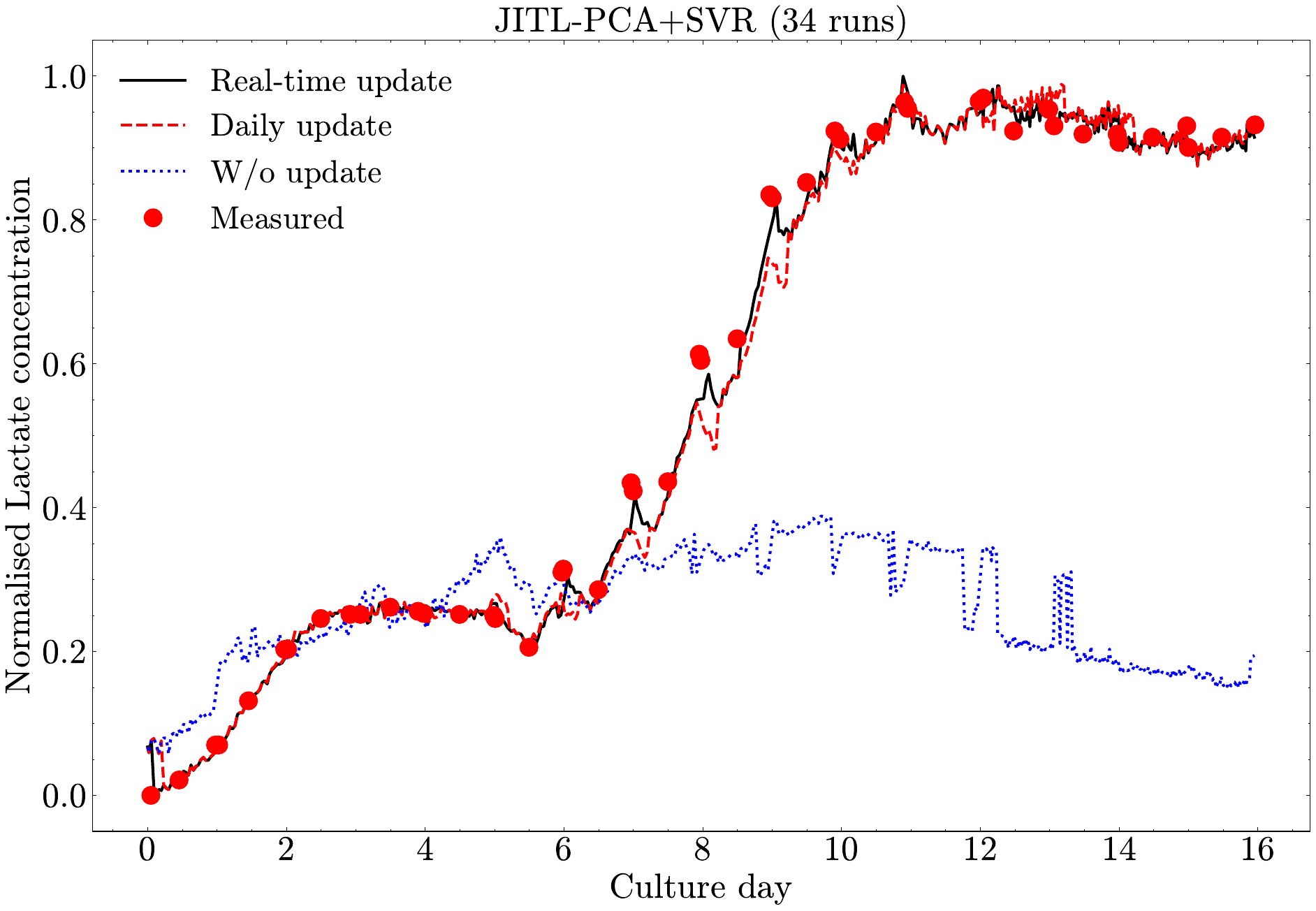}}
\end{subfloat}
\begin{subfloat}[Predictions using the PCA+SVR \label{lactate_update_pred_pretrain_cs2a}]{
\includegraphics[width=0.36\textwidth]{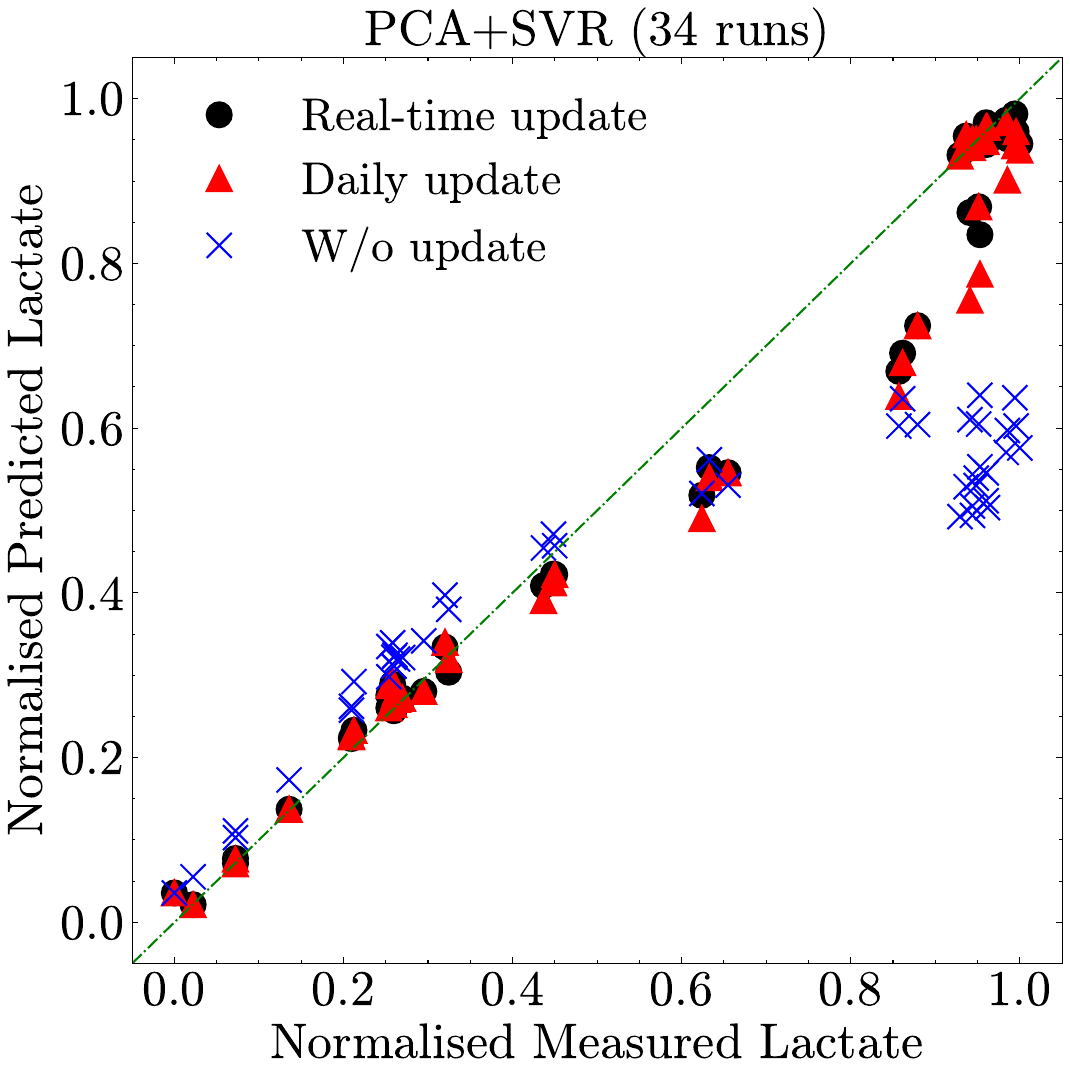}}
\end{subfloat}
\begin{subfloat}[Real-time monitoring using the PCA+SVR \label{lactate_update_realtime_pred_pretrain_cs2a}]{
\includegraphics[width=0.6\textwidth, height=0.36\textwidth]{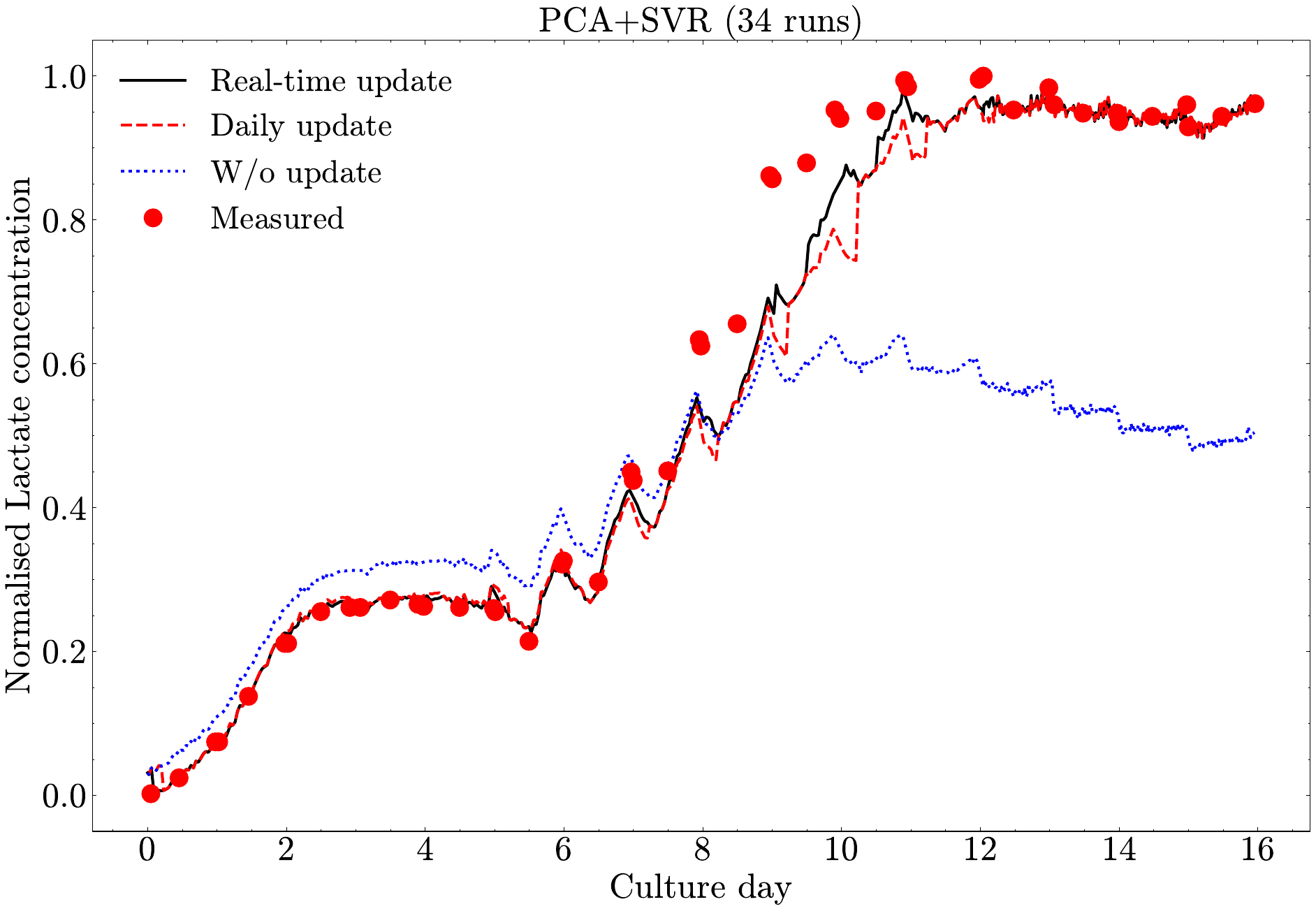}}
\end{subfloat}
\caption{The important roles of updating ML models for predictions of \textit{Lactate} in case study \textbf{2} with Cell Line \textbf{A}.} \label{model_update_role_cs2a_lactate}
\end{figure}

Fig. \ref{model_update_role_cs2b} presents the prediction performance and real-time glucose concentration monitoring results of the recursive PLSR model, the JITL model using PLSR as a local model, and the PLSR model pretrained on all 34 historical runs, both with and without updates, for the bioreactor run using cell line B. The outcomes for prediction and real-time monitoring of lactate and ammonium concentrations (cell line B) are presented in Figs. \url{S3} and \url{S4} in the Supplemental materials. Similarly, Fig. \ref{model_update_role_cs2a_lactate} illustrates lactate prediction results for the online SVR model, the JITL model employing PCA+SVR as a local model, and the PCA+SVR model pretrained on all 34 historical runs, with and without updates, for the bioreactor run with cell line A. Additional results for glucose and ammonium (cell line A) are provided in Figs. \url{S5} and \url{S6} in the supplemental materials.

In contrast to case study 1, pretrained models based on 34 historical runs without updates were unable to derive good predictive performance in this case. All machine learning models without updating failed to generalise metabolite concentration predictions for bioreactor runs conducted under experimental conditions distinct from those in the training data. However, the implementation of updates significantly enhanced prediction accuracy across all ML methods compared to the absence of updates. Incorporating newly acquired offline analytical values from the current run into the models or data library enabled the deployed ML agents to quickly adapt to the dynamic changes in cellular behaviors, including nutrient consumption and secondary metabolite production. These findings underscore the paramount importance of updating deployed ML agents with the latest offline analytical measurements to maintain robust monitoring performance for bioreactor runs conducted under experimental conditions that differ from the historical training data.

\begin{table} [!ht]
\centering
\caption{Performance of various ML models for case study 2 (\textbf{Cell Line A}).}\label{case_study_2a_table}
\resizebox{\textwidth}{!}{
\begin{tabular}{llccccccccc}
\toprule
\textbf{} &
  \textbf{} &
  \multicolumn{2}{c}{\textbf{Glucose}} &
  \multicolumn{2}{c}{\textbf{Lactate}} &
  \multicolumn{2}{c}{\textbf{Ammonium}} \\ 
  \cmidrule(lr){3-8}
\textbf{Model} &
  \textbf{Criteria} &
  \textbf{NMAE (\%)} &
  \textbf{$\mathbf{R}^2$ (\%)} &
  \textbf{NMAE (\%)} &
  \textbf{$\mathbf{R}^2$ (\%)} &
  \textbf{NMAE (\%)} &
  \textbf{$\mathbf{R}^2$ (\%)} \\ 
\midrule
\multirow{3}{*}{\textbf{\makecell[l]{RPLSR \\ (Pretrain: 1 run)}}} &
  \cellcolor[HTML]{D9EAD3}Daily update &
  \cellcolor[HTML]{D9EAD3} 6.86&
  \cellcolor[HTML]{D9EAD3} 86.72&
  \cellcolor[HTML]{D9EAD3} 3.57&
  \cellcolor[HTML]{D9EAD3} 98.11&
  \cellcolor[HTML]{D9EAD3} 6.02&
  \cellcolor[HTML]{D9EAD3} 90.94\\
 &
  \cellcolor[HTML]{FDE9D9}Real-time update &
  \cellcolor[HTML]{FDE9D9} 6.4&
  \cellcolor[HTML]{FDE9D9} 87.9&
  \cellcolor[HTML]{FDE9D9} 2.49&
  \cellcolor[HTML]{FDE9D9} 99.09&
  \cellcolor[HTML]{FDE9D9} 4.57&
  \cellcolor[HTML]{FDE9D9} 94.19\\
 &
  No update &
  11.60 &
  61.59 &
  22.60 &
  26.40 &
  30.67 &
  -59.22 \\ \midrule
\multirow{3}{*}{\textbf{\makecell[l]{OSVR \\ (Pretrain: 1 run)}}} &
  \cellcolor[HTML]{D9EAD3}Daily update &
  \cellcolor[HTML]{D9EAD3} 5.52 &
  \cellcolor[HTML]{D9EAD3} 90.20 &
  \cellcolor[HTML]{D9EAD3} 5.16 &
  \cellcolor[HTML]{D9EAD3} 96.60 &
  \cellcolor[HTML]{D9EAD3} 9.16 &
  \cellcolor[HTML]{D9EAD3} 73.49 \\
 &
  \cellcolor[HTML]{FDE9D9}Real-time update &
  \cellcolor[HTML]{FDE9D9} \textbf{5.06} &
  \cellcolor[HTML]{FDE9D9} \textbf{91.49} &
  \cellcolor[HTML]{FDE9D9} 3.69 &
  \cellcolor[HTML]{FDE9D9} 98.03 &
  \cellcolor[HTML]{FDE9D9} 7.90 &
  \cellcolor[HTML]{FDE9D9} 78.71 \\
 &
  No update &
  9.41 &
  75.45 &
  47.17 &
  -186.69 &
  60.88 &
  -547.55 \\ \midrule
\multirow{3}{*}{\textbf{\makecell[l]{JITL \\ (Data library: 34 runs)}}} &
  \cellcolor[HTML]{D9EAD3}Daily update data &
  \cellcolor[HTML]{D9EAD3} 5.56 &
  \cellcolor[HTML]{D9EAD3} 89.02 &
  \cellcolor[HTML]{D9EAD3} 2.84 &
  \cellcolor[HTML]{D9EAD3} 98.75 &
  \cellcolor[HTML]{D9EAD3} 4.22 &
  \cellcolor[HTML]{D9EAD3} 94.60 \\
 &
  \cellcolor[HTML]{FDE9D9}Real-time update data &
  \cellcolor[HTML]{FDE9D9} 5.46 &
  \cellcolor[HTML]{FDE9D9} 88.72 &
  \cellcolor[HTML]{FDE9D9} \textbf{2.20} &
  \cellcolor[HTML]{FDE9D9} \textbf{99.25} &
  \cellcolor[HTML]{FDE9D9} \textbf{3.95} &
  \cellcolor[HTML]{FDE9D9} \textbf{95.75}\\
 &
  No update data &
  18.18 &
  15.90 &
  34.98 &
  -77.01 &
  32.77 &
  -174.73 \\ \midrule
\multirow{3}{*}{\textbf{\makecell[l]{Retraining \\ (Pretrain: 34 runs)}}} &
  \cellcolor[HTML]{D9EAD3}Daily retraining &
  \cellcolor[HTML]{D9EAD3} 7.73 &
  \cellcolor[HTML]{D9EAD3} 78.03 &
  \cellcolor[HTML]{D9EAD3} 4.27 &
  \cellcolor[HTML]{D9EAD3} 95.92 &
  \cellcolor[HTML]{D9EAD3} 5.92 &
  \cellcolor[HTML]{D9EAD3} 91.67\\
 &
  \cellcolor[HTML]{FDE9D9}Real-time retraining &
  \cellcolor[HTML]{FDE9D9} 7.21 &
  \cellcolor[HTML]{FDE9D9} 81.78 &
  \cellcolor[HTML]{FDE9D9} 3.47 &
  \cellcolor[HTML]{FDE9D9} 97.28 &
  \cellcolor[HTML]{FDE9D9} 4.19 &
  \cellcolor[HTML]{FDE9D9} 95.63 \\
 &
  No retraining &
  20.78 &
  -6.64 &
  20.05 &
  44.56 &
  37.35 &
  -196.46 \\ 
\bottomrule
\end{tabular}
}
\end{table}

\begin{table}[!ht]
\centering
\caption{Performance of various ML models for case study 2 (\textbf{Cell Line B}).}\label{case_study_2b_table}
\resizebox{\textwidth}{!}{
\begin{tabular}{llccccccccc}
\toprule
\textbf{} &
  \textbf{} &
  \multicolumn{2}{c}{\textbf{Glucose}} &
  \multicolumn{2}{c}{\textbf{Lactate}} &
  \multicolumn{2}{c}{\textbf{Ammonium}} \\ 
  \cmidrule(lr){3-8}
\textbf{Model} &
  \textbf{Criteria} &
  \textbf{NMAE (\%)} &
  \textbf{$\mathbf{R}^2$ (\%)} &
  \textbf{NMAE (\%)} &
  \textbf{$\mathbf{R}^2$ (\%)} &
  \textbf{NMAE (\%)} &
  \textbf{$\mathbf{R}^2$ (\%)} \\ 
\midrule
\multirow{3}{*}{\textbf{\makecell[l]{RPLSR \\ (Pretrain: 1 run)}}} &
  \cellcolor[HTML]{D9EAD3}Daily update &
  \cellcolor[HTML]{D9EAD3} 8.12 &
  \cellcolor[HTML]{D9EAD3} 87.30 &
  \cellcolor[HTML]{D9EAD3} 5.78 &
  \cellcolor[HTML]{D9EAD3} 95.66 &
  \cellcolor[HTML]{D9EAD3} 11.65 &
  \cellcolor[HTML]{D9EAD3} 60.78 \\
 &
  \cellcolor[HTML]{FDE9D9}Real-time update &
  \cellcolor[HTML]{FDE9D9} 6.58 &
  \cellcolor[HTML]{FDE9D9} 91.14 &
  \cellcolor[HTML]{FDE9D9} \textbf{4.10} &
  \cellcolor[HTML]{FDE9D9} 97.27 &
  \cellcolor[HTML]{FDE9D9} 8.36 &
  \cellcolor[HTML]{FDE9D9} 75.79 \\
 &
  No update &
  11.46 &
  76.74 &
  26.95 &
  21.04 &
  31.55 &
  -115.23 \\ \midrule
\multirow{3}{*}{\textbf{\makecell[l]{OSVR \\ (Pretrain: 1 run)}}} &
  \cellcolor[HTML]{D9EAD3}Daily update &
  \cellcolor[HTML]{D9EAD3} 6.14 &
  \cellcolor[HTML]{D9EAD3} 92.62 &
  \cellcolor[HTML]{D9EAD3} 7.65 &
  \cellcolor[HTML]{D9EAD3} 92.07 &
  \cellcolor[HTML]{D9EAD3} 15.33 &
  \cellcolor[HTML]{D9EAD3} 28.53 \\
 &
  \cellcolor[HTML]{FDE9D9}Real-time update &
  \cellcolor[HTML]{FDE9D9} 4.99 &
  \cellcolor[HTML]{FDE9D9} \textbf{94.75} &
  \cellcolor[HTML]{FDE9D9} 6.06 &
  \cellcolor[HTML]{FDE9D9} 95.18 &
  \cellcolor[HTML]{FDE9D9} 12.68 &
  \cellcolor[HTML]{FDE9D9} 39.98 \\
 &
  No update &
  9.08 &
  84.95 &
  24.03 &
  26.12 &
  90.87 &
  -1265.64 \\ \midrule
\multirow{3}{*}{\textbf{\makecell[l]{JITL \\ (Data library: 34 runs)}}} &
  \cellcolor[HTML]{D9EAD3}Daily update data &
  \cellcolor[HTML]{D9EAD3} 5.73 &
  \cellcolor[HTML]{D9EAD3} 92.04 &
  \cellcolor[HTML]{D9EAD3} 5.42 &
  \cellcolor[HTML]{D9EAD3} 95.87 &
  \cellcolor[HTML]{D9EAD3} 9.04 &
  \cellcolor[HTML]{D9EAD3} 73.24 \\
 &
  \cellcolor[HTML]{FDE9D9}Real-time update data &
  \cellcolor[HTML]{FDE9D9} \textbf{4.75}&
  \cellcolor[HTML]{FDE9D9} 94.20&
  \cellcolor[HTML]{FDE9D9} \textbf{4.10} &
  \cellcolor[HTML]{FDE9D9} 97.30 &
  \cellcolor[HTML]{FDE9D9} \textbf{7.87} &
  \cellcolor[HTML]{FDE9D9} \textbf{79.86}\\
 &
  No update data &
  13.19 &
  74.39 &
  22.55 &
  35.69 &
  26.11 &
  -52.65 \\ \midrule
\multirow{3}{*}{\textbf{\makecell[l]{Retraining \\ (Pretrain: 34 runs)}}} &
  \cellcolor[HTML]{D9EAD3}Daily retraining &
  \cellcolor[HTML]{D9EAD3} 8.73 &
  \cellcolor[HTML]{D9EAD3} 80.00 &
  \cellcolor[HTML]{D9EAD3} 5.34 &
  \cellcolor[HTML]{D9EAD3} 96.26 &
  \cellcolor[HTML]{D9EAD3} 12.12 &
  \cellcolor[HTML]{D9EAD3} 64.70 \\
 &
  \cellcolor[HTML]{FDE9D9}Real-time retraining &
  \cellcolor[HTML]{FDE9D9} 6.48 &
  \cellcolor[HTML]{FDE9D9} 90.42 &
  \cellcolor[HTML]{FDE9D9} 4.38 &
  \cellcolor[HTML]{FDE9D9} \textbf{97.69} &
  \cellcolor[HTML]{FDE9D9} 9.13 &
  \cellcolor[HTML]{FDE9D9} 79.13 \\
 &
  No retraining &
  17.33 &
  44.71 &
  23.09 &
  49.61 &
  47.26 &
  -294.95 \\ 
\bottomrule
\end{tabular}
}
\end{table}

The performance of various ML models, with and without updating mechanisms, for different metabolite measurements is summarised in Table \ref{case_study_2a_table} for cell line A and Table \ref{case_study_2b_table} for cell line B. Detailed results of predicted values and real-time monitoring outcomes for glucose and lactate concentrations using ML models with real-time updates are illustrated in Fig. \ref{all_off_results_cs2}. Additional results can be found in the supplemental materials (Fig. \url{S7}). Notably, the real-time monitoring results demonstrate that online ML models and JITL models, trained on approximately 30 samples and updated with newly acquired offline measurements during the run, consistently outperformed models retrained on all 34 historical runs combined with the new offline analytical measurements. This discrepancy is likely due to the slower adaptation of ML models trained on the entire historical dataset to the new experimental conditions, particularly when the base and feed medium compositions in the current run differ entirely from those in historical runs. These findings emphasise the importance of training ML models on a smaller subset of samples that are most relevant to the new experimental conditions. Such an approach allows the models to rapidly adapt and accurately capture emerging trends in nutrient consumption and secondary metabolite production, thereby enhancing their applicability in bioreactors operating under novel working contexts.

\begin{figure}
\centering
\begin{subfloat}[Glucose prediction (Cell line A) \label{glucose_cs2a}]{
\includegraphics[width=0.33\textwidth]{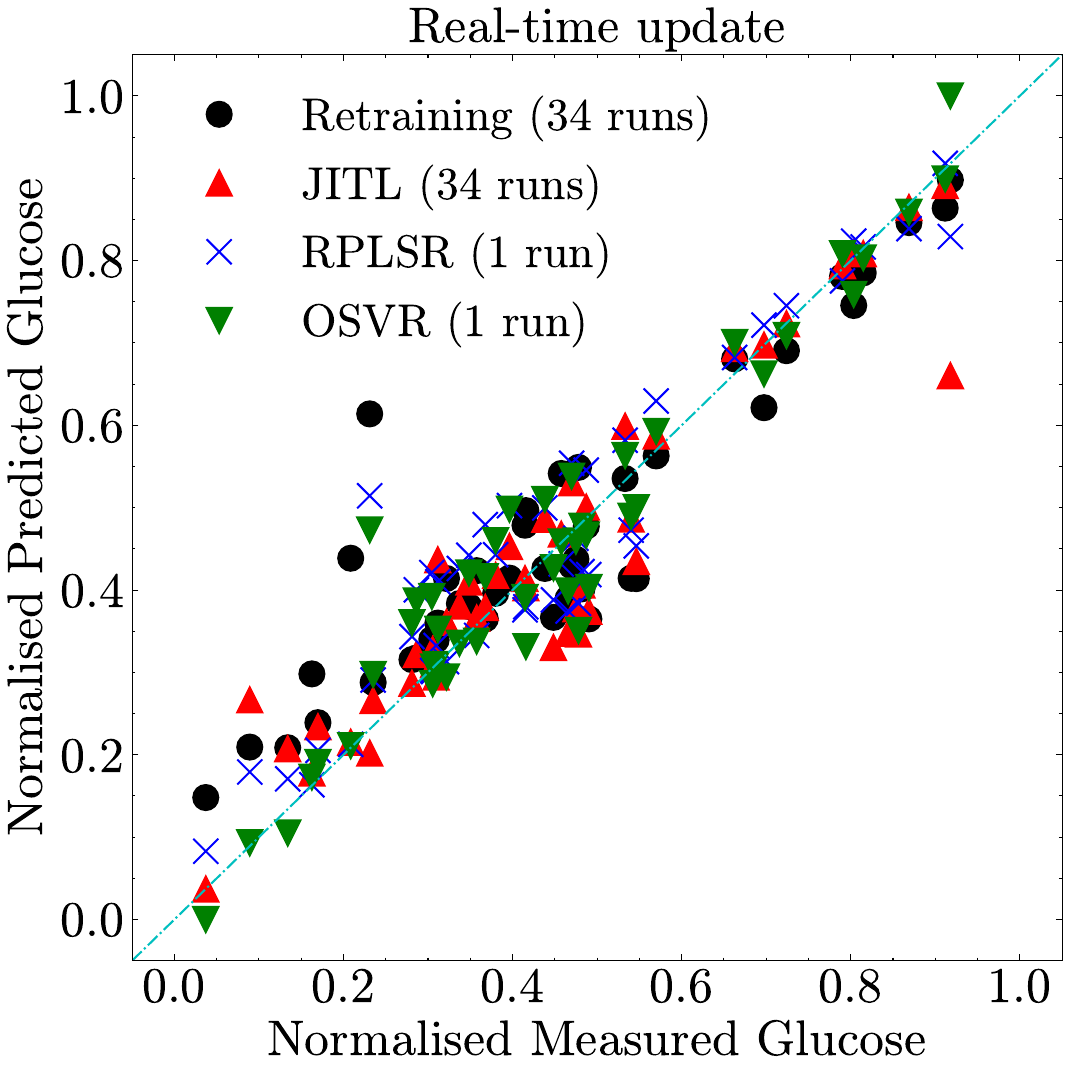}}
\end{subfloat}
\begin{subfloat}[Real-time Glucose monitoring (Cell line A) \label{glucose_monitoring_cs2a}]{
\includegraphics[width=0.6\textwidth, height=0.33\textwidth]{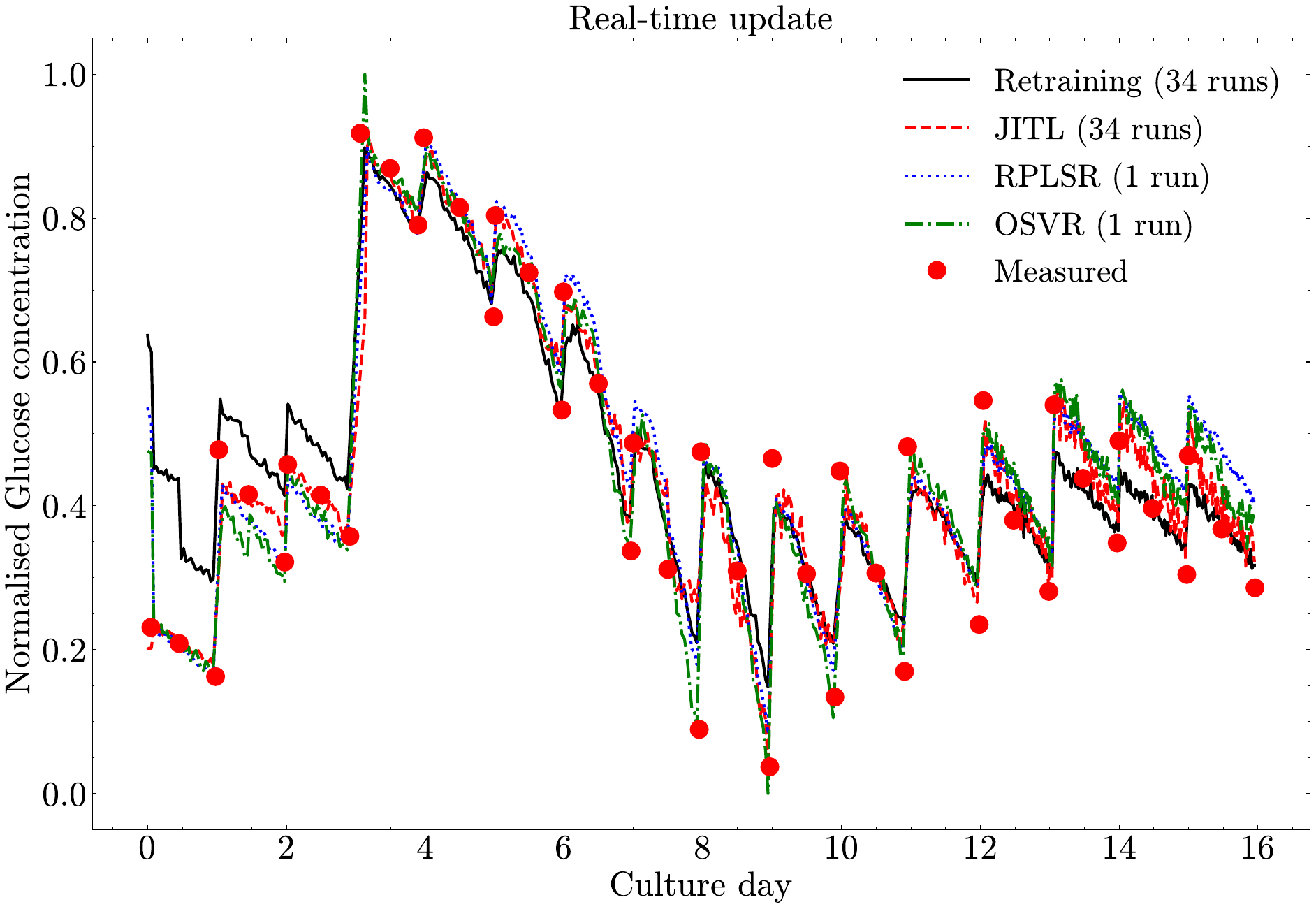}}
\end{subfloat}
\begin{subfloat}[Lactate prediction (Cell line A) \label{lactate_cs2a}]{
\includegraphics[width=0.33\textwidth]{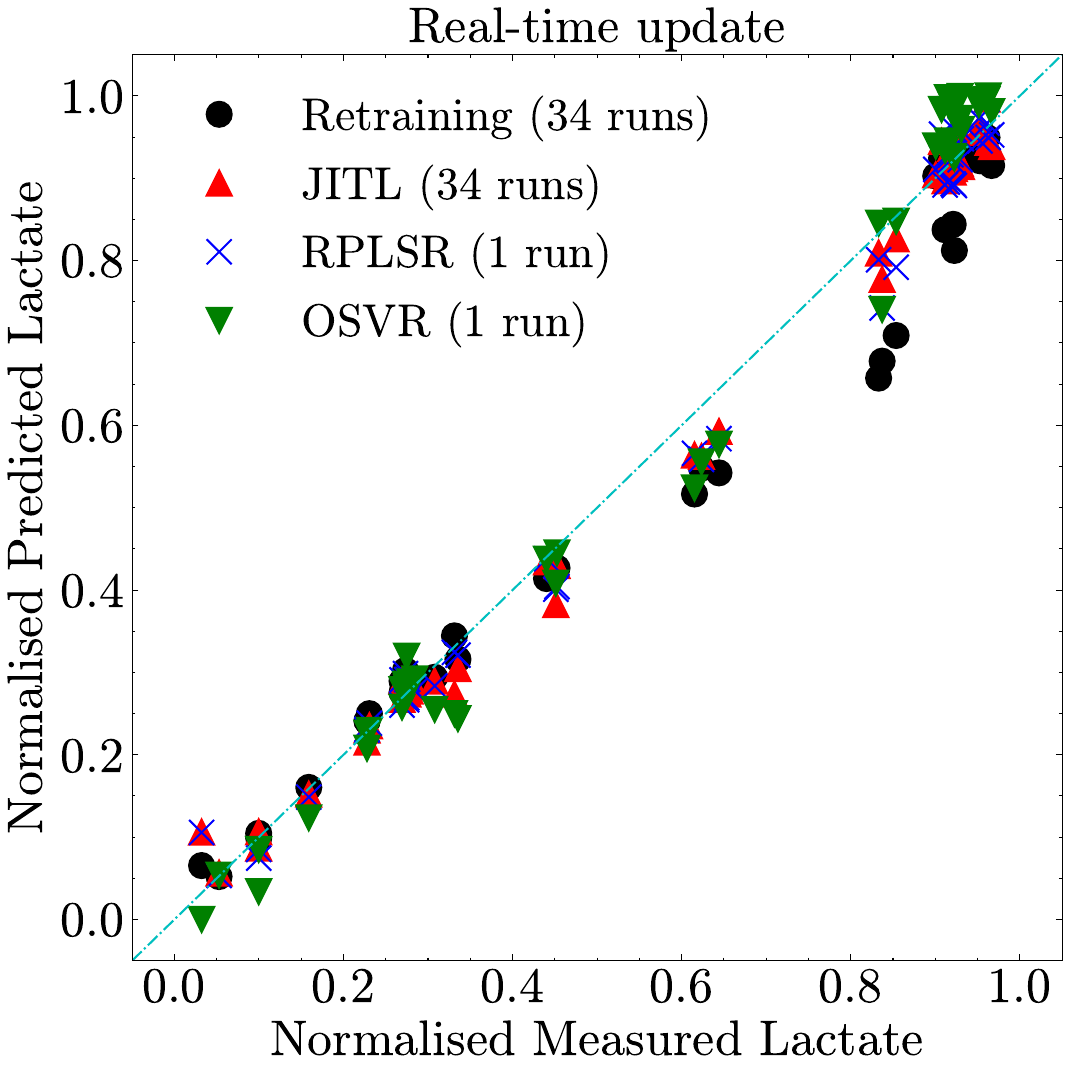}}
\end{subfloat}
\begin{subfloat}[Real-time Lactate monitoring (Cell line A)\label{lactate_monitoring_cs2a}]{
\includegraphics[width=0.6\textwidth, height=0.33\textwidth]{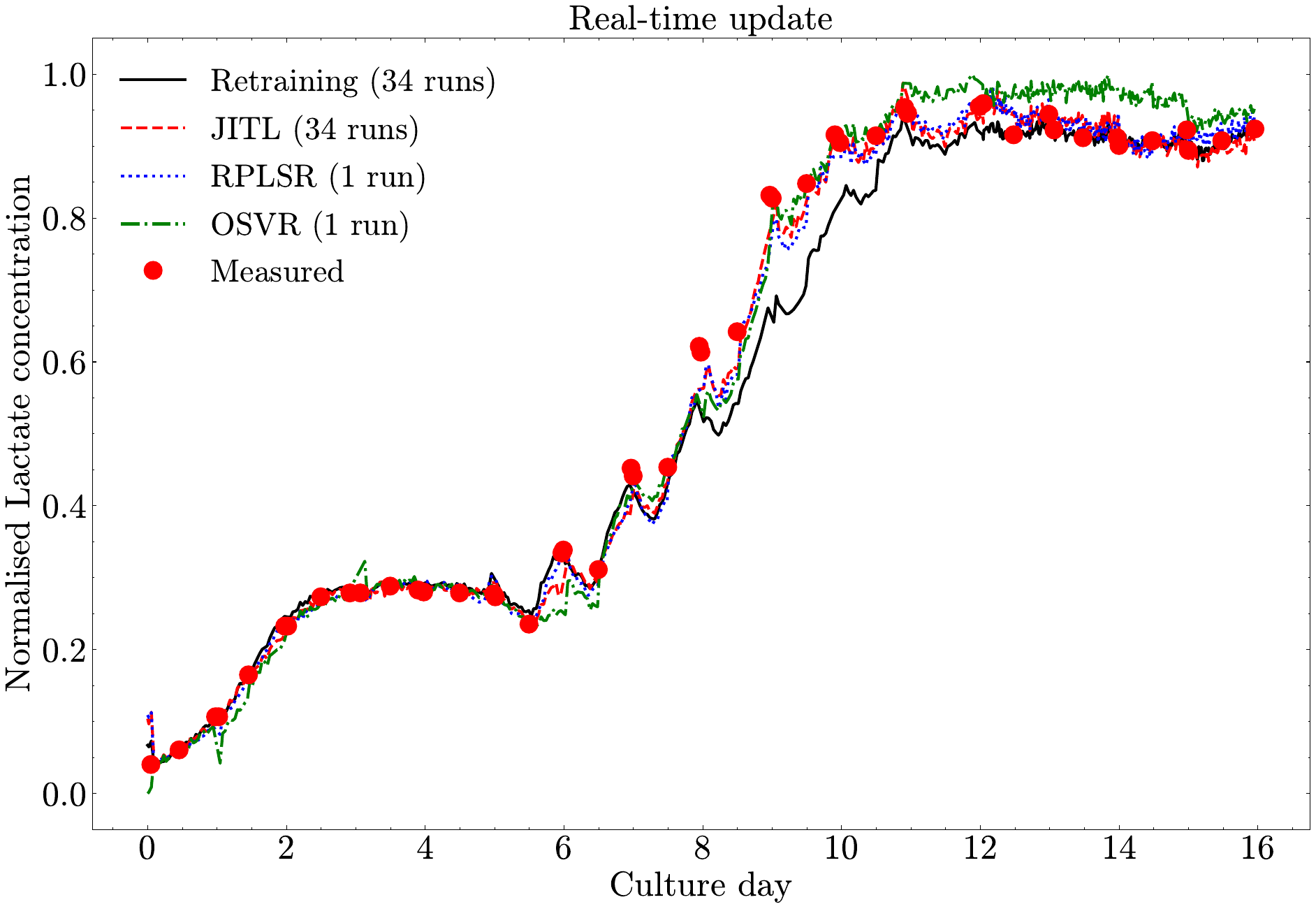}}
\end{subfloat}
\begin{subfloat}[Lactate prediction (Cell line B) \label{lactate_cs2b}]{
\includegraphics[width=0.33\textwidth]{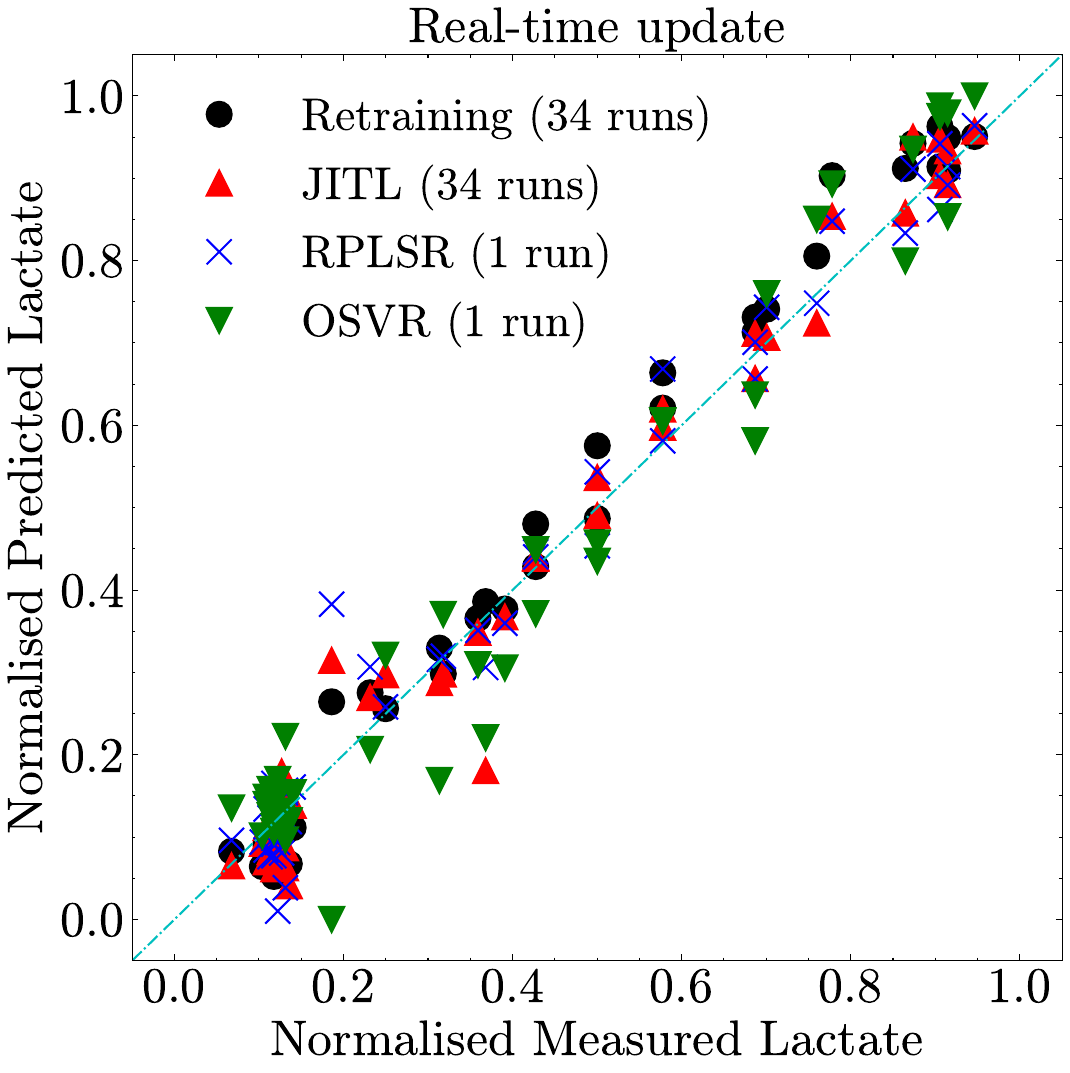}}
\end{subfloat}
\begin{subfloat}[Real-time Lactate monitoring (Cell line B)\label{lactate_monitoring_cs2b}]{
\includegraphics[width=0.6\textwidth, height=0.33\textwidth]{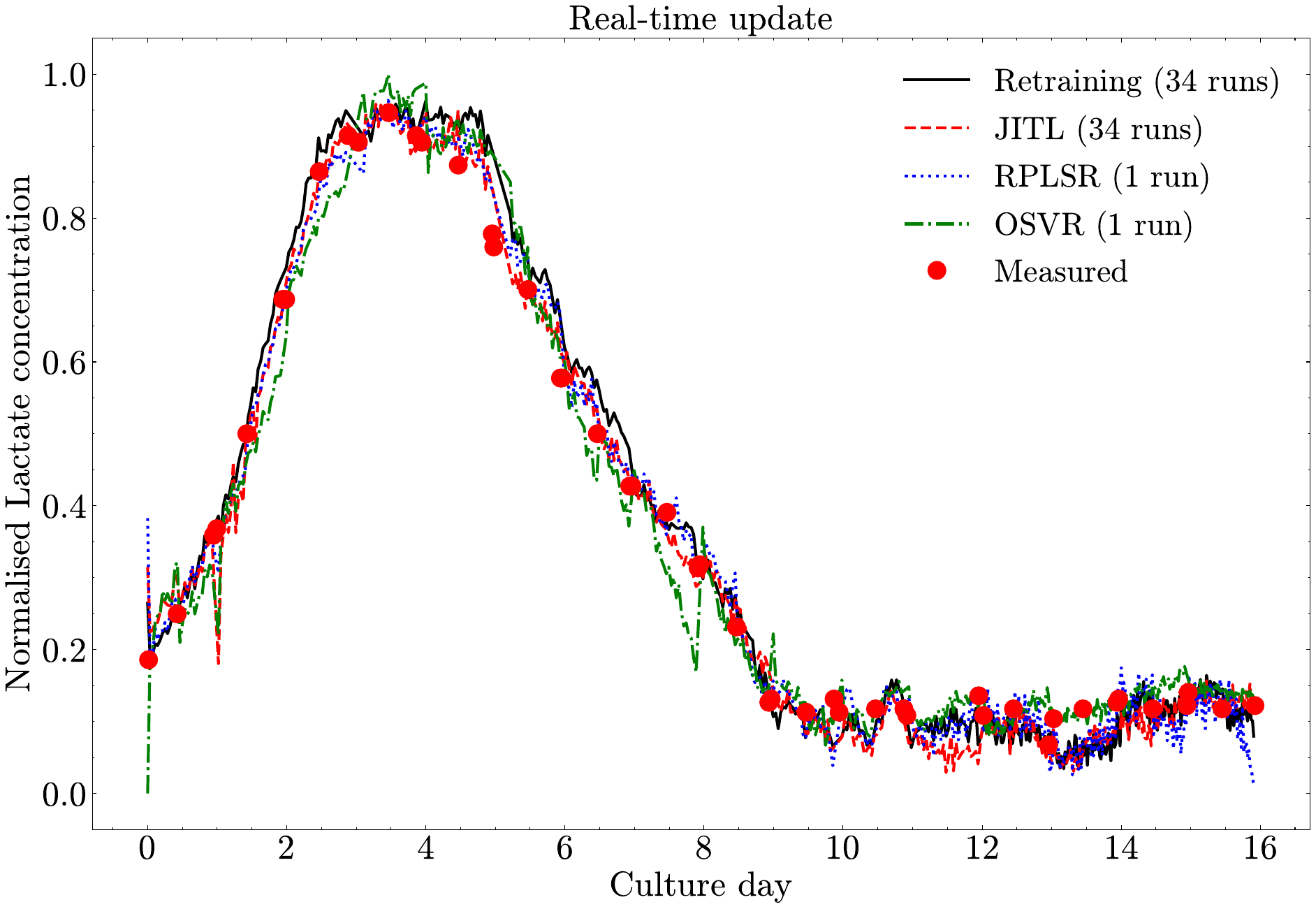}}
\end{subfloat}
\caption{Performance of all ML models with real-time update for different metabolite measurements in case study \textbf{2}.} \label{all_off_results_cs2}
\end{figure}

For glucose predictions, the OSVR model with real-time updates generated the best predictive performance for cell line A and also achieved the highest $\mathbf{R}^2$ value for cell line B. The JITL model, which adopted PLSR as a local model and employed real-time updates from the existing data library, exhibited the highest predictive accuracy (measured by NMAE) for the bioreactor with cell line B. Additionally, it achieved competitive performance for cell line A, with an NMAE of 5.46\%, slightly below that of the OSVR model. Overall, the OSVR model outperformed other machine learning approaches, including recursive PLSR, JITL models employing PLSR as a local model, and PLSR models pretrained on all 34 runs and subsequently retrained during operation. This superior performance is likely attributable to the non-linear relationship between Raman signals and glucose concentrations in this case study, which favors the use of non-linear models like OSVR over linear models such as PLSR.

For lactate predictions, JITL using PCA+SVR as a local model with real-time updates achieved the best performance for both cell lines A and B, closely aligned with the underlying process changes. RPLSR with real-time updates achieved the second best place and competitive with the real-time updating JILT models in both two cell lines. The retraining of PCA+SVR models on the training dataset including all 34 historical runs and newly collected lactate concentration during current run also achieved good results, with $\mathbf{R}^2$ values above 95\%, though slightly behind JITL and online learning models. However, all ML models without updates performed poorly which were confirmed by negative values or low values of $\mathbf{R}^2$.

For lactate predictions, the JITL model employing PCA+SVR as a local model with real-time updates gave the best performance for both cell lines A and B, effectively capturing the underlying process dynamics. Recursive PLSR with real-time updates achieved the second-highest performance, showing competitive results compared to the JITL models with real-time updates for both cell lines. Additionally, retraining PCA+SVR models using a training dataset that incorporated all 34 historical runs along with newly collected lactate concentrations during the current run also produced strong results, with $\mathbf{R}^2$ values exceeding 95\%. However, the performance of these retrained models remained slightly lower than that of the JITL and online learning models. In contrast, all machine learning models without updates exhibited poor performance, as evidenced by negative or low $\mathbf{R}^2$ values, underscoring the critical importance of real-time updates for maintaining predictive accuracy.

For ammonium predictions, the JITL model using the KPCA+SVR as a local model with real-time updates showed the highest accuracy for both cell lines A and B. This result highlights that models trained on a subset of Raman spectra most similar to the input Raman spectrum generally derive superior predictive performance. Real-time retraining of the KPCA+SVR models using data from all 34 historical runs combined with newly acquired offline measurements from the current run resulted in a better predictive accuracy compared to online ML models pretrained on a single historical run. This improvement can be attributed to the complex and nonlinear relationships between Raman signals and ammonium concentrations, which are challenging to capture with models trained on limited datasets.

Overall, across all biochemical markers, real-time updates consistently delivered superior predictive performance, underscoring the crucial role of dynamic model adaptation in addressing domain shifts. Machine learning models using daily updates illustrated slightly lower performance than those with real-time updates but remained effective. In contrast, models without updates exhibited poor performance, characterised by high error rates. Among the tested approaches, JITL with real-time updates proved to be the most robust, delivering the highest predictive accuracy across all markers. Although RPLSR and OSVR also performed well with updates, their performance dropped significantly in the absence of updates, highlighting their dependence on frequent learning. While retraining approaches produced satisfactory results, their computational demands and time requirements may limit their feasibility for real-time applications. For practical real-time scenarios, JITL and online ML models with updates emerge as the most suitable options. Furthermore, the substantial domain shifts observed in this case study necessitated frequent model updates to maintain predictive accuracy. Models without updates were unable to generalise to new operating conditions, rendering them ineffective for such applications.

\section{Results of the mixture-of-experts through two case studies} \label{results_moe}
In this study, four machine learning agents including RPLSR, OSVR, JITL, and pretrained models were implemented for real-time monitoring of glucose, lactate, and ammonium concentrations using Raman spectral data as input. The results, presented in Section \ref{results_4_agents}, demonstrate that for the same input Raman spectrum, some models tend to underpredict while others overpredict the target concentrations. This observation motivated the development of a mixture-of-experts approach, where predictions from all four models are aggregated through averaging. This section evaluates the empirical performance of the proposed mixture-of-experts solution in comparison to the individual learning agents.

\subsection{Case study 1}
 
\begin{table}[!ht]
\centering
\caption{Performance of the mixture-of-experts model and individual ML agents using real-time update for case study 1.}\label{moe_cs1_table}
\resizebox{\textwidth}{!}{
\begin{tabular}{L{5cm}ccccccccc}
\toprule
\textbf{} &
  \multicolumn{2}{c}{\textbf{Glucose}} &
  \multicolumn{2}{c}{\textbf{Lactate}} &
  \multicolumn{2}{c}{\textbf{Ammonium}} \\ 
  \cmidrule(lr){2-7}
\textbf{Model} &
  \textbf{NMAE (\%)} &
  \textbf{$\mathbf{R}^2$ (\%)} &
  \textbf{NMAE (\%)} &
  \textbf{$\mathbf{R}^2$ (\%)} &
  \textbf{NMAE (\%)} &
  \textbf{$\mathbf{R}^2$ (\%)} \\ 
\midrule
  \textbf{RPLSR} (Pretrain: 1 run) &
  2.17 &
  97.04 &
  5.59 &
  92.27 &
  11.44 &
  66.39 \\
\midrule
  \textbf{OSVR} (Pretrain: 1 run) &
  3.01 &
  94.34 &
  6.34 &
  92.09 &
  12.22 &
  43.04 \\
\midrule
  \textbf{JITL} (Data library: 34 runs) &
  2.22 &
  95.20 &
  3.59 &
  97.80 &
  9.66 &
  68.49 \\
\midrule
\textbf{Retraining} (Pretrain: 34 runs) &
  \textbf{1.98} &
  \textbf{98.86} &
  \textbf{2.60} &
  \textbf{98.72} &
  9.81 &
  67.90 \\ 
\midrule
  \cellcolor[HTML]{FDE9D9}\textbf{Mixture-of-experts} &
  \cellcolor[HTML]{FDE9D9}1.99 &
  \cellcolor[HTML]{FDE9D9}96.93 &
  \cellcolor[HTML]{FDE9D9}3.37 &
  \cellcolor[HTML]{FDE9D9}98.07 &
  \cellcolor[HTML]{FDE9D9}\textbf{7.79} &
  \cellcolor[HTML]{FDE9D9}\textbf{82.26} \\ 
\bottomrule
\end{tabular}
}
\end{table}

\begin{figure}[!ht]
\centering
\begin{subfloat}[Lactate \label{moe_realtime_monitoring_lactate_cs1}]{
\includegraphics[width=0.55\textwidth]{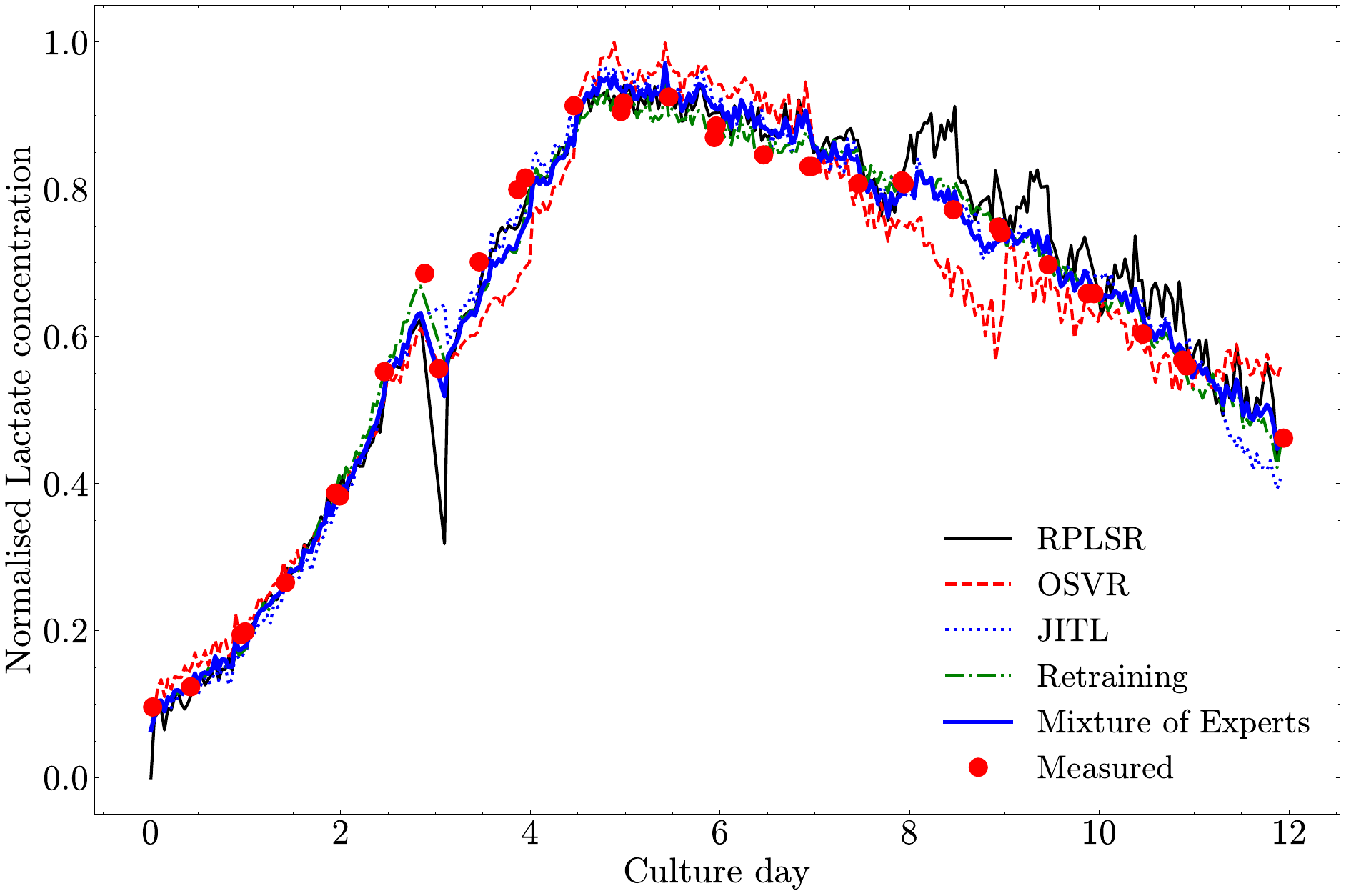}}
\end{subfloat}
\begin{subfloat}[Ammonium \label{moe_realtime_monitoring_ammonium_cs1}]{
\includegraphics[width=0.55\textwidth]{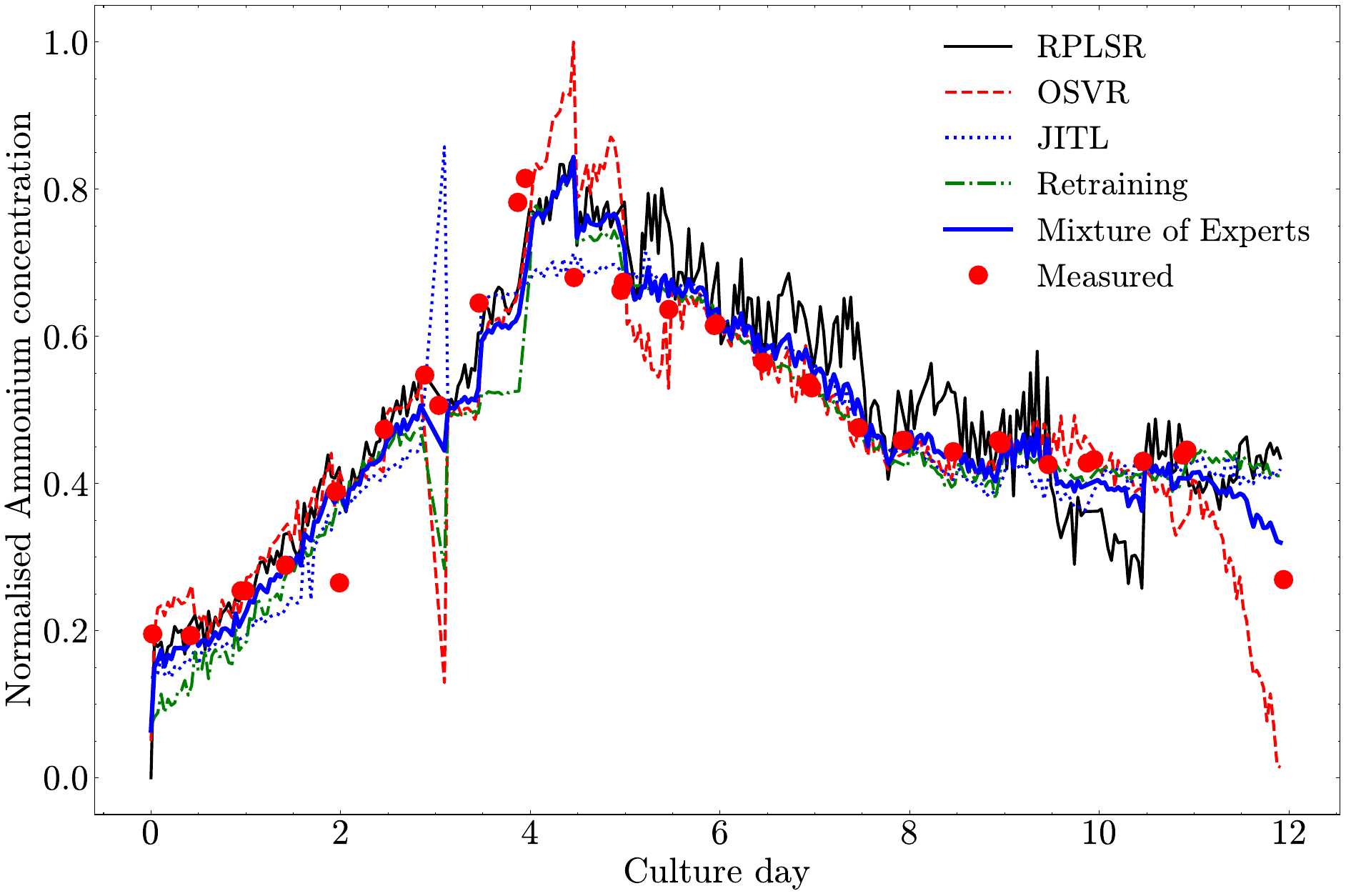}}
\end{subfloat}
\caption{The outcomes of Mixture-of-experts models and other ML models with real-time update for real-time monitoring of \textit{Lactate} and \textit{Ammonium} in case study \textbf{1}.} \label{moe_cs1}
\end{figure}

Table \ref{moe_cs1_table} presents the predictive accuracy of the mixture-of-experts model in comparison to individual machine learning agents with real-time updates. The real-time monitoring outcomes of these models are illustrated in Fig. \ref{moe_cs1} and Fig. \url{S8} (Supplemental document). In case study 1, the performance of the mixture-of-experts model for glucose and lactate prediction did not exceed that of the retrained models, which used a training dataset comprising all 34 historical runs combined with newly collected offline analytical measurements. However, its performance was comparable to the retrained models. This result can be attributed to the limited training data available to the online learning and JITL models, which were trained on only one run or a subset of 30 similar samples (in the case of the JITL model). These models exhibited considerably lower performance than the well-pretrained models trained on data from 34 runs. Consequently, the average performance of all four individual agents was inferior to that of retrained models developed using bioreactor data from the same feed medium compositions and feeding strategies as the testing bioreactor in this case study.

For ammonium prediction, while individual agents demonstrated relatively low performance, the mixture-of-experts model achieved satisfactory results. As shown in Fig. \ref{moe_realtime_monitoring_ammonium_cs1}, certain agents overpredicted while others underpredicted ammonium concentrations for the same Raman spectral input. The averaging of predictions from all individual agents mitigated these variations, deriving predictions closer to the actual values. Consequently, the mixture-of-experts model outperformed the individual agents in ammonium prediction. From Fig. \ref{moe_cs1}, it is evident that although the mixture-of-experts model may not outperform all individual agents across all cell culture process parameters, it provides a much smoother real-time predictive trend compared to individual learning agents.

\subsection{Case study 2}

\begin{figure}[!ht]
\centering
\begin{subfloat}[Glucose \label{moe_realtime_monitoring_glucose_cs2b}]{
\includegraphics[width=0.55\textwidth]{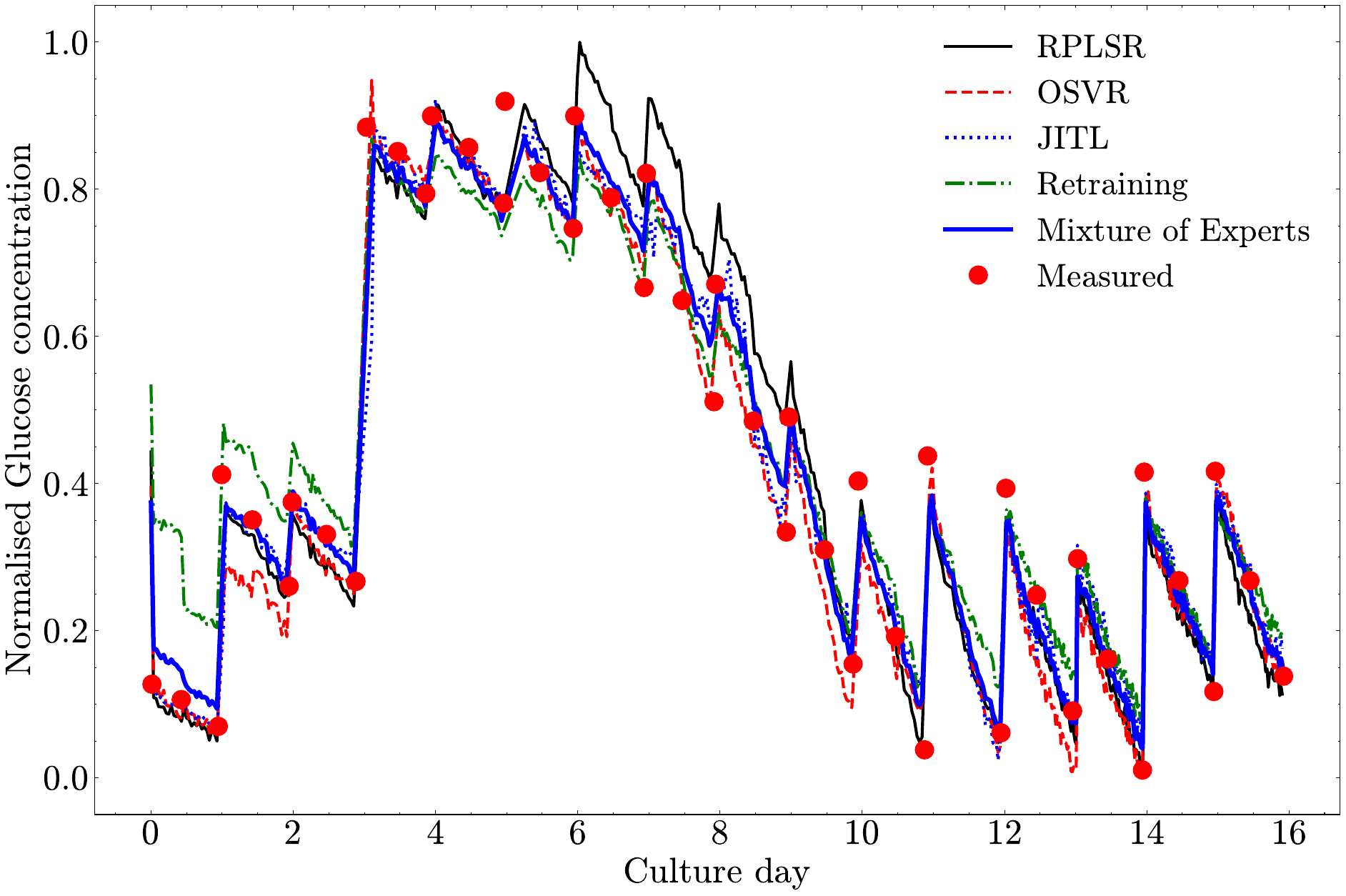}}
\end{subfloat}
\begin{subfloat}[Ammonium \label{moe_realtime_monitoring_ammonium_cs2b}]{
\includegraphics[width=0.55\textwidth]{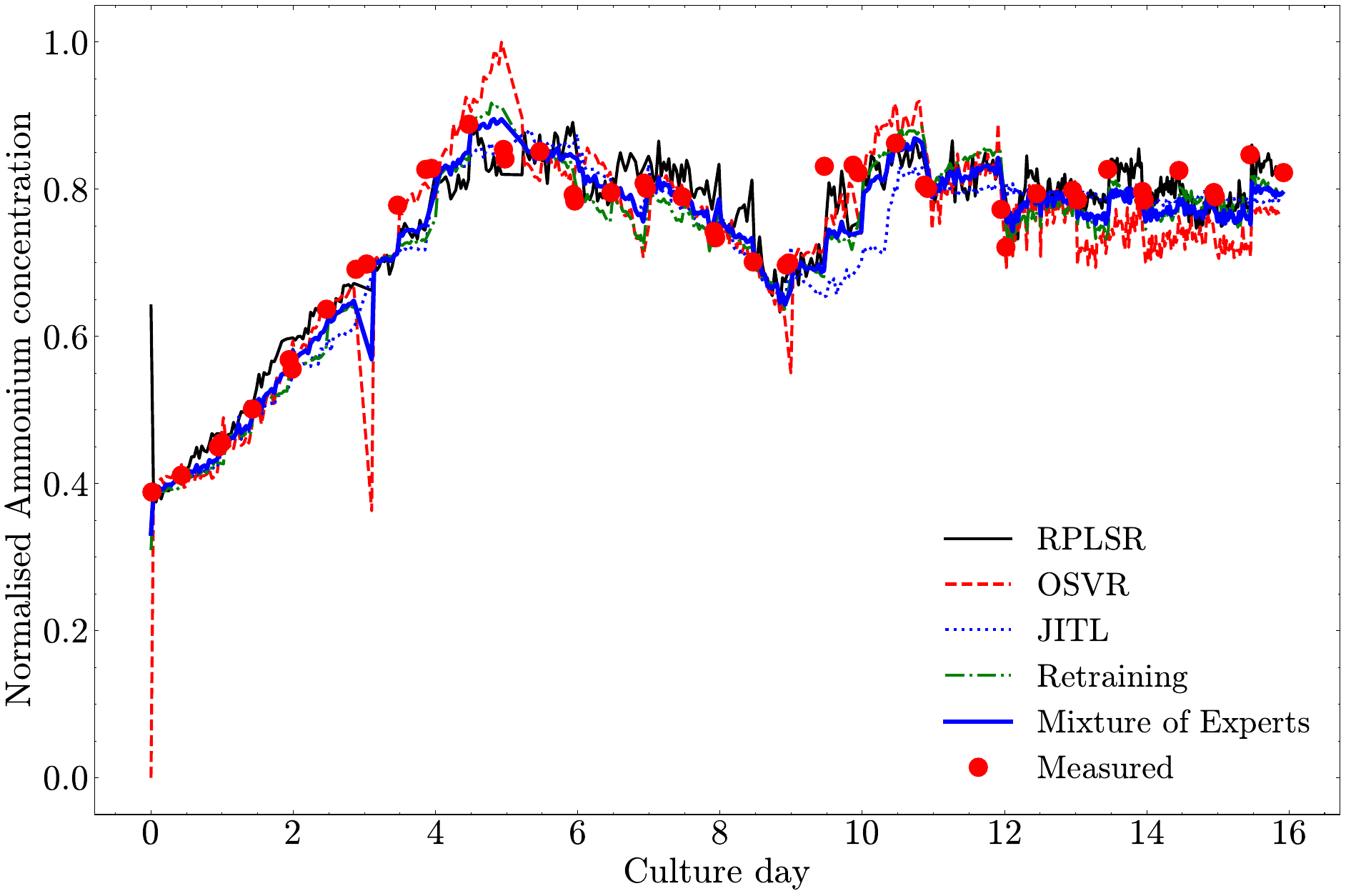}}
\end{subfloat}
\caption{The outcomes of Mixture-of-experts models and other ML models with real-time update for real-time monitoring of \textit{Glucose} and \textit{Ammonium} in case study \textbf{2} (Cell line B).} \label{moe_cs2b}
\end{figure}

Table \ref{moe_cs2b_table} presents the performance of the mixture-of-experts model and individual learning agents with real-time updates for predicting cell culture process indicators in the experimental bioreactor using cell line B. The real-time monitoring results for glucose and ammonium concentrations obtained from various machine learning models are shown in Fig. \ref{moe_cs2b} and Figs. \url{S9} and \url{S10} (Supplemental document). Unlike the findings from Case study 1, the mixture-of-experts model, which combines predictions from four individual learning agents updated or retrained whenever offline data becomes available, demonstrated superior performance across all three biochemical indicators. This was observed even under conditions involving entirely new culture feed medium compositions and feeding strategies. Furthermore, the real-time monitoring trends for all bioprocess indicators using the mixture-of-experts approach were significantly smoother compared to those produced by individual models. 

\begin{table}
\centering
\caption{Performance of the mixture-of-experts model and individual ML agents using real-time update for case study 2 (Cell line B).}\label{moe_cs2b_table}
\resizebox{\textwidth}{!}{
\begin{tabular}{L{5cm}ccccccccc}
\toprule
\textbf{} &
  \multicolumn{2}{c}{\textbf{Glucose}} &
  \multicolumn{2}{c}{\textbf{Lactate}} &
  \multicolumn{2}{c}{\textbf{Ammonium}} \\ 
  \cmidrule(lr){2-7}
\textbf{Model} &
  \textbf{NMAE (\%)} &
  \textbf{$\mathbf{R}^2$ (\%)} &
  \textbf{NMAE (\%)} &
  \textbf{$\mathbf{R}^2$ (\%)} &
  \textbf{NMAE (\%)} &
  \textbf{$\mathbf{R}^2$ (\%)} \\ 
\midrule
  \textbf{RPLSR} (Pretrain: 1 run) &
  6.58 &
  91.14 &
  4.1 &
  97.27 &
  8.36 &
  75.79 \\
\midrule
  \textbf{OSVR} (Pretrain: 1 run) &
  4.99 &
  94.75 &
  6.06 &
  95.18 &
  12.68 &
  39.98 \\
\midrule
  \textbf{JITL} (Data library: 34 runs) &
  4.75 &
  94.20 &
  4.1 &
  97.30 &
  7.87 &
  79.86\\
\midrule
\textbf{Retraining} (Pretrain: 34 runs) &
  6.48 &
  90.42 &
  4.38 &
  97.69 &
  9.13 &
  79.13\\ 
\midrule
  \cellcolor[HTML]{FDE9D9}\textbf{Mixture-of-experts} &
  \cellcolor[HTML]{FDE9D9}\textbf{4} &
  \cellcolor[HTML]{FDE9D9}\textbf{96.32} &
  \cellcolor[HTML]{FDE9D9}\textbf{3.36} &
  \cellcolor[HTML]{FDE9D9}\textbf{98.54} & 
  \cellcolor[HTML]{FDE9D9}\textbf{7.56} &
  \cellcolor[HTML]{FDE9D9}\textbf{84.14} \\ 
\bottomrule
\end{tabular}
}
\end{table}

As illustrated in Fig. \ref{moe_cs2b}, when cultivation conditions deviate from the historical training data, machine learning models retrained on a extensive dataset comprising all 34 historical runs and newly collected offline analytical measurements exhibited slower adaptation to emerging trends in cellular behaviors, such as nutrient consumption and secondary metabolite production. In contrast, online learning and JITL models, trained on smaller but more contextually relevant datasets, demonstrated a faster response to these changes in cultivation conditions. However, the rapid adaptation characteristic of single instance-based online learning models can sometimes lead to over-reliance on the most recent data point, resulting in increasing variability in model performance, as observed with the OSVR models for ammonium prediction. To address this, the mixture-of-experts approach averages the predictions of all adaptive models, effectively balancing their respective strengths and weaknesses. This integration enhances the robustness of real-time predictive performance under operational conditions that differ from the prior training data.    

\section{Discussions of lessons learned} \label{discussion}
As highlighted in \cite{khba24}, cell culture bioreactors are commonly monitored using machine learning models pretrained on diverse historical datasets encompassing various operational conditions, cell lines, and culture medium compositions. Once deployed, these models are typically not updated with newly collected offline analytical measurements from the current bioreactor. However, this deployment approach presents several limitations. Firstly, pretrained models built on global datasets often fail to deliver high performance across all operational conditions. This challenge arises due to the trade-off between robustness and accuracy: while increased variation in training datasets enhances the robustness of learning model to diverse conditions, it simultaneously reduces overall accuracy. Consequently, data from different cell lines, culture media, and cell lines are rarely combined into a single model without encountering significant performance issues \citep{tuwa20}. Secondly, pretrained models struggle to adapt to abrupt real-time process changes, such as new cultivation conditions, experimental setup modifications, or abnormal operations (e.g., nutrient pump failures). The inability to address these issues stems from the fixed structure and accuracy of global pretrained models, which are trained offline and lack the adaptability required for real-time process fluctuations \citep{tuwa20}.

To address these challenges, this study investigates various strategies to enable learning models to adapt and update in response to newly collected offline analytical measurements during the operational phase of a bioreactor. The goal is to enhance the performance of real-time monitoring of metabolite concentrations within the bioreactor using Raman spectral data as inputs. The proposed strategies include: (1) employing online machine learning agents capable of incremental learning from individual data points, (2) implementing just-in-time learning mechanisms to train local models for each incoming Raman spectral sample, and (3) retraining deployed models using a combination of historical datasets and newly collected offline measurements. The effectiveness of these methods was evaluated through two industrial case studies. The first case study used the same base and feed medium compositions and feeding strategies as those in the training data. However, an abnormal event caused a significant spike in nutrient concentrations. This case study aimed to evaluate the predictability and adaptability of learning models to changes in nutrient concentrations under consistent culture conditions. In the second case study, a completely new feed medium composition and feeding strategies were introduced to the bioreactor, differing from the training data. These changes indirectly altered cellular behaviours, such as nutrient consumption and secondary metabolite production, as reflected in the Raman spectra. The objective of this case study was to assess the predictive performance of pretrained models and the adaptability of learning mechanisms in dynamically changing environments. The results presented in Sections \ref{results_4_agents} and \ref{results_moe} provide critical insights into the deployment of machine learning agents for real-time monitoring of industrial-scale cell culture processes.

The findings from the first case study indicate that if new cell culture experiments use similar feeding strategies, base and feed medium compositions, and cell lines as those represented in the historical training data, well-pretrained models built on diverse datasets can still maintain high performance in real-time monitoring of glucose and lactate concentrations without the need for retraining. However, if performance improvements are required, retraining can be undertaken, albeit at the expense of additional time and computational resources. Furthermore, online learning models such as RPLSR, which are pretrained on data from a single bioreactor run, demonstrate the ability to quickly adapt to changes in cell growth behaviours, maintaining high predictive accuracy. For bioprocess indicators like ammonium concentrations, where the relationship between input Raman spectral features and the target variable is complex, the JITL models prove to be an effective choice. These models leverage the most relevant training samples for the current query pattern, offering more informative learning compared to using a large and diverse dataset.

In the second case study, a completely new feed culture medium composition and a novel feeding strategy with daily interventions were implemented, resulting in a significant high level of glucose concentrations in the bioreactor compared to the training data. These changes in feed medium materials significantly influenced the Raman spectral patterns, leading to shifts in data distribution and alterations in the mapping relationships between Raman spectra and target variables learned by the pretrained models. Additionally, variations in cell growth behaviours, such as nutrient consumption and secondary metabolite production driven by the new culture media and feeding operations, caused substantial deviations in the trends observed in the current bioreactor run compared to those in the extensive historical dataset. As a result, retraining the models on both the large historical dataset and a few newly collected values from the current run failed to effectively capture these novel cell growth behaviours. In contrast, online learning models pretrained on smaller datasets and JITL models, which construct new local models using the 30 most relevant samples for each input Raman query, demonstrated superior performance compared to the retrained models. The ability of JITL solution to build a local model for each new Raman query allowed it to adapt more effectively to evolving cell growth behaviors than online machine learning models relying on incremental learning from each new individual instance. In summary, JITL and online machine learning models are more suitable than retrained models for new experiments involving entirely novel feed culture media and feeding strategies.

The findings from both case studies suggest that updating the deployed ML agents with newly collected offline analytical measurements enables them to quickly adapt to changes in cell growth behaviours and cultivation conditions, which are common in biopharmaceutical manufacturing processes. Additionally, updating or retraining models when offline data becomes available demonstrates superior performance compared to fixed-time daily updates. Offline measurements in cell culture bioreactors are typically verified by well-trained operators to ensure accuracy, thereby minimizing the impact of outliers in these manufacturing practices. Consequently, the use of adaptive ML agents, leveraging Raman spectral data as inputs, is highly recommended for real-time monitoring of critical process indicators. 

Based on the results obtained from individual deployed learning agents, a mixture-of-experts approach was proposed to enhance predictive performance by averaging the outputs of all individual agents. Empirical evaluations demonstrated that this mixture-of-experts model, with real-time updating, provides more stable real-time predictions compared to individual models. In scenarios where a bioreactor run employs base and feed culture media distinct from the training data, the predictions from various learning models tend to vary. The mixture-of-experts approach mitigates this variation among individual models, thereby improving overall performance. For bioreactor runs employing the same base and feed culture media, as well as feeding strategies consistent with the training data, the mixture-of-experts approach ensures stable real-time predictions, although its performance may be marginally lower than that of a well-pretrained model updated during the cell culture process. Nevertheless, the mixture-of-experts solution excels in delivering smoother and more accurate real-time monitoring of metabolite concentration trends compared to individual learning agents.

\section{Conclusion} \label{conclusion}
This paper presented critical insights gained from the practical deployment of adaptive ML agents for real-time monitoring of cell culture process parameters using Raman spectral data as input features. These findings were validated through two industrial case studies representing commonly encountered scenarios in biomanufacturing. The first scenario involves cultivation conditions and experimental setups consistent with historical runs, but with incidents in feeding operations causing significant increases in nutrient concentrations. The second scenario introduces entirely new experimental setups and cultivation conditions to evaluate a novel cell culture process, differing from available historical data. These scenarios challenge the robustness and adaptability of the deployed ML agents, generating empirical results that provide valuable guidance for ML practitioners in selecting appropriate solutions for real-time bioreactor monitoring. As detailed in Section \ref{discussion}, adaptive ML methods, particularly just-in-time learning, demonstrate significant advantages in monitoring bioreactor processes under novel operational conditions. The pretrained models perform effectively in stable operating environments, but dynamic conditions necessitate more flexible retraining strategies. These findings offer practical recommendations for deploying ML in biomanufacturing, underscoring the importance of maintenance strategies depending on process variability.

Future research will focus on developing more sophisticated methods to integrate and balance the strengths of pretrained and adaptive models rather than using only simple mixture-of-experts solution. This includes leveraging advanced data augmentation techniques to enhance model adaptability and performance in dynamic bioprocessing environments. Additionally, the proposed approaches should be applied to monitor large-scale industrial cell culture bioreactors to evaluate the scalability and robustness of these solutions in real-world biomanufacturing processes. Furthermore, real-time predictive values generated by the models could provide useful information for the design and implementation of automated control strategies for feeding procedures, with the ultimate goal of maximizing product titer at the end of the upstream cell cultivation process.

\section*{CRediT authorship contribution statement}
\textbf{Thanh Tung Khuat:} Conceptualization, Methodology, Investigation, Software, Validation, Visualization, Writing – original draft. \textbf{Johnny Peng:} Methodology, Software, Validation, Visualization, Writing – review \& editing. \textbf{Robert Bassett:} Methodology, Investigation, Software, Data Curation, Writing – review \& editing. \textbf{Ellen Otte:} Conceptualization, Investigation, Validation, Supervision, Writing – review \& editing. \textbf{Bogdan Gabrys:} Conceptualization, Methodology, Investigation, Validation, Project administration, Funding acquisition, Writing – review \& editing. 

\section*{Declaration of Competing Interest}
Robert Bassett and Ellen Otte are employees of CSL Innovation Pty Ltd. Thanh Tung Khuat, Johnny Peng, and Bogdan Gabrys declare no competing interests, including no known competing financial interests or personal relationships that could have appeared to influence the work reported in this paper.

\section*{Acknowledgements}
This research was supported under the Australian Research Council's Industrial Transformation Research Program (ITRP) funding scheme (project number IH210100051). The ARC Digital Bioprocess Development Hub is a collaboration between The University of Melbourne, University of Technology Sydney, RMIT University, CSL Innovation Pty Ltd, Cytiva (Global Life Science Solutions Australia Pty Ltd) and Patheon Biologics Australia Pty Ltd.

\section*{Data availability}
The data used in this study is the intellectual property of CSL Limited and is therefore not shared publicly.

\section*{Declaration of generative AI and AI-assisted technologies in the writing process}
During the preparation of this work the authors used Copilot in order to improve the readability and language of the manuscript. After using this tool, the authors reviewed and edited the content as needed and take full responsibility for the content of the published article.

\bibliography{ref}

\end{document}